\documentclass[aps,pre,reprint,superscriptaddress]{revtex4-1}

\usepackage{color}
\usepackage{latexsym}
\usepackage{amssymb}
\usepackage{amsmath}  
\usepackage{graphicx}
\usepackage{empheq} 

\usepackage{soul}  
\usepackage{empheq} 
\usepackage{tabulary}
\usepackage{tabularx}

\begin{document}


\title{Ultrafiltration modeling of non-ionic microgels}


\author{Rafael Roa}
\email[]{r.roa@fz-juelich.de}
\affiliation{Forschungszentrum J\"ulich, Institute of Complex Systems (ICS-3), 52425, J\"ulich, Germany}

\author{Emiliy K. Zholkovskiy}
\affiliation{Ukrainian Academy of Sciences, Institute of Bio-Colloid Chemistry, 03142, Kiev, Ukraine}

\author{Gerhard N\"agele}
\affiliation{Forschungszentrum J\"ulich, Institute of Complex Systems (ICS-3), 52425, J\"ulich, Germany}


\date{\today}

\begin{abstract}

Membrane ultrafiltration (UF) is a pressure driven process allowing for the separation and enrichment of protein solutions and dispersions of nanosized microgel particles. The permeate flux and the near-membrane concentration-polarization (CP) layer in this process is determined by advective-diffusive dispersion transport and the interplay of applied and osmotic transmembrane pressure contributions. The UF performance is thus strongly dependent on the membrane properties, the hydrodynamic structure of the Brownian particles, their direct and hydrodynamic interactions, and the boundary conditions. We present a macroscopic description of cross-flow UF of non-ionic microgels modeled as solvent-permeable spheres. Our filtration model involves recently derived semi-analytic expressions for the concentration-dependent collective diffusion coefficient and viscosity of permeable particle dispersions [Riest \emph{et al.}, \emph{Soft Matter}, 2015, 11, 2821]. These expressions have been well tested against computer simulation and experimental results. We analyze the CP layer properties and the permeate flux at different operating conditions and discuss various filtration process efficiency and cost indicators. Our results show that the proper specification of the concentration-dependent transport coefficients is important for reliable filtration process predictions. We also show that the solvent permeability of microgels is an essential ingredient to the UF modeling. The particle permeability lowers the particle concentration at the membrane surface, thus increasing the permeate flux.

\end{abstract}


\maketitle

\section{Introduction}

Microgels are colloidal dispersions of nano- to micro-sized visco-elastic gel particles (typically cross-linked polymer networks) that can swell or deswell in a controlled way in response to external stimuli such as temperature, pH value, salinity and particle concentration \cite{Holmqvist:2012hv,Pan:2015bb,FernandezNieves:2011uv}. The microgel phase behavior is richer than that of classical colloidal hard spheres due to their size variability and viscoelasticity. This makes microgels an ideal model system to study, for example, the effect of intrinsic particle viscoelasticity on fundamental physical processes like crystallization kinetics and glass formation \cite{Siebenburger:2012ex}. Microgels are used in a wide range of applications including drug delivery and advanced oil recovery, and for pharmaceutical or cosmetic products \cite{FernandezNieves:2011uv,Siebenburger:2012ex,Saunders:1999ec}.

In the production process of microgels, the reactor effluent consists of a complex mixture of microgels and unreacted polymers dispersed in a solvent. Thus, various separation and concentration steps are necessary. Membrane UF is a  suitable method to remove traces of free polymers and to concentrate the sample for later practical applications \cite{Siebenburger:2012ex}. UF is a pressure-driven process widely used in commercial large-scale plants by which smaller Brownian particles (like nanosized colloids and proteins) are concentrated and separated from smaller ingredients \cite{Blatt:1970hq,Bowen:1995do}. Its major advantage is the low energy requirement. UF has been traditionally applied to waste water treatment \cite{Shannon:2008bk,Nakamura:2013ck}, fruit juice concentration and clarification \cite{Noble:1995tc}, and protein concentration \cite{Noble:1995tc,Kozinski:1972fy,Belfort:1993if}. Inside the human body, UF takes place in Bowman's capsule at the kidneys, where water and other small molecules are separated from blood \cite{Grimellec:1975ib}.

UF must be distinguished from membrane microfiltration of larger micro-sized particles which are also pressure driven. However, the applied transmembrane pressures are smaller and the permeation fluxes are substantially larger than in UF. Moreover, and different from UF, Brownian motion and transmembrane osmotic pressure are small effects only. Consequently, shear-induced particle diffusion and migration are the leading hydrodynamic mechanisms in microfiltration. For cross-flow setup in microchannels, shear-induced migration enhances the back migration away from the membrane, thus reducing the formation of a stagnant cake layer \cite{Zydney:1986hv,Romero:1988vl,Pelekasis:1998jj,Kromkamp:2005kc,Vollebregt:2010jk}. Brownian particle motion dominates in UF, in contrast to microfiltration, with the local osmotic equilibrium being only weakly perturbed by the pressure driven flow. 

The efficiency of UF is influenced by the CP layer, which is the particles-enriched region next to the membrane surface created by advective-diffusive transport \cite{Bowen:1995do,Belfort:1993if}. The CP layer formed by mobile particles effectively increases the resistance to solvent transport and the local osmotic pressure, hence reducing the permeate flux. 

Membrane fouling is an undesirable irreversible modification of the membrane caused by specific physical and chemical interactions between the membrane and the particles \cite{Bacchin:1995hi,Song:1998jg,Bacchin:2006es,Buetehorn:2011dv}. It gives rise to a stagnant cake layer deposited on the membrane and to the intrusion or clogging of particles inside the membrane pores. Fouling substantially lowers the permeate flux because of the added hydrodynamic resistance and can also change the membrane selectivity. Moreover, the filtration channel becomes constricted by the growth of the cake layer. In order to avoid or reduce fouling effects in large volume concentration processes, UF is often performed in a cross-flow setup where, under steady-state conditions, the feed dispersion flow is directed parallel to the membrane with inlet and outlet ports. 

For small values of the applied transmembrane pressure (TMP), the permeate flux is found to be proportional to the TMP, while at larger TMP the flux increases sublinearly. Eventually, a limiting flux is reached, insensitive to any further increase in the TMP.  Blatt \textit{et al.} investigated the permeate flux in cross-flow UF using thin-film theory (gel-layer model) under conditions of limiting flux \cite{Blatt:1970hq}. The thin-film theory has been subsequently improved within a boundary layer approximation by regarding explicitly the CP layer mechanism  \cite{Shen:1977hm,Trettin:1980dm,Gill:1988hp}. In these later works, the concentration dependence of the collective diffusion coefficient and/or the dispersion viscosity was considered using phenomenological expressions for the transport coefficients, with applications to specific systems like BSA proteins, sucrose, and dextran. Another model refinement was obtained by including the osmotic pressure buildup at the membrane surface opposing the influence of the applied TMP \cite{Kozinski:1972fy,Wijmans:1984fd,Jonsson:1984ee}.

In addition, numerical solutions of coupled advection-diffusion, Navier-Stokes, and continuity equations by finite difference methods have been used to analyze CP in cross-flow UF and hence to predict the permeate flux and the particle concentration profile at the membrane surface \cite{Ilias:1993gc,Elimelech:1998jm,Bhattacharjee:1999cz,Bowen:2001ib,Bacchin:2002fm}. In most of these works, the concentration-dependent osmotic pressure, collective diffusion coefficient and dispersion viscosity were taken into account using approximate expressions for the special cases of solvent-impermeable colloidal hard spheres and charged-particles models describing sucrose, polyvinyl alcohols, and BSA and lactoferrin proteins. Typically employed approximate expressions for the sedimentation coefficient have been the truncated virial expansion result by Batchelor \cite{Batchelor:1972ii} and the cell model result by Happel \cite{Happel:1958ip}. The semi-phenomenological expressions by Eilers and Krieger-Dougherty \cite{Mewis:2011bo} have been used to describe the zero-frequency viscosity of neutral solvent-impermeable hard-sphere dispersions. The inaccuracy of these transport coefficient expressions has its impact in the respective UF model predictions. 

Recently, Bouchoux \textit{et al.} studied dead-end filtration of milk casein micelles \cite{Bouchoux:2014kb}. These  soft, deformable and solvent-permeable micelles can be regarded to some extent as biological siblings of the synthetic microgels. One of the major issues of their work was to determine the permeability of the dispersion, which was obtained by a combination of osmotic stress and filtration measurements. Different from cross-flow, the dead-end or frontal filtration is an intrinsically non-steady process where the thickness of the immobile cake layer increases during the process. 	

The present work includes the first comprehensive theoretical study of particle permeability effects in the concentration process of non-ionic microgel dispersions by cross-flow UF. The microgel particles are modeled as solvent-permeable spheres interacting by a hard-core potential. In contrast to previous works, we use accurate analytic expressions for the concentration-dependent collective diffusion coefficient and viscosity of permeable hard-sphere dispersions, which have been well tested against computer simulation results and experimental data on non-ionic PNIPAM microgels \cite{Abade:2010gt,Abade:2010ev,Abade:2012em,Riest:2015jo,Eckert:2008hp}. These expressions have been obtained using a hydrodynamic radius model wherein the internal hydrodynamic particle structure is mapped on a hydrodynamic radius parameter for unchanged direct (non-hydrodynamic) interactions. Riest {\it et al.} have shown recently that this simple particle model captures quantitatively the static and dynamic properties of non-ionic PNIPAM microgel dispersions in an organic solvent \cite{Riest:2015jo,Eckert:2008hp}. This description is accurate up to large particle concentrations where the effect of solvent permeability is quite significant. Using a boundary layer analysis of the advective-diffusion transport in the CP layer \cite{Shen:1977hm,Trettin:1980dm}, we calculate the concentration profile and permeate flux at the membrane for different operation conditions. Our focus is on particle permeability effects using concentration-dependent dispersion properties. Various efficiency and energy cost indicators for the filtration process are introduced and numerically evaluated to check the efficiency and sustainability of the UF filtration process.

The paper is structured as follows: in Sec. \ref{sec:modeling}, we describe the macroscopic cross-flow UF model and the similarity solution scheme for the CP boundary layer. We also discuss a transcendental equation that simplifies the problem for constant dispersion transport properties. Under such conditions, this equation leads to upper and lower bounds for the concentration profiles, which can be profitably used for a quick check on unwanted cake formation. The employed analytic expressions for the concentration-dependent collective diffusion coefficient, shear viscosity and osmotic compressibility of permeable Brownian particle dispersions are discussed in Sec. \ref{sec:properties}. Numerical results for the CP profile and the permeate flux are analyzed in Sec. \ref{sec:results} regarding their dependence on the TMP, shear rate, feed concentration, particle size, and in particular the particle permeability. The significance of the microgel permeability is assessed by comparison with results for impermeable particles. We also evaluate the UF process in terms of process efficiency and cost indicators. Our conclusions are presented in Sec. \ref{sec:conclusions}.

\section{Macroscopic UF model}\label{sec:modeling}

\subsection{Cross-flow transport}

\begin{figure}[t!]
\begin{center}
\includegraphics[width=1\linewidth]{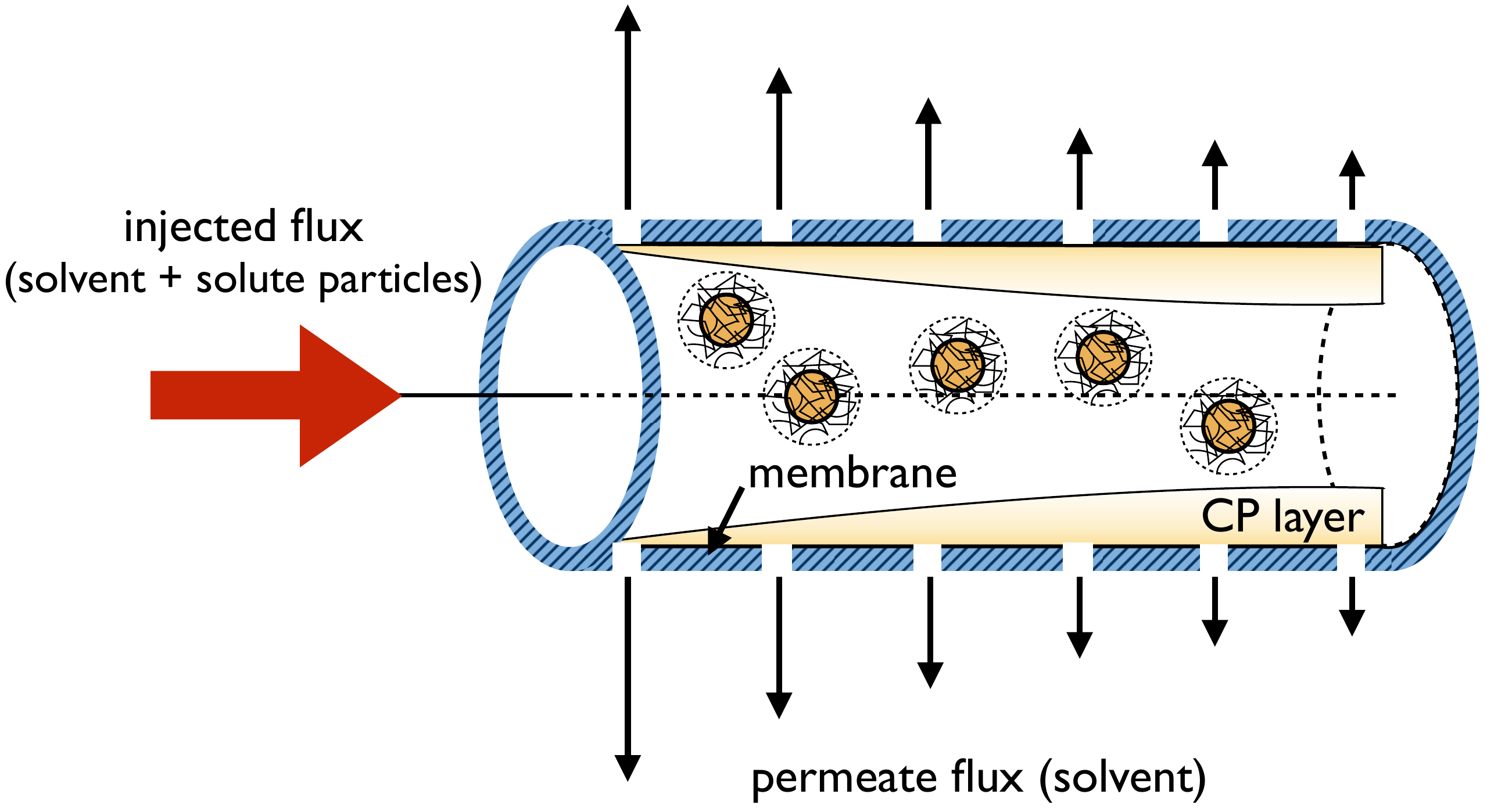}
\caption{Sketch of inside-out cross-flow UF in a hollow cylindrical fiber membrane.}
\label{fig1}
\end{center}
\end{figure}

We consider a homogeneous feed dispersion of non-ionic Brownian microgels particles of hard-core radius $a$ dispersed in a Newtonian solvent at  volume fraction $\phi_0$. As illustrated in Fig. \ref{fig1}, the dispersion is longitudinally and steadily pumped through a hollow cylindrical fiber membrane of length $L$ and inner radius $R$. Due to an applied (transversal) transmembrane pressure, $\Delta P$, a small fraction of the axially in-flowing solvent permeates the membrane to the outside of the fiber. We consider the membrane to be fully particle retaining. In this so-called inside-out cross-flow UF setup, the particle advection towards the membrane is balanced at steady-state by the diffusive back transport away from the membrane. 
%

A coarse-grained continuum mechanics description can be used for $R \gg a$ where the granular nature of the colloidal Brownian particles, assumed to be monodisperse and neutrally buoyant, and the morphology of the membrane pores are not resolved. The momentum balance of the dispersion-averaged flow is then described by the effective Navier-Stokes equation, 
\begin{eqnarray}\label{navierstokes0}
\rho \left( \frac{\partial \mathbf{v}}{\partial t} + \mathbf{v} \cdot \nabla\mathbf{v} \right) = &-& \nabla p  +  \eta\Delta \mathbf{v}  \nonumber \\
&+& \nabla\eta\cdot \left[ \nabla\mathbf{v} + (\nabla\mathbf{v})^T \right]\;\!.
\end{eqnarray}

We have used that the dispersion is incompressible on the macroscopic scale, implying that $\nabla\cdot\mathbf{v}=0$. In Eq. (\ref{navierstokes0}), $\rho$ is the constant dispersion mass density and $\eta(\phi)$ is the effective dispersion shear viscosity depending on the particle volume fraction $\phi = \left(4\pi/3\right)n a^3$, with $n$ denoting the mean particle number density. The dispersion-averaged velocity, $\mathbf{v}(\mathbf{r},t)$, and pressure, $p(\mathbf{r},t)$, depend on the spatial position, $\mathbf{r}$, and time $t$. Here, $\nabla p = \nabla p_F +\nabla \Pi$, wherein $p_F$ is the fluid-phase pressure contribution whose gradient adjusts itself such that the flow remains  incompressible throughout the dispersion, and $\Pi$ is the equilibrium osmotic pressure due to the Brownian particulate phase \cite{Brady:1993hu}. The third term on the right-hand side of Eq. (\ref{navierstokes0}), invoking $\nabla \eta$  and the dyadic tensor gradient $\nabla\mathbf{v}$ plus its transpose $(\nabla\mathbf{v})^T$, originates from the concentration-dependent viscosity, $\eta(\phi)$, in the dispersion viscous stress tensor. This term contributes in an inhomogeneous dispersion region such as the CP layer in UF. 

The mass balance (particle conservation) is described by the continuity equation,
\begin{equation}\label{continuity0}
\frac{\partial n}{\partial t}+\nabla\cdot\mathbf{J}=0\;\!,
\end{equation}
where $n(\mathbf{r},t)$ and $\mathbf{J}(\mathbf{r},t)$ are the coarse-grained local particle number concentration and advective-diffusive particle flux, respectively. The latter is the sum,
\begin{equation}\label{flux}
\mathbf{J}=\mathbf{J}^{ad}+\mathbf{J}^{D}=n\:\!\mathbf{v} - D\:\!\nabla n \;\!,
\end{equation}
of advection and Brownian diffusion fluxes. Here, $D(\phi)$ is the concentration-dependent long-time collective or gradient diffusion coefficient. For non-zero particle concentration and dominantly repulsive interactions, $D(\phi)$ is in general larger than the single-particle translational diffusion coefficient, $D_0$, with the latter measured at infinite dilution. We note that 
\begin{equation}\label{diffusion-osmotic pressure}
\mathbf{J}^{D} = - D(\phi)\:\!\nabla n = - \frac{D_0 K(\phi)}{k_B T} \nabla \Pi \;\!,
\end{equation}
is valid for a monodisperse dispersion of electroneutral particles at temperature $T$, with $k_B$ denoting Boltzmann's constant. Thus, the diffusion flux is driven by the gradient in the osmotic pressure $\Pi(\phi)$, which in turn is proportional to the gradient in the particles chemical potential. The transport property $K(\phi)$ is the long-time macroscopic sedimentation coefficient. It is defined as the ratio of the mean sedimentation velocity (measured relative to the center of volume rest frame) of a weakly settling homogeneous dispersion of volume fraction $\phi$ to the settling velocity of an isolated particle. 

Substituting the particle flux into Eq. (\ref{continuity0}) and using the local volume fraction rather than $n({\bf r},t)$, the macroscopic advection-diffusion equation,
\begin{equation}\label{advdiff0}
\frac{\partial \phi}{\partial t} +\mathbf{v}\cdot\nabla \phi =\nabla\cdot(D\nabla \phi) \;\!,
\end{equation}
is obtained. Analytic expressions for the collective diffusion coefficient, $D(\phi)$, and the zero-frequency dispersion viscosity, $\eta(\phi)$, of solvent-permeable Brownian particles are discussed in Sec. \ref{sec:properties}. 

Eqs. (\ref{navierstokes0})-(\ref{advdiff0}) apply to UF where Brownian particle dispersions are considered 
under laminar flow conditions for which the single-particle shear P\'eclet number, $Pe_{\dot{\gamma}}^a = \tau_D^a / \tau_{\dot{\gamma}} \lesssim 0.01$, is so small that the system is only slightly perturbed away from thermal equilibrium  (linear response regime). Non-Newtonian effects, such as normal stress differences and shear-thinning or thickening, are then absent or negligibly small. Here, $\tau_D^a=a^2/D_0$ is the Brownian diffusion time for diffusion across the particle radius and $\tau_{\dot{\gamma}}=1/\dot{\gamma}$ is the shear flow advection time for the characteristic shear rate $\dot{\gamma}$. Assuming fully developed Poiseuille feed flow at the cylindrical fiber entrance, consistent with no-slip hydrodynamic boundary condition at the membrane surface, the characteristic shear rate is \cite{Probstein:2005vy}
\begin{equation} \label{bulkshear}
  \dot{\gamma}= \frac{2\;\!u_m}{R} = \frac{\Delta p_L\;\!R}{2\;\!\eta(\phi_0)\;\!L}\;\!,
\end{equation}
where $\Delta p_L$ is the longitudinally applied pressure difference across the fiber of length $L$, to be  distinguished from the TMP, and $u_m$ is the inflow dispersion velocity at the fiber center (axis)  equal to twice the cross-section averaged velocity $\langle u \rangle$. 

In typical microfiltration experiments $Pe_{\dot{\gamma}}^a \gtrsim 1$ and Brownian diffusion is hence not important. The flow-driven particle transport at larger concentrations is dominated instead by non-equilibrium many-particle hydrodynamic interaction (HI) effects, most notably by anisotropic collective hydrodynamic diffusion and cross-streamlines particle migration. These HI effects must be properly taken into account in the expressions for the coarse-grained particle current and dispersion-averaged stress tensor. The latter includes now normal stress differences,  a shear-rate dependent non-equilibrium effective viscosity, and an effective osmotic-type pressure contribution \cite{Vollebregt:2010jk,Yurkovetsky:2008ev}, all depending on the local shear rate in addition to $\phi$. Microfiltration is not treated in this work.

\subsection{Steady-state boundary layer}

Cross-flow UF is commonly performed at laminar flow conditions where the Reynolds number associated with cylindrical fiber flow fulfills that $Re_R=R\left(u_m/\nu\right) \lesssim 2000$, with $\nu=\eta/\rho$ denoting the kinematic viscosity. This implies $Re_a = (a/R) Re_R \ll 1$ for the particle Reynolds number in terms of the bulk shear rate, guaranteeing that particle inertia is negligible. Moreover, and quite importantly, the permeate velocity is small compared with the longitudinal cross-flow velocity in the bulk of the fiber, $u_m \gg v_w^0$, where $v_w^0$ is the maximal permeate velocity measured for a clean membrane at given TMP using pure solvent as feed. Due to the strong axial advection, and after a short transient \cite{Romero:1990ei,De:1996hq}, a stationary thin CP layer develops on the membrane surface where diffusive-advective particle transport takes place. The thickness, $\delta$, of the diffuse boundary layer is much smaller than the inner radius, $R$, of the hollow fiber. This layer is characterized by marked axial concentration gradients. In the bulk region outside the boundary layer, the particles are transported essentially by axial advection with preserved uniform concentration $\phi_0$ and unperturbed flow profile. 

\begin{figure}[b!]
\begin{center}
\includegraphics[width=1\linewidth]{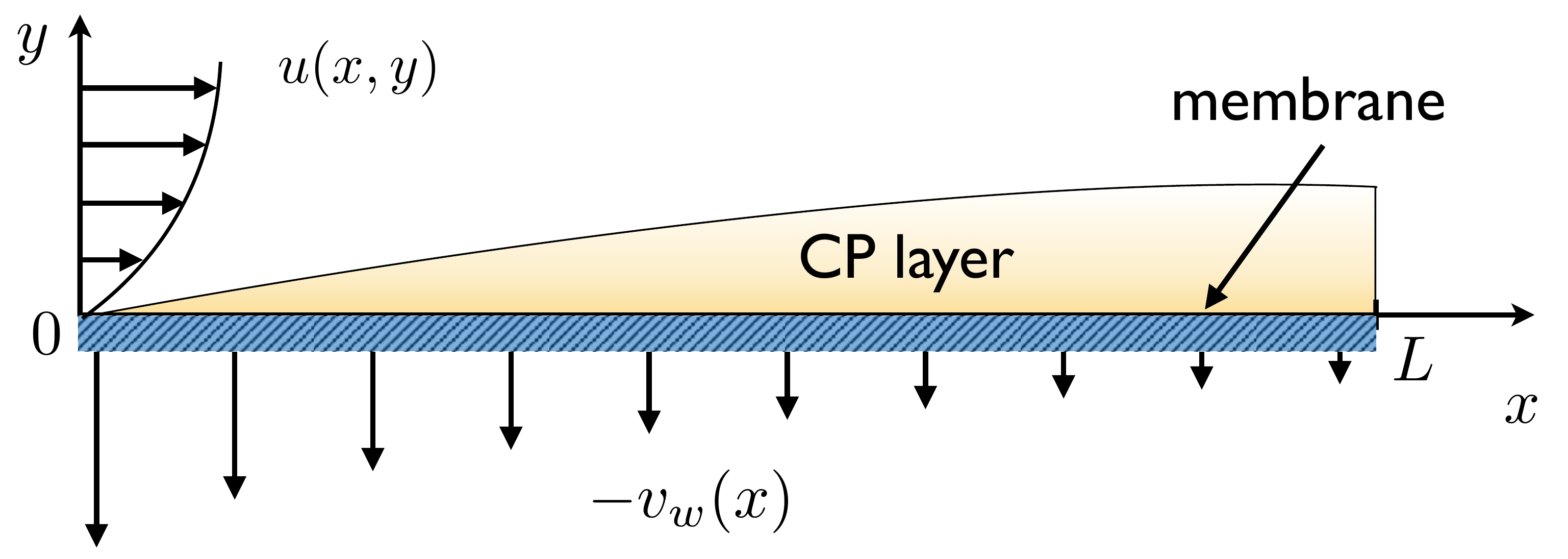}
\caption{Locally flat CP layer next to the surface, $y=0$, of the hollow cylindrical fiber membrane of length $L$ in steady-state cross-flow UF. The osmotic-pressure-regulated permeate transverse flow profile, $v(x,0)=-v_w(x)$, and the axial bulk flow profile, $u(x,y)$, are sketched.}
\label{fig2}
\end{center}
\end{figure}

The locally planar CP boundary layer is conveniently analyzed using the local Cartesian coordinates shown in Fig. \ref{fig2}, where the $x$-axis with the unit vector ${\bf e}_x$ extends longitudinally along the membrane surface and the $y$-axis with the unit vector ${\bf e}_y$ stretches out radially towards the fiber axis. The steady-state flow velocity in this local frame, 
\begin{equation} \label{localvelocity}
 \mathbf{v}(\mathbf{r})=u(x,y)\;\!{\bf e}_x+v(x,y)\;\!{\bf e}_y\;\!,
\end{equation}
fulfills the incompressibility condition 
\begin{equation}\label{incompress}
\frac{\partial u}{\partial x} + \frac{\partial v}{\partial y}=0\;\!.
\end{equation}

Using $\partial /\partial x \sim 1/L$ and $\partial /\partial y \sim 1/\delta$ with $\delta \ll L$ in Eq. (\ref{advdiff0}), the steady-state advection-diffusion equation reduces in the CP layer to
\begin{equation}\label{advdiff}
u(x,y)\frac{\partial \phi}{\partial x}+v(x,y)\frac{\partial \phi}{\partial y}=\frac{\partial}{\partial y}\left(D(\phi)\frac{\partial \phi}{\partial y}\right)\;\!,
\end{equation}
where the axial diffusion flux contribution has been neglected in comparison to the transversal one. A similar analysis largely reduces the effective Navier-Stokes Eq. (\ref{navierstokes0}) in the CP layer to the remaining expression, 
\begin{equation}\label{momentum0}
\frac{\partial}{\partial y}\left(\eta(\phi)\frac{\partial u}{\partial y}\right)=0\;\!,
\end{equation}
describing constant axial dispersion shear stress across the CP layer. The stress is independent of the axial position $x$ and is obtained from the asymptotic matching to the bulk region as \cite{Shen:1977hm,Romero:1990ei}
\begin{equation}\label{momentum}
\eta(\phi)\frac{\partial u}{\partial y}=\eta(\phi_0)\frac{\partial u}{\partial y}\Big |_{y\rightarrow\infty}=\eta(\phi_0)\dot{\gamma}\;\!.
\end{equation}

Note here that the time scale for which steady-state viscous flow has developed is much shorter than the time scale for the development of the steady-state CP layer. This is due to the very large Schmidt number, $Sc=\nu/D_0$, in colloidal dispersions, given by the ratio of fluid vorticity and particle thermal diffusion coefficients \cite{De:1996hq,Probstein:2005vy}.

The steady-state and positive valued permeate velocity through the uniformly porous membrane with its inner wall at $x=0$,
\begin{equation}\label{bcpermv}
v_w(x) = -v(x)\;\!,
\end{equation}
is described in UF using Darcy's law by \cite{Probstein:2005vy}
\begin{equation}\label{bcdarcy}
v_w(x)=v_w^0\left[1- \frac{\Pi\left(\phi_w(x)\right)}{\Delta P}\right]\;\!,
\end{equation}
where  
\begin{equation} \label{fluidflux}
 v_w^0 = L_p\;\!\Delta P
\end{equation}
is the permeate velocity at zero particle concentration, for a given TMP $\Delta P>0$. Here, $L_p=1/(\eta_0R_m)$ is the hydraulic permeability of the clean membrane, and $R_m$ is the associated hydraulic resistance.   
According to Eq. (\ref{bcdarcy}), the permeate velocity is determined by the difference between TMP and the osmotic pressure, $\Pi(\phi_w)$, at the inner membrane wall. Here,  
\begin{equation} 
\phi_w(x)=\phi(x,0)
\end{equation}
is the particle volume fraction at the inner membrane surface which increases with increasing distance $x$ from the fiber entrance. In obtaining Eq. (\ref{bcdarcy}), we have assumed (i) that the membrane is impermeable to the particles, (ii) zero colloid concentration in the permeate, (iii) stationary, fully developed permeate flow inside the membrane, and (iv) the absence of fouling effects which effectively enlarge the hydraulic resistance above the value $R_m$ of a clean membrane. We also neglect the axial pressure drop which in typical UF applications is far smaller than the TMP. 

The remaining boundary conditions used in solving Eqs. (\ref{incompress})-(\ref{momentum0}) are as follows: we use  the no-slip boundary condition for the axial velocity component at the membrane surface,
\begin{equation}\label{bcu0}
u(x,0)=0\;\!,
\end{equation}
and assume that the particle concentration right at the fiber entrance and distant from the membrane surface 
matches the bulk (feed) concentration,
\begin{equation}\label{bcbulkconc}
\phi(0,y)=\phi(x,y\rightarrow\infty)=\phi_0\;\!.
\end{equation}

Finally, the assumed impermeability of the membrane regarding the colloidal particles is considered using the zero normal particle flux condition, 
\begin{equation}\label{bcnoflux}
D(\phi)\frac{\partial \phi}{\partial y}\bigg |_{y=0}+v_w(x)\phi_w(x)=0\;\!,
\end{equation}
at the membrane surface. In using this reflecting boundary condition, particle adsorption by the membrane is excluded which adds to fouling.     

For given dispersion properties $D(\phi)$, $\eta(\phi)$, and $\Pi(\phi)$, the non-linear coupled partial differential Eqs. (\ref{incompress})-(\ref{momentum0}) in conjunction with the boundary conditions in Eqs. (\ref{momentum})-(\ref{bcnoflux}) form a two-dimensional boundary value problem. It can be solved only numerically for the searched for CP concentration and velocity profiles $\phi(x,y)$ and ${\bf v}(x,y)$, respectively. The only exception is the limiting case, $v_w^0=0$, of an impermeable fiber wall where $\phi=\phi_0$ holds uniformly throughout the fiber and the Poiseuille velocity profile extends right up to the membrane wall. The solution of the boundary value problem for $v_w^0 >0$ is largely simplified using a similarity analysis described in the following.

\subsection{Similarity solution}\label{subsec:blayer}

The similarity solution scheme used in this work follows that by Shen and Probstein for gel polarization \cite{Shen:1977hm}. We account here in addition for the osmotic pressure contribution to UF and give additional insight into the obtained solution.

We first estimate the thickness, $\delta=\delta(x)$, of the CP layer. To this end, we use that the transverse advection term, $v {\partial \phi}/{\partial y}$, in Eq. (\ref{advdiff}) is negligible at the outer diffuse edge of this layer in comparison to the axial one, since $u_m \gg v_w^0$ and since $\phi$ has decayed basically to $\phi_0$. Right at the membrane surface, however, the axial advection term is zero instead due to the no-slip boundary condition. Using 
that $\partial /\partial y \sim 1/\delta$, the estimate 
\begin{equation} \label{delta}
 \delta=\left(\frac{3D_bx}{\dot{\gamma}}\right)^{1/3}\;\!,
\end{equation} 
is obtained, with $D_b=D(\phi_0)$ denoting the bulk collective diffusion coefficient. Eq. (\ref{delta}) 
describes an axial growth of the CP layer thickness proportional to $x^{1/3}$. It follows that $\delta(L)/R \sim (L/R)^{1/3} / \left(Pe_{\dot{\gamma}}^R\right)^{1/3}$. The axial advection P\'eclet number $Pe_{\dot{\gamma}}^R = \tau_D^R / \tau_\textrm{ad}$, with $\tau_D^R=R^2/D_0$ and $\tau_\textrm{ad}=R/u_m$, is in UF orders of magnitude larger than one, consistent with the thin boundary layer assumption we have started from. 

Eq. (\ref{delta}) implies the existence of a similarity solution, $\phi(\lambda)$, for the CP profile which for each axial position $x$ is a function of the dimensionless similarity variable \cite{Shen:1977hm,Trettin:1980dm},
\begin{equation}\label{position-dimensionless}
\lambda=\frac{y}{\delta} \;\!,
\end{equation}
only with $\lambda \propto y/x^{1/3}$. The dispersion incompressibility condition in Eq. (\ref{incompress}) and the shear stress Eq. (\ref{momentum}) are expressed in terms of $\lambda$ as
\begin{eqnarray}
\frac{d v}{d \lambda}&=&-\frac{\delta}{3x}\lambda\frac{d u}{d \lambda}  \label{inc1} \\
\frac{1}{\dot{\gamma}\delta}\frac{d u}{d \lambda}&=&\frac{1}{\hat{\eta}(\phi)} \label{mom1}\;\!,
\end{eqnarray}
respectively, where $\hat{\eta}(\phi)=\eta(\phi)/\eta(\phi_0)$ is the reduced viscosity. 

Dimensionless velocities are introduced accordingly by
\begin{eqnarray}
 V(\lambda) &=& -\frac{3x}{\delta^2\dot{\gamma}}\;\!v  \label{dimV}\\
 U(x,y)     &=& \frac{u}{\dot{\gamma}\;\!\delta}  \label{dimW}\;\!,
\end{eqnarray}
where the axial velocity $u$ has been scaled with the fluid velocity, $\dot{\gamma}\;\!\delta$, at the outer edge of the CP layer, and the transverse velocity $v$ by the velocity $D_b/\delta=\delta^2\dot{\gamma}/(3x)$ characteristic of the  advection-diffusion balance. The negative sign in the definition of $V$ has been included to make it positive valued. Different from $V$, the reduced axial velocity $U$ is not a function of $\lambda$ only. This follows from considering the special  case, $\hat{\eta}=1$, of constant viscosity where $u=\dot{\gamma} y$ is the solution of Eq. (\ref{mom1}) satisfying the no-slip condition at the membrane surface. The corresponding transverse velocity is obtained from the  incompressibility condition and Eq. (\ref{bcpermv}) as $v(x,y)= - v_w(x)$, being constant across the CP layer for a given $x$.    

The non-dimensional velocity related to $U$ defined as  
\begin{equation}\label{dimG}
 G(\lambda)= U-\lambda\;\!,
\end{equation}
is, however, an explicit function of $\lambda$. Substitution of Eq. (\ref{dimG}) into Eq. (\ref{mom1}) leads to
\begin{equation}\label{mombaldim}
\frac{d G}{d \lambda}=\frac{1}{\hat{\eta}(\phi)}-1\;\!,
\end{equation}
which is the non-dimensionalized shear stress equation. The dimensionless incompressibility condition follows from Eqs. (\ref{inc1}) and (\ref{dimG}) as
\begin{equation}\label{incompressdim}
\frac{d V}{d \lambda}=G(\lambda)-\lambda\frac{d G}{d \lambda} \;\!.
\end{equation}

Using Eqs. (\ref{position-dimensionless}), (\ref{dimW}), (\ref{dimG}) and (\ref{incompressdim}), the advection-diffusion Eq. (\ref{advdiff}) is expressed in terms of $\lambda$, $V$ and $G$ as 
\begin{equation}\label{advdiffdim}
\frac{d}{d \lambda}\left[\hat{D}(\phi)\frac{d \phi}{d \lambda}\right]=
-\frac{d\phi}{d \lambda}\Big\{\lambda \left[\lambda+G(\lambda)\right]+V(\lambda)\Big\}\;\!,
\end{equation}
where $\hat{D}(\phi)=D(\phi)/D_b$. It is noticed here that with $\phi$ and $V$, also $G$ is a function of $\lambda$.  

In terms of $\lambda$, $V$ and $G$, the boundary conditions in Eqs. (\ref{bcpermv}), (\ref{bcu0})-(\ref{bcnoflux}) are expressed as
\begin{eqnarray}
G(\lambda=0)&=&0 \label{bcG0} \\
V(\lambda=0)&=&V_w \label{bcW0} \\
\phi(\lambda\rightarrow\infty)&=&\phi_0 \label{bcFinf} \\
\hat{D}(\phi_w)\frac{d \phi}{d \lambda}(\lambda=0)+ V_w\phi_w&=&0 \label{bcF0} \;\!.
\end{eqnarray}

Here, $V_w=({3x}/{\delta^2\dot{\gamma}})v_w$ is the dimensionless permeate velocity, with $v_w$ given by Eq.  (\ref{bcdarcy}). Using the boundary conditions at $\lambda=0$, Eqs. (\ref{mombaldim}) and (\ref{incompressdim}) are integrated,   
\begin{eqnarray}
G(\lambda)&=& \int_0^{\lambda} \frac{d\lambda'}{\hat{\eta}\left(\phi(\lambda')\right)} \label{G_functional} \\
V(\lambda) - V_w&=&2 \int_0^{\lambda} d\lambda'\;\!G(\lambda') - \lambda\;\!G(\lambda)  \label{V_functional}\;\!,
\end{eqnarray}
with $G$ and $V$ expressed now as functionals of the concentration profile.
 
Eqs. (\ref{mombaldim})-(\ref{advdiffdim}) form a set of coupled nonlinear ordinary differential equations which in  combination with the boundary conditions in Eqs. (\ref{bcG0})-(\ref{bcF0}), can be solved numerically for the searched for profile $\phi(\lambda)$ and the reduced velocities $G(\lambda)$ and $V(\lambda)$. Since $V_w$ in Eq. (\ref{bcdarcy}) varies with the axial position along membrane surface, due to the likewise varying osmotic pressure, the set $\{\phi,G,V\}$ must be calculated for each $x$ value separately. We have performed these calculations with MATLAB's routine bvp4c \cite{Kierzenka:2001bsb}, using analytic expressions for collective diffusion coefficient, effective dispersion viscosity and osmotic pressure presented in the following section. 

The general behavior of the concentration profile follows from the integrated Eq. (\ref{advdiffdim}), 
\begin{equation} \label{slope_of_phi}
 \frac{d \phi}{d \lambda} = -\frac{V_w \phi_w}{\hat{D}(\phi(\lambda))} \exp\Big\{ - \int_0^\lambda d\lambda' \frac{ \left[ \lambda'\left(\lambda' + G\right) + V \right] }{\hat{D}(\phi(\lambda'))}   \Big\} \;\!,
\end{equation}
where the zero flux condition at the membrane wall has been used. In accord with physical expectation, it  describes the strictly monotonic compressed exponential decay of $\phi(\eta)$ from the wall value $\phi_w$ at $\lambda=0$ towards the large-$\lambda$ bulk value $\phi_0$ where $d\phi/d\lambda\to0$.

\subsection{Solution for constant viscosity and diffusivity}\label{subsubsec:constantprop}

An additional simplification follows for a concentration-independent reduced viscosity $\hat{\eta}$ in the CP layer not necessarily equal to the bulk value of one. In this case, the solutions of Eqs. (\ref{mombaldim}) and \ref{incompressdim})  are $G=\lambda(1/\hat{\eta}-1)$ and $V=V_w$, respectively, for all $\lambda\geq0$. Assuming in addition a constant reduced collective diffusion coefficient $\hat{D}$, Eq. (\ref{slope_of_phi}) can be integrated using Eq. (\ref{bcFinf}), resulting in \cite{Trettin:1980dm}
\begin{equation}\label{phiwnew}
\phi_w=\frac{\phi_0}{1-(V_w/\hat{D}) \int_0^\infty d\lambda \;\! \mathrm{exp}\big\{ {-\lambda^3}/(3\hat{\eta}\hat{D})-\lambda V_w/\hat{D}\big\}} \;\!,
\end{equation}
where $V_w=V_w(\phi_w)$. This is a transcendental equation for the membrane wall concentration, $\phi_w(x)$, at a given axial distance $x$ from the feed inlet. It can be solved using an iteration method. For given osmotic pressure expression, the permeate velocity $v_w(x)$ follows then from substituting the solution $\phi_w(x)$ into Eq. (\ref{bcdarcy}). In Sec. \ref{sec:results}, Eq. (\ref{phiwnew}) is profitably used to generate a good approximation of the profile $\phi_w(x)$ for non-constant collective diffusion and viscosity coefficients.

\section{Permeable particles dispersions}\label{sec:properties}

We model non-ionic microgel dispersions generically as dispersions of uniformly solvent-permeable colloidal hard spheres \cite{Abade:2010gt,Abade:2010ev,Abade:2012em,Riest:2015jo}. As shown in a recent theoretical-experimental work by Riest {\it et al.} \cite{Riest:2015jo}, this simplifying model provides an excellent description of static and dynamic light scattering data, and hence the pair distribution function and collective diffusion properties, of non-ionic PNiPAM microgels \cite{Eckert:2008hp}.

\subsection{Osmotic pressure}

The Brownian particles accumulated in the CP layer give rise, for $Pe_{\dot{\gamma}}^a \ll 1$, to an equilibrium osmotic pressure profile $\Pi(\phi)$. According to Eq. (\ref{diffusion-osmotic pressure}), the osmotic pressure gradient drives a thermal diffusion current pointing away from the membrane into the bulk. On the other hand, the osmotic pressure at the membrane wall, $\Pi(\phi_w)$, reduces the permeate flux as quantified by Eq. (\ref{bcdarcy}).  Since the microgels are modeled as permeable hard spheres, which are characterized statically by the hard-core radius $a$ with associated volume fraction $\phi$, the osmotic pressure in the fluid phase is described to excellent accuracy by the Carnahan-Starling (CS) equation of state \cite{Mewis:2011bo}, 
\begin{equation}\label{Zcarnahanstarling}
Z(\phi)=\frac{1+\phi+\phi^2-\phi^3}{(1-\phi)^3}\;\!,
\end{equation}
where $Z(\phi)=\Pi(\phi)/\left(n(\phi)k_BT\right)$ is defined as the ratio of osmotic and ideal gas pressures. Eq. (\ref{Zcarnahanstarling}) applies not only to the fluid-phase branch extending up to the freezing concentration, $\phi_f\approx 0.494$, but decently well also to part of the non-crystalline metastable phase branch from $\phi_f$ up to  the melting volume fraction $\phi_m\approx 0.545$. The CS $Z(\phi)$ does not reproduce the osmotic pressure divergence, $ \Pi\sim 1/(\phi -\phi_\textrm{rcp})$, of the metastable non-crystalline phase at the random close packing concentration $\phi_\textrm{rcp} \approx 0.64$ \cite{Rintoul:1996fs,Brady:1993hu}. This divergence is of no concern in this work where stagnant cake layers formed by particle jamming are not considered.

\subsection{Hydrodynamic particle model}
 
The mesh-size averaged flow inside a microgel particle is described in our model by the Brinkman-Debye-Bueche (BDB) equation \cite{Brinkman:1947wz,Debye:1948if}, wherein the amount of fluid permeability is characterized by the non-dimensional parameter
\begin{equation} \label{chikappa}
\chi=\kappa a \;\!, 
\end{equation}
equal to the inverse of the hydrodynamic penetration depth, $\kappa^{-1}$, in units of the hard-core radius $a$. The penetration depth is equal to the square root of the Darcy permeability of a microgel particle, and it roughly equals its mean  mesh size. Values of $\chi$ can vary in principle between $\chi=\infty$, corresponding to an impermeable hard sphere with no-slip surface boundary condition, down to low values characteristic of strongly permeable particles. For the continuum picture underlying the BDB equation to be valid, the mean pore size should be no larger than one tenth of the particle radius ($\chi \gtrsim 10$). Typical values in experimentally studied dispersions are $\chi \sim 20 -40$ or larger \cite{Duits:2001ij,Riest:2015jo}.

Transport properties of permeable particle dispersions such as $D$ and $\eta$ depend on $\chi$ in addition to $\phi$. At infinite dilution, $D(\chi,\phi)$ reduces to the single-particle translational diffusion coefficient of a permeable sphere,
\begin{equation}\label{D0SE}
D_0(\chi)=\frac{k_BT}{6\pi\eta_0a_h(\chi)}\;\!,
\end{equation}
with the hydrodynamic particle radius, $a_h$, given by \cite{Abade:2010gt}
\begin{equation} \label{ahyd}
\frac{a_h(\chi)}{a}=\frac{2\chi^2\left[\chi-\tanh(\chi)\right]}{2\chi^3+3\left[\chi-\tanh(\chi)\right]} \;\!.
\end{equation}

The hydrodynamic radius becomes equal to $a$ for $\chi \to \infty$. It decreases with increasing permeability 
(decreasing $\chi$) since the viscous drag is reduced when more fluid is allowed to penetrate the particle. For realistic parameter values, $\chi \gtrsim10$, the particle-fluid interface can be considered as locally flat. 
The ratio $a_h/a$ is then well approximated by the flat interface value
\begin{equation}\label{ahyd_flat}
\frac{a_{h,f}(\chi)}{a} = 1 - \frac{1}{\chi}\;\!,
\end{equation}
with the error introduced in $a_h$ being of order ${\cal O}(1/\chi^2)$ small. The flat-interface radius, $a_{h,f}$, and the associated reduced slip length, $L^\ast_{h,f}=1 - a_{h,f}/a=1/\chi$, are independent of the single-particle transport property used for their definition. In other words, the same value for $L_{h,f}$ is obtained regardless whether the translational or rotational single-particle diffusion coefficients or the intrinsic viscosity are used \cite{Abade:2010ev,Riest:2015jo}.     
  
As explained in Refs. \cite{Abade:2010ev,Riest:2015jo}, diffusion and rheological properties of a large variety of dispersions of spherical colloidal particles  of quite different internal hydrodynamic structure and surface boundary conditions, including core-shell particles and particles with Navier slip boundary conditions, can be characterized hydrodynamically in terms of the single hydrodynamic parameter $a_h$. This key observation allows for the usage of the simplifying hydrodynamic radius model wherein the particles are described hydrodynamically as no-slip spheres, characterized by $a_h$ derived from a single-particle transport property, for unchanged direct particle interactions. Regarding the considered non-ionic microgels, this means an unchanged direct interaction radius $a$. The error introduced in using the hydrodynamic radius model for transport coefficient calculations, such as those for $D$ and $\eta$ is of ${\cal O}\left((L_{h,f}^\ast)^2\right)$ small \cite{Riest:2015jo}. Since Eqs. (\ref{ahyd}) and (\ref{ahyd_flat}) describe a one-to-one mapping between $\chi$ and the universal parameter $a_h$, the at first sight specific model of uniformly permeable spheres used in this work is in fact   
quite general.

\subsection{Collective diffusion}

For the considered macroscopic length and time scales, collective thermal diffusion or gradient diffusion refer to the relaxation of thermally induced local concentration gradients by the cooperative diffusive motion of Brownian particles which interact both directly and hydrodynamically. The long-time collective diffusion coefficient  appearing in Fick's law of macroscopic gradient diffusion of a non-ionic dispersion of monodisperse permeable spheres is given, 
according to Eq. (\ref{diffusion-osmotic pressure}), by
\begin{equation}\label{dcperm}
 D(\chi,\phi) = D_0(\chi) \frac{K(\chi,\phi)}{S(\phi)} \;\!,
\end{equation} 
namely by the single-particle diffusion coefficient $D_0(\chi)$ times the ratio of the long-time sedimentation coefficient, $K(\chi,\phi)$ and the thermodynamic coefficient $S(\phi)$. The latter is equal to the zero-scattering wavenumber limit of the static structure factor determined in scattering experiments \cite{Nagele:1996hr} and to the reduced isothermal osmotic compressibility,
\begin{equation} \label{compressibility factor}
\frac{1}{S(\phi)} = \left( \frac{\partial \left[\phi Z(\phi)\right]}{\partial \phi} \right)_{\mu_s,T}\;\!,
\end{equation}
where $\mu_s$ denotes the chemical potential of the dispersing fluid. 

An issue not addressed in the UF literature is that the long-time collective diffusion coefficient $D$, and the associated long-time sedimentation coefficient $K$, in Eqs. (\ref{diffusion-osmotic pressure}) and (\ref{dcperm}) are smaller than their respective short-time counterparts. This is due to the, at long times, slowing influence of the micro-structural environment, which is perturbed from the equilibrium isotropic state by the sedimenting or collectively diffusing particles \cite{Nagele:2013co}. Different from self-diffusion and viscosity, where the respective short-time and long-time transport coefficients are strongly different in concentrated dispersions, the short-and long-time collective diffusion and sedimentation coefficients differ by less than $7\;\!\%$ only, even in concentrated dispersion of no-slip hard spheres where the hydrodynamic interactions are particularly strong \cite{Wajnryb:2004di}. The relative difference in the latter two quantities is of purely hydrodynamic origin and caused by the non-pairwise additive near-distance part of the hydrodynamic particle interactions \cite{Dhont:1996ub}. We emphasize that the present discussion applies to Brownian particles for low sedimentation P\'eclet number characteristic of UF. In the large P\'eclet number regime, micro-structural changes and their effect on the concentration dependence of the sedimentation velocity become highly significant \cite{Benes:2007kv}.

\begin{figure}[t!]
\begin{center}
\includegraphics[width=1\linewidth]{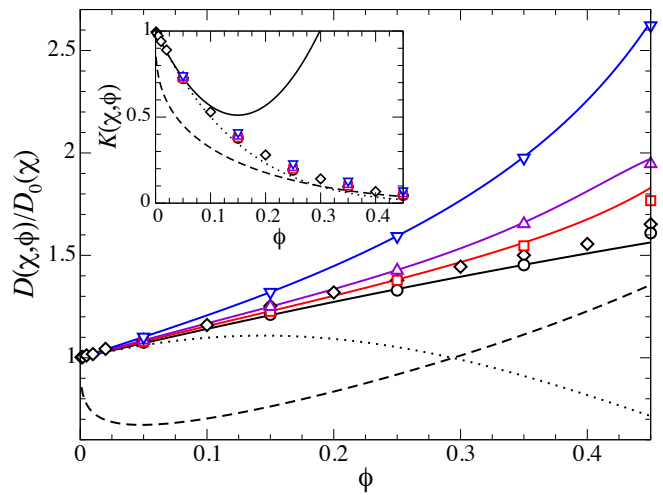}
\caption{
Inset: concentration dependence of the short-time sedimentation coefficient, $K(\chi,\phi)$, of permeable hard-sphere dispersions. Main figure: corresponding reduced short-time collective diffusion coefficient, $D(\chi,\phi)/D_0(\chi)$, obtained from $K(\chi,\phi)$ by division with the CS $S(\phi)$. $\Diamond$ and $\circ$ are simulation results for impermeable spheres ($\chi=\infty$) by Ladd \cite{Ladd:1990gr} and Abade \textit{et al.} \cite{Abade:2010gt}. $\Box$,  $\bigtriangleup$, and $\bigtriangledown$ are simulation results by Abade \textit{et al.} \cite{Abade:2010gt} for $\chi=$ 100, 50, and 20, respectively. Colored lines are polynomial fits to the respective simulation data. Dotted and dashed lines: Eq. (\ref{Batchelor}) for $\alpha=6.55$ and the cell model prediction by Happel \cite{Happel:1958ip}, respectively, both for impermeable spheres. The cell model predictions  for $K$ and $D$ are non-analytic at $\phi=0$. Black solid lines in main figure and inset: second-order virial  expansion results for impermeable spheres given in Eqs. (\ref{difuvirialcichocki}) and (\ref{sedvirialcichocki}), respectively.
}
\label{fig3}
\end{center}
\end{figure}

In this work, we use the short-time form as a substitute for the long-time $K(\chi,\phi)$ in Eq. (\ref{dcperm}) since first, the calculation of the short-time sedimentation coefficient is far simpler and second, precise simulation data are available for the short-time coefficient in its dependence both on $\phi$ and $\chi$ \cite{Abade:2010gt,Abade:2012em,Riest:2015jo}. The wall concentrations in our UF results are such that $\phi_w <0.4$ so that the long-time coefficients $D$ and $K$ are only slightly overestimated by this substitution. The inset of Fig. 3 shows simulation results for the concentration dependence of the short-time $K(\chi,\phi)$, for values of the inverse reduced permeability given by $\chi=$ 20, 50, 100, and $\infty$. These values correspond, according to Eq. (\ref{ahyd_flat}), to reduced hydrodynamic radius $a_h/a =0.95$, 0.98, 0.99, and 1.    

For impermeable no-slip hard spheres, the numerically precise second-order virial (concentration) expansion result,
\begin{equation}\label{sedvirialcichocki}
K(\phi)=1-6.546\phi+21.918\phi^2+\mathcal{O}(\phi^3)\;\!,
\end{equation}
for the short-time sedimentation coefficient has been obtained by Cichocki {\it et al.} \cite{Cichocki:2002ft}. As seen from the inset of Fig. \ref{fig3}, this expression describes the simulation data \cite{Ladd:1990gr,Abade:2010gt} well for $\phi < 0.1$ only (solid line). The division of $K(\phi)$ in Eq. (\ref{sedvirialcichocki}) by the second-order virial expansion result of the hard-sphere osmotic compressibility, $S(\phi)$, leads to the second-order virial expansion expression \cite{Banchio:2008gt}
\begin{equation}\label{difuvirialcichocki}
D(\phi)/D_0=1+1.454\phi-0.45\phi^2+\mathcal{O}(\phi^3)\;\!,
\end{equation}
for the short-time collective diffusion coefficient. This quadratic polynomial describes the weakly monotonic increase of $D(\phi)$ with increasing $\phi$ astonishingly well even up to $\phi\approx0.5$ where the non-sheared dispersion starts to solidify. See here the solid black line in the main part of Fig. \ref{fig3}. This finding points to a mutual cancellation of higher-order terms in the virial expansions of $K$ and $S$. The simulation data for $D$ shown in the main part of the figure have been obtained from dividing the simulation data for $K$ by the CS $S(\phi)$ obtained from Eqs. (\ref{Zcarnahanstarling}) and (\ref{compressibility factor}).   

Previous UF works dealing with impermeable colloidal particles have commonly used either simply a constant $D=D_0$,  or the semi-empirical expression \cite{Russel:1989vm,Bhattacharjee:1999cz}
\begin{equation}\label{Batchelor}
   K(\phi)= \left(1 - \phi\right)^\alpha\;\!,
\end{equation}
with the exponent $\alpha = 6.55$ selected to match Batchelor's \cite{Batchelor:1972ii} first-order virial coefficient result (Eq. (\ref{sedvirialcichocki}) and the dotted line), or the cell model result for $K(\phi)$ by Happel \cite{Happel:1958ip} (dashed line in inset) in conjunction with the CS $S(\phi)$ as divisor \cite{Ilias:1993gc,Elimelech:1998jm,Bhattacharjee:1999cz,Bowen:2001ib,Bacchin:2002fm}. While Eq. (\ref{Batchelor}) describes the simulation data for impermeable spheres depicted in the inset overall decently well, the decay of $K(\phi)$ at larger $\phi$ is overestimated to such an extent that, after division by $S(\phi)$, the resulting $D$ is decreasing for larger $\phi$, in conflict with simulation data and experimental findings for colloidal hard spheres \cite{Banchio:2008gt}. Notice here that  $K(\phi)$ and $S(\phi)$ are both monotonically decreasing with increasing $\phi$ in a concave-shaped form. Their ratio is thus very sensitive to small changes in both quantities. Using the exponential form for $K$ in Eq. (\ref{Batchelor}), Bhattacharjee {\it et al.} \cite{Bhattacharjee:1999cz} combined an approximate long-distance treatment of the hydrodynamic mobilities of two no-slip spheres with the Percus-Yevick solution for the pair distribution function leading to a concentration-dependent exponent $\alpha(\phi)$. While their modified approach predicts the monotonic increase of $D(\phi)$ with increasing $\phi$, the increase is significantly overestimated in comparison to the simulation data. 

The cell model predictions for $K$ and $D$ (dashed lines) are ruled out since they violate the exact analytic low-$\phi$ forms of these quantities noted in Eqs. (\ref{sedvirialcichocki}) and (\ref{difuvirialcichocki}), respectively. In particular, an infinite negative slope of $D(\phi)$ at $\phi \to 0$ is predicted by this model. In summarizing the present discussion on collective diffusion of impermeable particles, our  conclusion is that Eq. (\ref{difuvirialcichocki}) should be used as a good description of the $D(\phi)$ of impermeable colloidal hard spheres. 

A thorough analysis of the sedimentation coefficient of colloidal dispersions of hard spheres with internal hydrodynamic structure (non-zero permeability) was given in \cite{Riest:2015jo}. This article provides an approximate analytic expression for $K(\gamma,\phi)$, where $\gamma=a_h/a$, describing the simulation data quite well for all $\phi <0.5$ and $\gamma \geq 0.8$ using Eq. (\ref{ahyd_flat}). We only quote here the resulting low-$\phi$ expression
\begin{equation}
 \frac{D(\chi,\phi)}{D_0(\chi)} \approx 1 + \left(1.454 +\frac{8.592}{\chi} \right)\phi + {\cal O}(\phi^2)\;\!, 
\end{equation}
valid for $\chi \gtrsim 10$, with ${\cal O}(1/\chi^2)$ corrections disregarded. It quantifies the initially linear increase of $K(\chi,\phi)$ at low $\phi$ displayed in Fig. \ref{fig3}, with increasing slope with increasing permeability.

The remaining smaller differences between the analytic expression for $K$ in \cite{Riest:2015jo} and the simulation data are enhanced at larger $\phi$ after the division by $S(\phi)$. In our UF calculations for permeable particles, we therefore use a direct fit of the simulation data for $D(\chi,\phi)$ \cite{Abade:2010gt} in the form of a fifth order polynomial (colored lines). 

The simulation data for $K$ and $D$ for a given concentration increase with increasing permeability (decreasing $\chi$). The respective increase is more pronounced for larger $\chi$ since the main effect of permeability is to lower the hydrodynamic interactions between the particles. Permeable hard spheres sediment and diffuse collectively faster than impermeable ones for two related reasons: first, the hydrodynamic interactions are weaker and second, $D_0$ is larger due to $a_h < a$. The sedimentation coefficient decreases with increasing volume fraction owing to an increasing fluid backflow and enhanced hydrodynamic interactions \cite{Russel:1989vm,Nagele:2013co}. For a given $\chi$, $D$ increases with increasing $\phi$ which reflects that the diminution in $K$ is overcompensated by the diminution in the osmotic compressibility. The monotonic increase of $D$ with increasing $\phi$ is characteristic of colloidal hard spheres but not of particles with longer-ranged repulsive interactions or particles having shorter-ranged attractions. For example, in lower-salinity dispersions of charge-stabilized colloids and protein solutions, $D$ behaves non-monotonically, passing through a pronounced maximum with increasing $\phi$ \cite{Heinen:2012iy}.

\subsection{Effective dispersion viscosity}

The effective dispersion viscosity, $\eta$, in a steadily sheared dispersion is the ratio between the dispersion shear stress and the rate of strain \cite{Mewis:2011bo}. It depends on the concentration and internal hydrodynamic structure of the Brownian particles and their direct and HIs. Since $Pe_{\dot{\gamma}}^a \ll 1$ in UF, there are no shear-thinning effects. The zero-frequency low-shear-rate limiting effective viscosity of a permeable particles dispersion is the sum 
\begin{eqnarray}\label{visczerofreq}
\eta(\chi,\phi)=\eta_\infty(\chi,\phi) \left[1+\frac{\Delta\eta(\chi,\phi)}{\eta_\infty(\chi,\phi)} \right]\;\!,
\end{eqnarray}
of the high-frequency viscosity contribution, $\eta_\infty(\chi,\phi)$, which is of purely hydrodynamic origin, and the  shear relaxation viscosity part, $\Delta\eta(\chi,\phi)$, originating from the relaxation of the shear-perturbed dynamic particle cages formed around each particle. The viscosity part, $\Delta\eta$, is influenced both by direct and HIs, with the consequence that, for concentrated dispersions, the long-time viscosity $\eta$ is substantially larger than the short-time viscosity $\eta_\infty$. This distinguishes the viscosity from collective diffusion where the difference between the associated short-time and long-time collective diffusion coefficients stays small.

As discussed in Refs. \cite{Riest:2015jo} and \cite{Brady:1993hu}, the term in brackets in Eq. (\ref{visczerofreq}) can be expected to be only weakly influenced by HIs. This suggests the no-HI factorization approximation
\begin{eqnarray}\label{no-HI approx}
\frac{\Delta\eta(\chi,\phi)}{\eta_\infty(\chi,\phi)} \approx 
\frac{\Delta\eta(\phi)}{\eta_0}\bigg |_\mathrm{no-HI}\;\!,
\end{eqnarray}
where $\Delta(\eta)|_\mathrm{no-HI}$ is the shear relaxation part without HI, with $\eta_\infty|_\mathrm{no-HI}$ being equal to the fluid viscosity $\eta_0$. In this approximation, the hydrodynamic particle structure quantified by $\chi$ and the HIs are assumed to affect $\eta$ only through the factored-out high-frequency viscosity in Eq. (\ref{visczerofreq}). This simplifies the calculation of $\Delta\eta$ considerably. An analytic estimate for the $\Delta\eta$ of colloidal hard spheres without HIs is given by 
\begin{equation}\label{viscnoHIbrady}
\frac{\Delta\eta(\phi)}{\eta_0}\bigg |_\mathrm{no-HI} \approx \frac{12}{5}\phi^2g(2a^+,\phi)\;\!,
\end{equation}
where $g(2a^+,\phi)={(Z(\phi)-1)}/{4\phi}$ is the contact value of the equilibrium radial-distribution function, and  $Z(\phi)$ is given for $\phi < 0.5$ by Eq. (\ref{Zcarnahanstarling}). This estimate combines the exact quadratic-order low concentration limit of $\Delta(\eta)|_\mathrm{no-HI}$ with its divergence at the random closed packing volume fraction, $\phi_\textrm{rcp}$, constituting the upper concentration limit of the non-crystalline metastable branch of the hard-sphere phase diagram. This divergence is triggered by the divergence of $g(2a^+,\phi)$ at $\phi_\textrm{rcp}$ \cite{Riest:2015jo,Brady:1993hu}.

The existing high-precision simulation data \cite{Abade:2010ev} for the high-frequency viscosity of permeable colloidal hard spheres in the fluid phase region $\phi < 0.5$  are well described by the the generalized Sait$\hat{\mathrm{o}}$ formula \cite{Abade:2010ev,Riest:2015jo},
\begin{equation}\label{viscgensaito}
\eta_\infty(\chi,\phi)=1+[\eta](\chi)\phi\;\!\frac{1+\hat{S}(\chi,\phi)}{1-\frac{2}{5}
\phi\;\![\eta](\chi)\left[1+\hat{S}(\chi,\phi)\right]}\;\!,
\end{equation}
with the Sait$\hat{\mathrm{o}}$ function $\hat{S}(\chi,\phi)=\phi\left(k_h(\chi)-{2}/{5}\right)[\eta](\chi)$. Both the intrinsic viscosity, given by $[\eta](\chi) \approx (5/2)\left(1 - 3/\chi\right)$ for $\chi >10$, and the Huggins coefficient, $k_h(\chi)$, depend on the particle permeability parameter $\chi$ only. Analytic fitting expressions of these two quantities are given in Ref.  \cite{Riest:2015jo}.

In this work, we use the factorization approximation in Eqs. (\ref{visczerofreq})-(\ref{viscgensaito}) as the analytic input for $\hat{\eta}(\chi,\phi)=\eta(\chi,\phi)/\eta(\chi,\phi_0)$ in our UF calculations. 

\begin{figure}[t!]
\begin{center}
\includegraphics[width=1\linewidth]{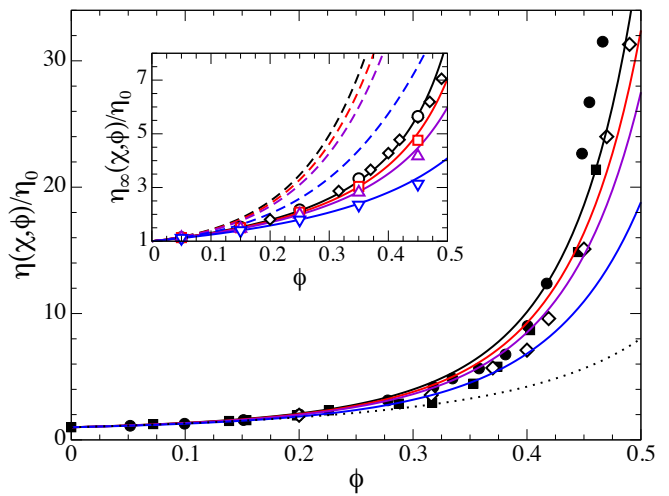}
\caption{
Concentration dependence of the steady-shear zero-frequency dispersion viscosity $\eta(\chi,\phi)$ (main figure), and the associated high-frequency viscosity $\eta_\infty(\chi,\phi)$ (inset), for permeable colloidal hard spheres. $\bullet$, $\blacksquare$ and $\Diamond$ are experimental data for $\eta$ by Segr\`e \textit{et al.} \cite{Segre:1995uu}, and Weiss \textit{et al.} \cite{Weiss:1998hd}, and simulation results by Foss and Brady \cite{Foss:2000kj}, respectively, all for $\chi=\infty$. Solid black, red, violet, and blue lines: factorization approximation predictions for $\eta$ in Eqs. (\ref{visczerofreq})-(\ref{viscgensaito}) at $\chi=\infty$, 100, 50, and 20, respectively. Black dotted line: Krieger-Dougherty formula for $\chi=\infty$ \cite{Mewis:2011bo}, using $\phi_\textrm{max}=\phi_\textrm{cp}$. In the inset: $\Diamond$ and $\circ$ are simulation data for $\eta_\infty$ by Foss and Brady \cite{Foss:2000kj} and Abade \textit{et al.} \cite{Abade:2010ev}, for impermeable spheres ($\chi=\infty$). $\Box$, $\bigtriangleup$, and $\bigtriangledown$ are simulations results by Abade \textit{et al.} \cite{Abade:2010ev} for $\chi=$ 100, 50, and 20, respectively. Correspondingly colored solid and dashed lines in the inset are the results by the generalized Sait$\hat{\mathrm{o}}$ formula in Eq. (\ref{viscgensaito}), and the cell model expression by Ohshima \cite{Ohshima:2009ij}, respectively.  
}
\label{fig4}
\end{center}
\end{figure}

The concentration dependence of $\eta(\chi,\phi)$ is plotted in the main part of Fig. \ref{fig4} together with the associated high-frequency viscosity, $\eta_\infty(\chi,\phi)$, shown in the inset. While both viscosities increase with increasing concentration, $\eta$ is substantially larger than $\eta_\infty$ for large concentrations, due to the at large $\phi$  dominating shear relaxation contribution $\Delta\eta$. Notice the different scales of the ordinate in the inset and the main figure part.

We consider first the high-frequency viscosity presented in the inset. The simulation data by Abade \textit{et al.} (open symbols) for permeable hard spheres  \cite{Abade:2010ev} show that $\eta_\infty$ decreases with increasing particle permeability (decreasing $\chi$) since viscous dissipation is reduced. The simulation data are quantitatively described by the generalized Sait$\hat{\mathrm{o}}$ formula in Eq. (\ref{viscgensaito}) for all considered permeabilities (solid lines in the inset). For comparison, also recent cell model results by Ohshima \cite{Ohshima:2009ij} for the $\eta_\infty$ of permeable spheres are shown (dashed lines). For impermeable particles, $\chi=\infty$, the cell model prediction reproduces an earlier result by Ruiz-Reina {\it et al.} \cite{RuizReina:2003dc}. As it can be noticed from the inset, the cell model method strongly overestimates the high-frequency viscosity at larger $\phi$ for all considered permeabilities.

Consider next the concentration dependence of the steady-shear viscosity, $\eta$, displayed in the main part of Fig. \ref{fig4}. The experimental data (filled black symbols, taken from \cite{Segre:1995uu,Weiss:1998hd}) and in particular the simulation data (open diamonds, from \cite{Foss:2000kj}) for impermeable hard spheres are overall well represented by the factorization approximation in Eqs. (\ref{visczerofreq})-(\ref{viscnoHIbrady}) combined with the generalized Sait$\hat{\mathrm{o}}$ formula in Eq. (\ref{viscgensaito}) for $\eta_\infty(\infty,\phi)$ (solid black line). A slight improvement over this factorization approximation result for $\eta$ has been discussed by Riest {\it et al.} \cite{Riest:2015jo}. This improvement, however, does not noticeably change our UF predictions for the CP layer and permeate flux profiles. 

Analogous to the corresponding behavior of $\eta_\infty(\chi,\phi)$, the steady-shear viscosity $\eta$ is lowered with increasing permeability due to the weakened HIs. This trend is described by the factorization approximation results for the three considered finite values of $\chi$ (differently colored solid lines). To our knowledge, no simulation or experimental zero-frequency viscosity data of permeable particles are available to date for a direct comparison with our theoretical predictions for $\eta(\chi,\phi)$. 

In previous UF works dealing with hard-sphere-like dispersions \cite{Ilias:1993gc,Elimelech:1998jm,Bhattacharjee:1999cz,Bowen:2001ib,Bacchin:2002fm}, either a concentration-independent value for $\eta$ or the phenomenological expressions by Eilers or Krieger-Dougherty have been used. The Krieger-Dougherty expression for the steady-shear viscosity reads \cite{Mewis:2011bo}
\begin{equation}
\frac{\eta}{\eta_0} = \left(1 - \frac{\phi}{\phi_\textrm{max}} \right)^{-\phi_\text{max}\;\![\eta](\chi)}\;\!,
\end{equation}
where the particle permeability enters only via the intrinsic viscosity in the exponent. Here, $\phi_\textrm{max}$ is the  maximal volume fraction at which the expression diverges. In previous UF works \cite{Bowen:2001ib,Bacchin:2002fm}, the closed packing fcc crystal value $\phi_\textrm{cp}=\sqrt{2}\pi/6 \approx 0.74$ was used for $\phi_\textrm{max}$. According to the dotted line in Fig. \ref{fig4} valid for $\chi=\infty$, the Krieger-Dougherty relation for $\phi_\textrm{max}=\phi_\textrm{cp}$ strongly underestimates the steady-shear viscosity at large concentrations. This underestimation is less severe if, as suggested by de Kruif {\it et al.} \cite{deKruif:1985eh} for low shear P\'eclet numbers, $\phi_\textrm{rcp}\approx 0.64$ is used instead of $\phi_\textrm{cp}$. The Eilers expression describes a viscosity curve (not shown in the figure) similar to the one representing the Krieger-Dougherty equation.

\section{Filtration model results}\label{sec:results}

We present in this section our boundary layer model results for cross-flow UF of permeable (more generally: hydrodynamically structured) Brownian particles dispersions. The selected system parameters are: membrane length $L=0.5$ m, inner radius $R=0.5$ mm, hydraulic membrane permeability $L_p=0.5 - 6.7\times10^{-10}$ mPa$^{-1}$s$^{-1}$, and feed particle volume fraction $\phi_0=1.0 - 1.5\times 10^{-3}$ which is small enough that the Einstein expression  $\eta\approx(5/2)\left(a_h/a\right)^3\phi_0$ for permeable particles, with $a_h/a$ according to Eq. (\ref{ahyd}), applies for the feed viscosity. The calculations are performed for the room temperature value $T=293.15$ K.

The explored parameters in our calculations are the applied TMP $\Delta P = 0.3 - 5.0$ kPa, particle excluded volume radius $a=10 - 30$ nm, and characteristic shear rate $\dot{\gamma}=30 - 130$ s$^{-1}$, in addition to the inverse permeability parameter $\chi=20,30,50$, $100$, and $\infty$. These are realistic operation conditions for the UF of permeable Brownian particles,  which we have selected such that $Pe_{\dot{\gamma}}^a \ll 1$, with Brownian diffusion dominating shear-induced hydrodynamic diffusion \cite{Belfort:1993if}. Moreover, since $R  \ll L$, the boundary layer condition $u_m \gg v_w^0$ and, with $Re_R \sim 1$, also the condition of laminar pipe flow are fulfilled. The pipe radius $R$ and the feed velocity $u_m$ enter only implicitly into the similarity solution scheme for the CP layer through Eq. (\ref{bulkshear}) for the characteristic shear rate. The particle radius $a$ affects the CP layer and permeate flux directly by the factor $D_0(\chi)$ in Eq. (\ref{dcperm}). We further note that the selected operating conditions are such that membrane fouling caused by particle jamming is avoided. Fouling is not considered in the present UF study which focuses on particle permeability effects on the CP layer and permeate flux, based on accurate expressions for the concentration and permeability dependent viscosity and collective diffusion coefficient. 

\subsection{Concentration-dependent transport properties}

Before discussing the influence of the particles permeability on cross-flow filtration, we analyze the effects of the concentration dependence of the transport properties $D(\phi)$ and $\eta(\phi)$. For later comparison, we consider in this subsection the limiting case, $\chi=\infty$, of impermeable hard spheres. 

\begin{figure}[t!]
\begin{center}
\includegraphics[width=1\linewidth]{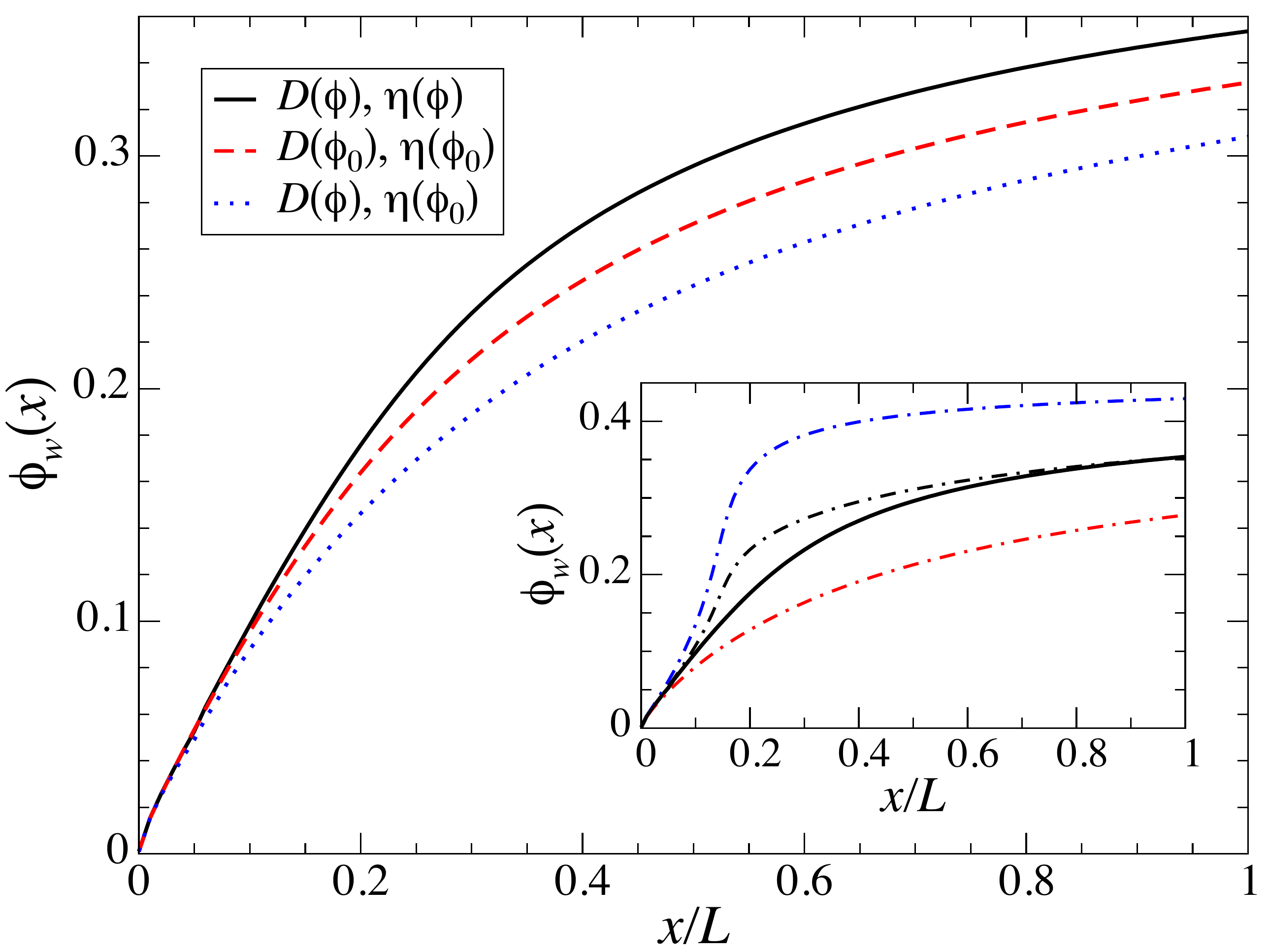}
\caption{Concentration profile, $\phi_w(x)$, of impermeable colloidal hard spheres at the membrane surface, as function of the axial distance, $x$, from the inlet, in units of the fiber length $L$. Black solid line: similarity solution scheme result for concentration-dependent $D(\phi)$ and $\eta(\phi)$ according to Eqs. (\ref{difuvirialcichocki}) and Eqs. (\ref{visczerofreq})-(\ref{viscgensaito}), respectively. Red dashed and blue dotted lines:  similarity solution results for constant $D(\phi_0) \approx D_0$ and $\eta(\phi_0)\approx \eta_0$, and for variable $D(\phi)$ and constant $\eta(\phi_0)$, respectively. In the inset, the similarity solution result for concentration-variable $D$ and $\eta$ (black solid line) is compared  with the solutions of the transcendental Eq. (\ref{phiwnew}) for constant $D(\phi_w)$ and $\eta(\phi_0)$ (red dashed-dotted line), and constant  $D(\phi_0)$ and $\eta(\phi_w)$ (blue dashed-dotted line). The black dashed-dotted line is the arithmetic average of the  latter two solutions. System parameters: $\Delta P=5$ kPa, $\dot{\gamma}=65$ s$^{-1}$, $a=10$ nm, $\phi_0=1.0\times10^{-3}$, and  $L_p=6.7\times 10^{-10}$ mPa$^{-1}$s$^{-1}$.}
\label{fig5}
\end{center}
\end{figure}

Figure \ref{fig5} depicts the CP layer profile at the membrane surface, $\phi_w(x)$, in its dependence on the reduced axial distance, $x/L$, from the fiber inlet (Fig. \ref{fig2}). The main figure part includes results of the similarity solution scheme discussed in Subsec. \ref{subsec:blayer}. The black solid line is the result for $\phi_w(x)$ based on Eq. (\ref{difuvirialcichocki}) for $D(\phi)/D_0$, and Eqs. (\ref{visczerofreq})-(\ref{viscgensaito}) for $\eta(\phi)/\eta_0$. It shows the typical UF behavior of $\phi_w(x)$ at non-fouling conditions, namely its monotonic concave-shaped increase with increasing $x$ caused by the steady-state advective-diffusive mass balance inside the CP layer. 
An asymptotic analysis based on Eqs. (\ref{advdiffdim}) and (\ref{bcdarcy}) shows that the surface concentration grows initially as 
\begin{eqnarray}\label{Ainitial}
  \frac{\phi_w(x)}{\phi_0} \approx 1 + A\;\!x^{1/3}\;\!, 
\end{eqnarray}
in the immediate vicinity of the fiber inlet (not resolved on the scale of Fig. \ref{fig5}), with $A = 1.857\times v_w^0/\left[\dot{\gamma}\;\!{D(\phi_0)}^2 \right]^{1/3}$. Thus, $\phi_w(x)$ shares the fractional $x$-dependence of the CP layer thickness $\delta(x)$ for very small $x/L$ . As expected, the initial growth of the surface concentration at the inlet is larger for larger TMP and larger membrane permeability $L_p$ (larger $v_w^0$) and smaller for larger characteristic shear rate $\dot{\gamma}$ and bulk diffusion coefficient $D(\phi_0)$. Since the curves for $\phi_w(x)$ are non-intersecting for different values of $\dot{\gamma}$, $D(\phi_0)$, and $v_w^0$, these trends remain valid for non-small values of $x/L$. 

In some earlier UF works \cite{Trettin:1980dm,Bhattacharjee:1999cz}, the concentration dependence of the collective diffusion coefficient, $D(\phi)$, for impermeable particles was accounted approximately, but the dispersion viscosity was treated as constant inside the thin boundary layer, of value equal to the bulk (feed) viscosity $\eta(\phi_0)$. It is noticed from the blue dotted line in Fig. \ref{fig5}, obtained for constant $\eta=\eta(\phi_0)$, that at larger distance from the inlet, the growth of $\phi_w(x)$ with increasing $x$ is underestimated in comparison to the surface concentration profile for variable viscosity, with the latter increasing for increasing concentration in accord with Fig. \ref{fig4}. The reason for this is, as quantified in Eq. (\ref{mom1}), that the axial particle transport lowering the CP layer formation is overestimated for a constant $\eta$ inside the layer taken equal to the minimal feed dispersion value. Commonly, $\phi_0 \ll 1$ so that the feed viscosity and collective diffusion coefficient of electrically neutral particle systems are practically equal to the fluid viscosity, $\eta_0$, and single-particle diffusion coefficient, $D_0$, respectively (see Figs. \ref{fig3} and \ref{fig4}).    

In most classical UF models, both $D$ and $\eta$ are described as being constant \cite{Blatt:1970hq,Jonsson:1984ee,Ilias:1993gc,Elimelech:1998jm,Sarkar:2008de} and set equal to the feed dispersion or infinite dilution values. The surface concentration profile (red dashed line in Fig. \ref{fig5}) for constant $D(\phi_0)$ and $\eta(\phi_0)$ is located in between that  for concentration-variable $D$ and $\eta$, and for variable $D(\phi)$ and constant $\eta=\eta(\phi_0)$ (blue dotted line), respectively, for the reason that the transverse diffusion flux of particles away from the membrane surface is underestimated when the minimal bulk dispersion value $D(\phi_0)$ is used instead of a concentration-variable $D(\phi)$ (see again Fig. \ref{fig3}). The main part of Fig. \ref{fig5} states that the precise modeling of the concentration dependence of $D$ and $\eta$ is key to a proper identification of operating conditions for which membrane fouling caused by the cake formation of jammed particles can be avoided. 

In the inset of Fig. \ref{fig5}, the similarity solution result for $\phi_w(x)$ with concentration-dependent $D(\phi)$ and $\eta(\phi)$ (black solid line) is compared with two semi-analytic solutions for constant $D$ and $\eta$ of the transcendental Eq. (\ref{phiwnew}) (colored dashed-dotted lines). The red dashed-dotted line is the result of the transcendental equation for constant $D(\phi_w)$ and $\eta(\phi_0)$ (largest value of $D$, given at the membrane surface, and smallest possible viscosity value, given in the bulk). Since the extent of CP layer formation is underestimated in this way, a lower bound for the surface concentration profile with concentration-dependent transport coefficients is obtained. An upper bound follows when the minimal diffusion coefficient $D(\phi_0)$ and the maximal viscosity $\eta(\phi_w)$  are used as constant input values in Eq.  (\ref{phiwnew}) (blue dashed-dotted line). The strong increase of the blue dashed-dotted line at small $x/L$ is a consequence of approximating $\eta$ by its maximal membrane surface value $\eta(\phi_w)$ throughout the CP layer. 

Sufficiently distant from the inlet, the concentration profile for concentration-variable $D$ and $\eta$ is well described by the arithmetic average of the two bounding red and blue dashed-dotted lines (black dashed-dotted line). Since the numerical solution of Eq. (\ref{phiwnew}) is fast and stable, the arithmetic average result is  convenient for disclosing unwanted operating conditions where cake formation by particle excluded volume solidification takes place.

\subsection{Microgel permeability effects}

\begin{figure}[b!]
\begin{center}
\includegraphics[width=1\linewidth]{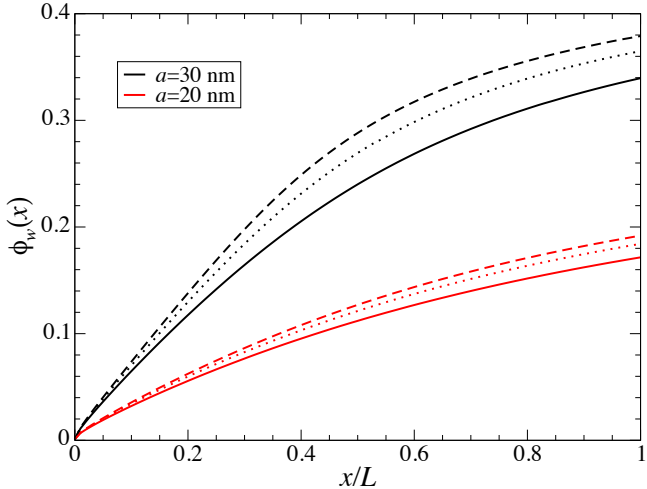}
\caption{Membrane surface concentration profile for permeable particles for two particle radii $a$ as indicated (black and red lines). Solid and dotted lines are similarity solution results for inverse permeability parameters $\chi=20$ and $\chi=50$, respectively. Dashed lines represent results for impermeable hard spheres, $\chi=\infty$. Other parameters: $\Delta P=300$ Pa, $\dot{\gamma}=75$ s$^{-1}$, $\phi_0=1.0\times10^{-3}$, and  $L_p=5\times10^{-9}$ mPa$^{-1}$s$^{-1}$.}
\label{fig6}
\end{center}
\end{figure}

We proceed by discussing the influence of the particle permeability on the CP layer, $\phi(x,y)$, and permeate flux,  $v_w(x)$, profiles. The results shown in the following have been obtained from the numerical solution of the similarity solution scheme in Subsec. \ref{subsec:blayer}. As input, we have used the factorization approximation results for $\eta(\chi,\phi)$, described by Eqs. (\ref{visczerofreq})-(\ref{viscgensaito}) and pictured in Fig. \ref{fig4}, and the polynomial fits to the high-precision simulation data of $D(\chi,\phi)$, depicted in Fig. \ref{fig3}.      

Figure \ref{fig6} shows how the membrane-surface concentration profile, $\phi_w(x)$, is affected by the particle permeability ($\chi=20$, $50$, and $\infty$), using two different particle radii $a$. We recall from Eq. (\ref{chikappa}) that $\chi=20$, for example, amounts to a hydrodynamic penetration depth equal to the twentieth part of the particle radius. According to Fig. \ref{fig6}, the membrane surface concentration decreases with increasing permeability (decreasing $\chi$). The reason for this is that for a given concentration $D(\chi,\phi)$ increases and $\eta(\phi)$ decreases with increasing permeability (see Figs. \ref{fig3} and \ref{fig4}). Both effects enlarge the number of particles leaving the CP layer, with $\phi_w$ being reduced accordingly. The diminution of $\phi_w(x)$ is most significant at the fiber outlet $x=L$. We also notice from Fig. \ref{fig6} a strongly enhanced surface concentration when the particle radius is enlarged from $20$ nm (red lines) to $30$ nm (black lines). This can be mainly attributed to the particle size dependence of the single-particle diffusion coefficient, $D_0(\chi)$, in Eq. (\ref{D0SE}), which is proportional to $1/a$ for constant $\chi$. Recall here the $\left[D(\chi,\phi_0)\right]^{-2/3}$ dependence of the coefficient $A$ in Eq. (\ref{Ainitial}), quantifying the surface concentration growth at the inlet.  

\begin{figure}[t!]
\begin{center}
\includegraphics[width=1\linewidth]{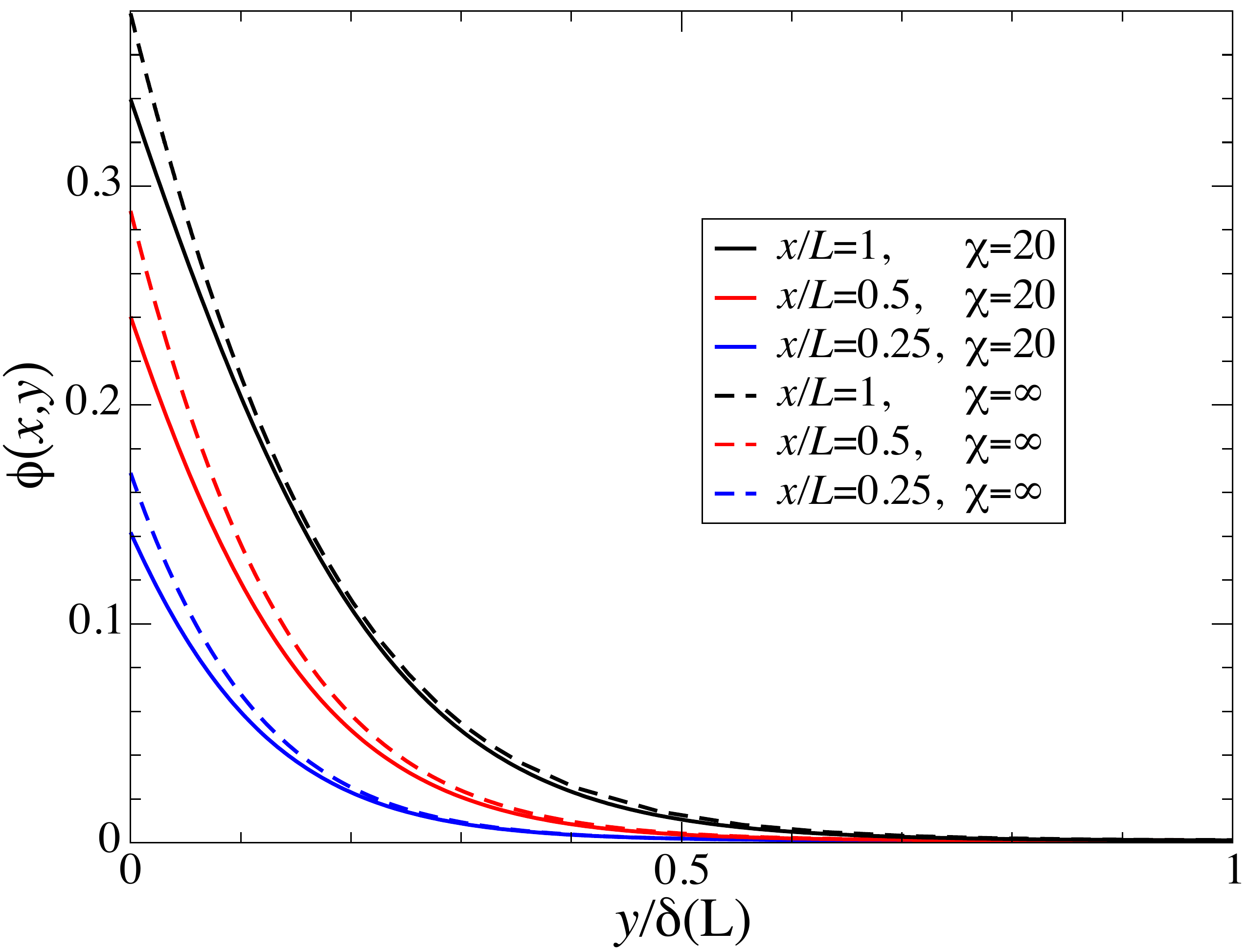}
\caption{Transversal boundary layer concentration profile, $\phi(x,y)$, at different axial positions $x$ as indicated (differently colored lines). The transverse distance, $y$, from the membrane surface is scaled by the CP layer thickness at the fiber outlet, $\delta(L)$. Solid lines: permeable particles with $\chi=20$. Dashed lines: impermeable particles, $\chi=\infty$. Other parameters as in Fig. \ref{fig6} for $a=30$ nm.}
\label{fig7}
\end{center}
\end{figure}

The transverse microgel concentration profile, $\phi(x,y)$, in the CP layer is shown in Fig. \ref{fig7} as a function of the distance $y$ from the membrane surface, for different reduced axial positions $x/L$ indicated in the figure. As described implicitly by Eq. (\ref{slope_of_phi}), the transverse profile decays strictly monotonically from the membrane surface value, $\phi_w(x)$, at $y=0$ towards the bulk concentration value, $\phi_0$, asymptotically reached for $y\gg \delta(x)$. Due to the gradual buildup of the CP layer along the fiber, the transverse particle concentration profile is larger for larger axial distance $x$ from the inlet. The effect of particle permeability is to decrease the transverse concentration profile. This decrease is most pronounced at the membrane surface (dashed and solid lines). 

\begin{figure}[t!]
\begin{center}
\includegraphics[width=1\linewidth]{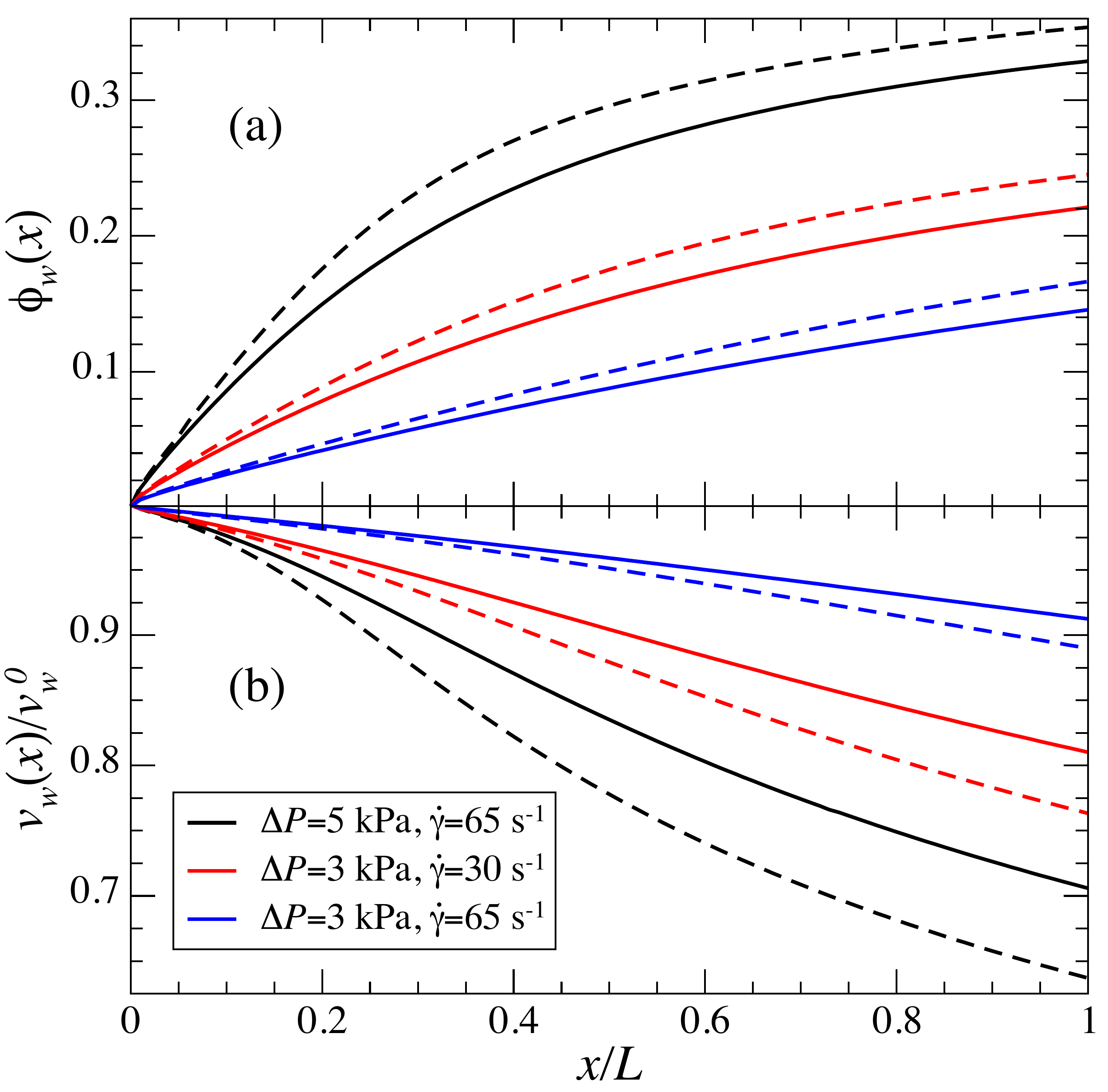}
\caption{(a) Surface concentration profile, $\phi_w(x)$, and (b) permeate velocity profile, $v_w(x)$, 
scaled by the pure solvent flux, $v_w^0$, for different TMPs, $\Delta P$, and shear rates, $\dot{\gamma}$, as indicated (differently colored lines). Solid lines: permeable particles with $\chi=20$. Dashed lines: impermeable particles, $\chi=\infty$. Other parameters as in Fig. \ref{fig5}.}
\label{fig8}
\end{center}
\end{figure}

Results for $\phi_w(x)$ and the permeate velocity profile, $v_w(x)$, along the membrane surface are shown in Figs. \ref{fig8}(a) and (b), respectively, for different combinations of $\Delta P$ and $\dot{\gamma}$ as indicated, and $\chi=20$ and $\infty$ (solid and dashed lines), respectively. The permeate velocity displayed in Fig. \ref{fig8} decreases monotonically with increasing $x$, due to the gradually sharpening CP layer structure: the particle concentration $\phi_w$ increases with the axial distance from the inlet, thus enhancing the osmotic pressure at the membrane surface, $\Pi(\phi_w)$, which in turn, according to Eq. (\ref{bcdarcy}), lowers the values of $v_w(x)$ for given TMP. Since $\phi_w$ increases as $x^{1/3}$ in the immediate vicinity of the fiber inlet, and since $\Pi(\phi_w) \propto \phi_w +{\cal O}(\phi_w^2)$ near the inlet where $\phi_w \approx \phi_0 \ll 1$, the permeate velocity decreases as $x^{1/3}$ very close to the fiber inlet. 

According to the black and blue lines in Fig. \ref{fig8}(a) for $\Delta P= 5$ kPa and $3$ kPa, respectively, an  enlargement of $\Delta P$ for fixed  $\dot{\gamma}$ causes an increase in $\phi_w$ due to the larger permeation drag experienced by the particles (larger $v_w^0$ in Eq. (\ref{fluidflux})). Quite interestingly, an increase of the TMP by $2$ kPa is overcompensated in the ratio $\Pi(\phi_w)/\Delta P$ in Eq. (\ref{bcdarcy}) by the associated enlargement of $\Pi(\phi_w)$, causing a decrease of the relative permeate velocity, $v_w/v_w^0$, for enlarged TMP as depicted in Fig. \ref{fig8}(b). While the relative permeate velocity is lowered, its absolute  value, $v_w$, is still enlarged for enlarged TMP. Different from this, if $\dot{\gamma}$, and thus the influx of the feed particles, increases for fixed $\Delta P$ (red and blue lines), the resulting enhanced axial drag gives rise to a smaller values of the membrane surface concentration and associated osmotic surface pressure and hence to a larger absolute permeate velocity. Notice here that $\delta(x) \propto \dot{\gamma}^{-1/3}$ according to Eq. (\ref{delta}). 

Figures \ref{fig8}(a) and (b) illustrate, in addition, the influence of the particle permeability: since $\phi_w$ decreases with increasing permeability, the permeate velocity, $v_w$, at large distances from the inlet is for particles with $\chi=20$ significantly larger than for impermeable ones (solid versus dashed lines).  

\begin{figure}[t!]
\begin{center}
\includegraphics[width=1\linewidth]{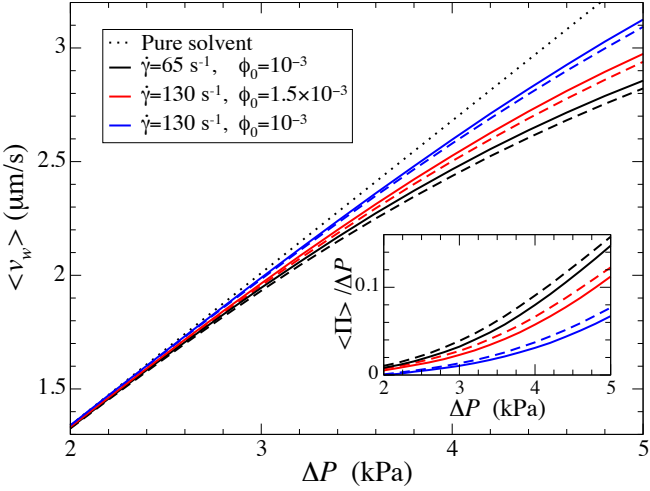}
\caption{Fiber-length-averaged permeate velocity, $\langle v_w\rangle$, as function of the applied TMP for different shear rates, $\dot{\gamma}$, and feed particle concentrations, $\phi_0$, as indicated (differently colored lines). Solid lines: permeable particles with $\chi=20$. Dashed lines: impermeable particles, $\chi=\infty$. The dotted line applies for pure solvent where $\langle v_w\rangle=L_p\Delta P$. 
Inset: fiber-length-averaged osmotic pressure, $\langle \Pi\rangle$, scaled by the TMP. Other parameters as in Fig. \ref{fig5}.}
\label{fig9}
\end{center}
\end{figure}

Consider next the fiber-length-averaged permeate velocity, 
\begin{equation}\label{vwaverage}
\langle v_w\rangle = \frac{1}{L}\int_0^L v_w(x) dx \;\!,
\end{equation}
which according to Eq. (\ref{bcdarcy}) is related to the fiber-length-averaged osmotic surface pressure, $\langle \Pi \rangle$, and the applied TMP by
\begin{equation}\label{opaverage}
\frac{\langle \Pi \rangle}{\Delta P} = 1- \frac{\langle v_w\rangle}{v_w^0}\;\!.
\end{equation}
 
The average permeate velocity is more easily accessible experimentally than the local axial velocity distribution $v_w(x)$. Results for $\langle v_w\rangle$ as a function of $\Delta P$ are presented in Fig. \ref{fig9} for different $\dot{\gamma}$ and feed concentrations $\phi_0$ as indicated. For small values of the TMP up to $\Delta P \approx 3$ kPa, the CP layer and the  associated osmotic surface pressure are so weak that $\langle v_w\rangle \approx v_w^0 = L_p\;\!\Delta P$ is valid. With further increasing TMP, the average permeate flux increases weaker than linearly in $\Delta P$. This is explained in the inset of Fig. \ref{fig9} which shows the pressure ratio $\langle \Pi \rangle/\Delta P$ as a function of $\Delta P$. The CP structure development along the membrane surface is such that $\langle \Pi \rangle$ increases more strongly than linearly with increasing $\Delta P$, giving rise to an average permeate velocity smaller than that for pure solvent feed, according to Eq. (\ref{bcdarcy}). Note further that $\langle v_w\rangle$ decreases with increasing $\phi_0$ since $\phi_w$ is increased. This effect becomes more significant with increasing $\Delta P$. Moreover, consistent with the observations made in Fig. \ref{fig8} for the axial permeate velocity distribution, $\langle v_w\rangle$ increases with increasing $\dot{\gamma}$ and with increasing particle permeability (dashed and solid lines).

The absence of a plateau region for $\langle v_w\rangle$ in Fig. \ref{fig9} at large TMP indicates that the operation  conditions for a limiting flux behavior at large TMP \cite{Probstein:2005vy} have not been approached. Operating UF under limiting flux conditions reduces the efficiency of the separation process. The discussion of $\langle v_w\rangle$ as a function of the applied TMP, such as the one in Fig. \ref{fig9}, is thus also of practical relevance. 

An important conclusion of the present subsection is that the particle permeability should be included in the UF modeling of non-ionic microgel dispersions. At large TMP, the permeate velocity (membrane surface concentration) of permeable particles is significantly larger (smaller) than the one of impermeable particles, for otherwise identical system parameters.

\subsection{Efficiency and cost indicators} \label{sec:indicators}

To assess the performance of a filtration process from different technical-economical viewpoints at given operating conditions, various efficiency and cost indicators can be introduced and evaluated. We discuss in the following several  indicators and their inter-relations. These indicators apply quite generally but are evaluated here in the context of the considered steady-state thin boundary layer cross-flow UF in a hollow cylindrical fiber membrane of length $L$ and inner radius $R$. A more detailed discussion of the process indicators is given in the supplementary information.    

The first efficiency indicator in the present list of indicators is the {\em Degree of Concentration Factor}, $\alpha \geq 1$, defined as the ratio of the particle volume fraction, $\phi_f$, in the finally obtained dispersion (retentate) to the feed dispersion volume fraction,
\begin{equation}\label{alpha}
\alpha=\frac{\phi_f}{\phi_0}=\frac{1}{1-\beta} \;\!.
\end{equation}

Here, $\beta < 1$ is the related so-called {\em Solvent Recovery} indicator which, by mass and dispersion volume conservation, is equal to the fraction of initial dispersion volume recovered in the permeate compartment as pure solvent. For fully developed homogeneous parabolic Poiseuille feed flow described by Eqs. (\ref{bulkshear}) and (\ref{bcbulkconc}), the second indicator is obtained in the supplementary information as 
\begin{equation}\label{beta}
\beta=\frac{4 L}{R}\frac{\langle v_w\rangle}{u_m} \;\!,
\end{equation}
where $u_m=\dot{\gamma} R/2$ is the feed inflow velocity at the center of the fiber.  

\begin{figure}[t!]
\begin{center}
\includegraphics[width=1\linewidth]{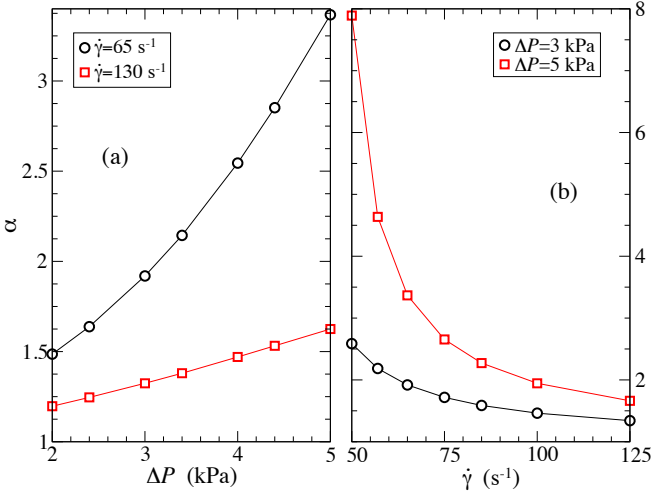}
\caption{Degree of Concentration Factor, $\alpha$, as a function of applied TMP (a) and shear rate (b). Other parameters as in Fig. \ref{fig5}.}
\label{fig10}
\end{center}
\end{figure}

According to Eq. (\ref{beta}), $\beta$ is proportional to $\langle v_w\rangle$, thus having an explicit $1/\dot{\gamma}$ dependence inherited from $u_m$. Considering the TMP and characteristic shear rate dependence of $\langle v_w\rangle$ displayed in Fig. \ref{fig9}, it follows that $\beta$ increases with increasing $\Delta P$, first linearly for small TMP values and subsequently sub-linearly for larger pressure values. In contrast, $\beta$ decreases monotonically with increasing $\dot{\gamma}$. Since $\alpha$ increases with increasing $\beta$ as quantified by Eq. (\ref{alpha}), the Degree of Concentration Factor $\alpha$ increases with increasing $\Delta P$ and decreases with increasing $\dot{\gamma}$, as expected. Figs. \ref{fig10}(a) and (b) show quantitative results for $\alpha$ as a function of $\Delta P$ and $\dot{\gamma}$, respectively, obtained from the similarity scheme solution for concentration-dependent transport properties.  

An important indicator for the material cost aspect of a baromembrane separation process is the {\em Productivity per Unit Membrane Area} indicator, $\theta$, defined as the dispersion volume flux at the outlet (retentate flux) divided by the membrane area. It can be expressed in terms of $\alpha$ and $\langle v_w\rangle$ as
\begin{equation}\label{theta}
\theta=\frac{\alpha^2 \,\! \langle v_w\rangle}{\alpha-1} \;\!.
\end{equation}

We refer to the supplementary information for the derivation of this equation. Note that $\theta$ has the physical dimension of a velocity. An alternative definition of the process productivity using the permeate rather than the retentate as the final product is discussed in Ref. \cite{Denisov:1994jh}. 

\begin{figure}[t!]
\begin{center}
\includegraphics[width=1\linewidth]{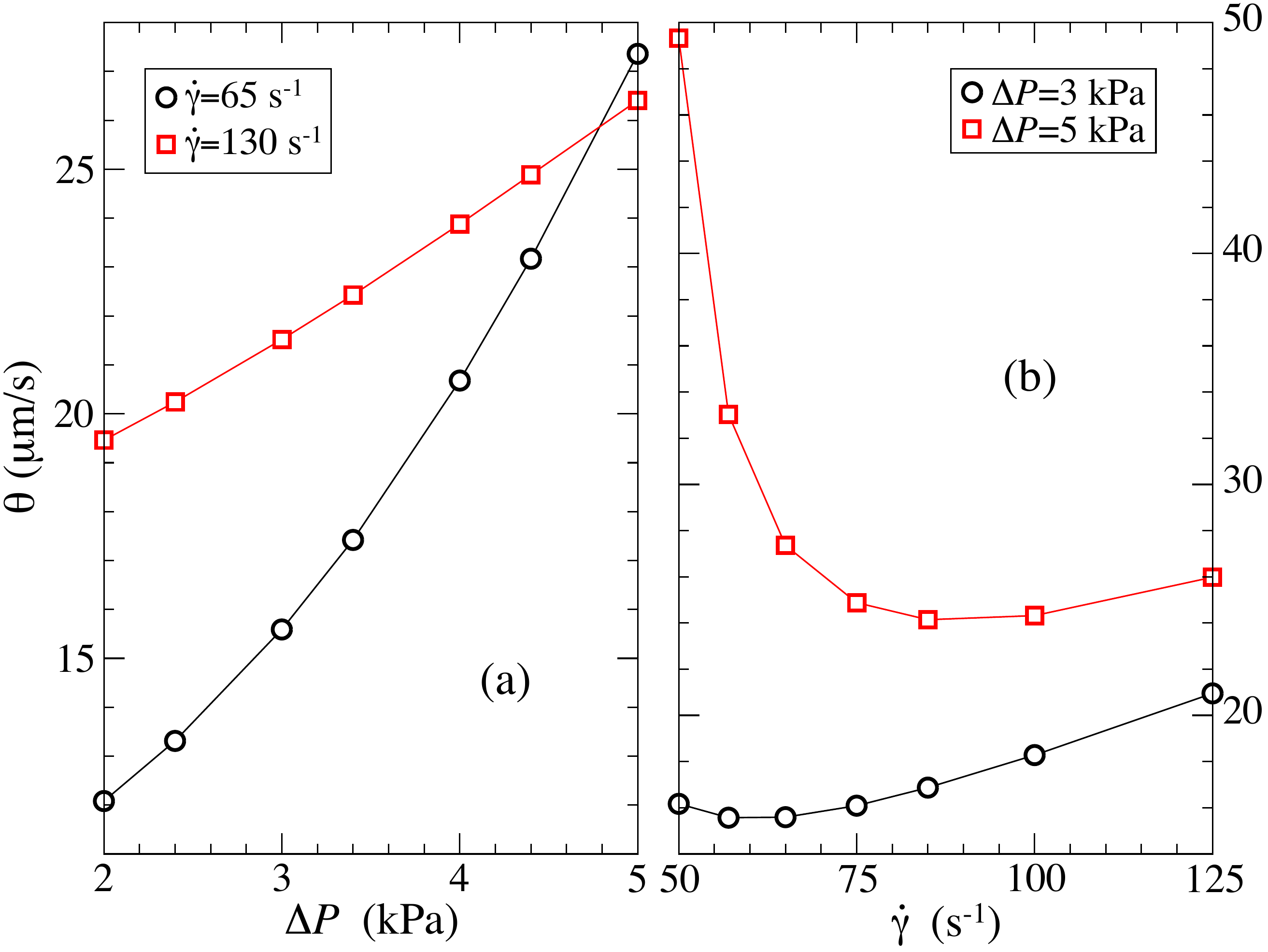}
\caption{Productivity per Unit Membrane Area, $\theta$, as a function of applied TMP (a) and shear rate (b). Other parameters as in Fig. \ref{fig5}.}
\label{fig11}
\end{center}
\end{figure}

As shown in Fig. \ref{fig11}(a), $\theta$ increases monotonically with increasing $\Delta P$, with a more pronounced  upturn for the smaller considered shear rate. Quite interestingly, the behavior of $\theta$ as a function of $\dot{\gamma}$ is non-monotonic, with a minimum value at a shear rate that increases with increasing TMP (Fig. \ref{fig11}(b)). Increasing the shear rate is associated with two opposite trends. The first one is a reduction in the CP layer that results in a decrease of the osmotic pressure opposing the TMP. This leads to a diminution of $\theta$. The second trend is an increase in the feed flow that leads to an increase in the outlet flow, with a correspondingly enlarged indicator $\theta$. At sufficiently low shear rates, the osmotic pressure noticeably compensates the TMP. Therefore, the first noted trend predominates thus lowering the CP layer profile accompanied by a stronger permeate flux and a lower outlet flow. When the shear rate is sufficiently high, the CP layer profile is reduced to such an extent that its further reduction does not substantially affect the permeate flux any more. Hence, the second noted trend leads to an increase in $\theta$ with increasing $\dot{\gamma}$.

We now discuss two energy cost indicators. The first one, termed {\em Specific Energy Consumption} indicator, $\omega$, is defined as the energy consumed for producing a unit volume of final product. In earlier related works \cite{Manth:2002fr,Zhu:2009et,Zhu:2009ir,Li:2010ho,Li:2011fr,Sharif:2009vk,Sarkar:2009de}, the final product has been identified with the permeate (for example, with desalinated water in case of desalination studies) whereas the final product is identified here with the retaining dispersion of concentration $\phi_f$. Moreover, in earlier works the CP layer has been either completely disregarded or treated in a simplified way using a semi-empirical parameter \cite{Sharif:2009vk}. Considering that basically the whole externally supplied power is consumed for pressing the solvent through the membrane, it is shown in the supplementary information that $\omega$ can be expressed as  
\begin{equation}\label{omega}
\omega=\left(\alpha-1\right)\Delta P \;\!,
\end{equation}
so that its shear rate dependence is inherited from that of $\alpha$. Consequently, $\omega$ decreases with increasing $\dot{\gamma}$. Moreover, it strongly increases with increasing $\Delta P$, quadratically in the pressure for small applied TMP or high shear rates, and more stronger than linearly but weaker than quadratically for larger TMP values.

While the indicators $\alpha$, $\theta$, and $\omega$ discussed thus far favor large values of the applied TMP for optimizing the one-stage cross-flow UF process, this conflicts with the second energy cost indicator considered here, namely the indicator $\epsilon \leq1$ of {\em Specific Energy Efficiency}. This indicator is defined \cite{Sharif:2009vk} as the ratio of the thermodynamically necessary minimal work per unit volume of final product, $\omega_\mathrm{min}$, to the energy per volume, $\omega$, in Eq. (\ref{omega}), consumed during the steady-state filtration process,
\begin{equation}\label{epsilon}
\epsilon=\frac{\omega_\mathrm{min}}{\omega}=\frac{\alpha\;\! \phi_0}{(\alpha-1)}
\int_{\phi_0}^{\phi_f}\left(\frac{\Pi(\phi)}{\Delta P}\right)\frac{d\phi}{\phi^2}\;\!.
\end{equation}

Explicitly, 
\begin{equation}\label{omegamin}
\omega_\mathrm{min}=-\frac{1}{V_f}\int_{V_0}^{V_f}\Pi\;\!dV=\phi_f\int_{\phi_0}^{\phi_f}
\Pi(\phi)\;\!\frac{d\phi}{\phi^2}
\end{equation}
is the minimal work per unit volume of retentate required in a adiabatically slow reversible thermodynamic process of isothermally compressing the dispersion from its initial volume $V_0$ to the final volume $V_f$ with the help of a fully particle-retentive non-adsorbing membrane employed as a plunger. We have used the relation $V=(4\pi/3)a^3N/\phi$, with $N$ the total number of mobile particles that remains constant during the compression. Note that the minimal work required in the adiabatically slow reverse-osmosis compression is equal to the change in free energy of the system minus the energy required for lowering the number of solvent molecules in the system during the compression. The integral in Eq. (\ref{epsilon}) can be performed analytically, using the CS Eq. (\ref{Zcarnahanstarling}) as input for the osmotic pressure $\Pi(\phi)$. The resulting expression for $\epsilon$ is given in the supplementary information.    

\begin{figure}[t!]
\begin{center}
\includegraphics[width=1\linewidth]{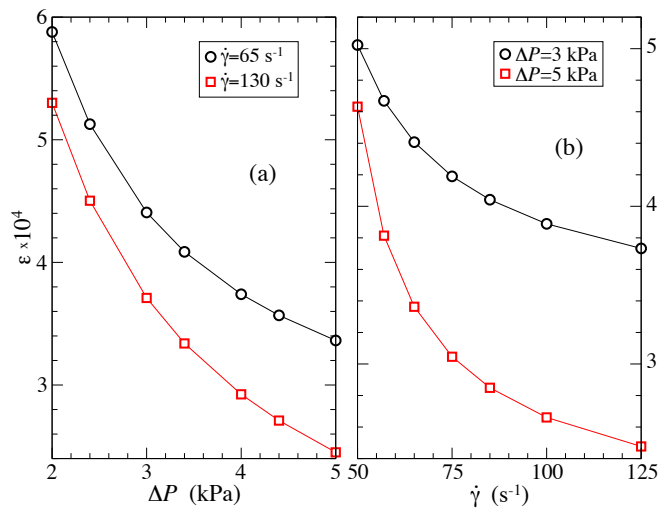}
\caption{Specific Energy Efficiency, $\epsilon$, as a function of applied TMP (a) and shear rate (b). Other parameters as in Fig. \ref{fig5}.}
\label{fig12}
\end{center}
\end{figure}

The behavior of the Specific Energy Efficiency as a function of TMP and characteristic shear rate is shown in Figs. \ref{fig12}(a) and (b), respectively. As expected, $\epsilon$ decreases with increasing $\dot{\gamma}$, but different from the earlier discussed indicators, $\epsilon$ decreases with increasing $\Delta P$. It should be noticed here that while the Degree of Concentration Factor $\alpha$ increases with increasing TMP, the factor $\alpha/(1-\alpha)$ in Eq. (\ref{epsilon}) decreases. We finally point to the small values of the Specific Energy Efficiency in Fig. \ref{fig12}. This is due basically to the comparatively large particle hard-core radius $a=10$ nm for which the osmotic pressure buildup along the membrane surface is small as described by the size dependence $\Pi \propto 1/a^3$.

\section{Conclusions}\label{sec:conclusions}

We have presented the first comprehensive theoretical study of particle permeability effects in the UF of non-ionic microgel dispersions modeled as solvent-permeable Brownian spheres. Based on a boundary layer analysis, we have provided a systematic study of inside-out cross-flow UF. The CP layer and permeate flux profiles have been calculated using accurate analytic expressions for the concentration-dependent collective diffusion coefficient, $D(\phi)$, and zero-frequency viscosity, $\eta(\phi)$, of dispersions of permeable particles. The employed expressions for these transport coefficients are well tested against computer simulation results and experiments \cite{Riest:2015jo}. They should be used instead of the substantially less accurate cell model results and the semi-empirical Krieger-Dougherty and Eilers expressions employed in earlier theoretical UF studies, which have dealt mostly with impermeable colloidal particles and proteins.  

We have shown that the concentration dependence both of $D$ and $\eta$ significantly influences the CP layer, in particular right at the membrane surface, with a correspondingly strong influence on the osmotic pressure profile and therefore on the permeate velocity. The usage of constant values for $D$ and $\eta$, if taken in particular for the common situation of low feed volume concentration $\phi_0$, results in the underestimation of the CP layer structure and hence in the overestimation of the permeate flux. The applicability of the transcendental Eq. (\ref{phiwnew}) for the membrane surface concentration profile, $\phi_w(x)$, with constant transport coefficients has been discussed (inset in Fig. \ref{fig5}). It can be used to generate an arithmetic average concentration profile in good agreement with calculations of $\phi_w(x)$ for concentration-dependent $D$ and $\eta$ at non-small distances $x$ from the fiber inlet. Since the numerical solution of the transcendental equation is quick and stable, the arithmetic-average profile can be profitably used, for example, in determining the axial range extending downstream from the inlet wherein the membrane is devoid of a stagnant cake layer \cite{Romero:1988vl,Romero:1990ei,vanderSman:2008gj}.

We have shown that the effect of non-zero microgel permeability is to reduce the particle concentration at the membrane surface with a consequently enlarged permeate velocity. It is worth to emphasize again that our study of permeability effects is relevant not only for non-ionic microgels, modeled here hydrodynamically as uniformly permeable spheres, but also for other hydrodynamically structured particles such as core-shell particles with a dry rigid core and an outer permeable polymer brush layer. As pointed out earlier and discussed in Ref. \cite{Riest:2015jo}, the reason for this is that even particles with a complicated internal structure are    hydrodynamically fully characterized, for quite general conditions, by their hydrodynamic radius.

Various efficiency and cost indicators assessing the steady-state UF process from different technical-economical viewpoints have been discussed in dependence of the process parameters and with full account of the CP layer structure. A detailed description of the indicators and their inter-relations is given in the supplementary information. Different indicators favor partially conflicting optimization strategies. For example, while an enlarged TMP leads to a larger Degree of Concentration factor, $\alpha$, and Productivity per Unit Membrane Area, $\theta$, the price to pay is a reduced Specific Energy Efficiency, $\epsilon$. The discussed indicators are of value going beyond the specific permeate-particles cross-flow UF process discussed in the paper. They can be used, for example, to evaluate different optimization strategies employing multistage processes \cite{Manth:2002fr,Zhu:2009et,Zhu:2009ir}, re-use of free energy accumulated in a filtration process \cite{Li:2010ho,Li:2011fr,Sharif:2009vk}, external electric fields \cite{Sarkar:2009de}, local production of flow instabilities \cite{Jaffrin:2012fo}, and the generation of high shear rates by moving parts \cite{Jaffrin:2012jx}. We expect our theoretical work to be useful in future experimental explorations of microgel cross-flow UF.              

The present paper was focused on identifying and analyzing hydrodynamic effects on UF arising from the particle permeability and concentration-permeability dependencies of $D$ and $\eta$, with fouling effects disregarded. The dynamics of cake layer formation by permeable particles can be described using cross-sectional averaged forms of the time-dependent advection-diffusion and stationary momentum balance equations \cite{Romero:1988vl,Romero:1990ei,Vollebregt:2010jk,vanderSman:2013fv}, with the local hydraulic resistance of the cake layer added to the clear membrane resistance $R_m$. 

In dealing with UF of microgel dispersions, also particle polydispersity and deformability play a role. Size polydispersity is expected to enlarge the particle concentration in the cake and CP layers. Moreover, there should be an enhanced tendency of membrane pore clogging by the small particles of the size distribution. There is also a tendency of size fractionation triggered by the transversal flow which, however, is counteracted by the strong Brownian motion in UF. The theoretical treatment of polydispersity in UF is a difficult task since it requires the additional consideration of collective cross-diffusion coefficients relating the diffusion flux of one component to the local concentration gradient of another one \cite{Nagele:1996hr,vanderSman:2012df,Wang:2015gf}. These coefficients are affected, in particular, by HIs causing a hydrodynamic dragging-along of small particles by larger ones.

For ionic microgels in polar solvents, the effective particle and membrane surface charges are additional important parameters \cite{Bowen:1995ft,Kim:2006en,Sarkar:2008de}. As shown in Ref. \cite{Holmqvist:2012hv}, while the permeability of ionic microgels is hydrodynamically less influential than that of uncharged ones, due to the longer-ranged electrostatic repulsion, the ionic microgels tend to shrink substantially with increasing concentration in the swollen-state temperature range. Moreover, the softness (single-particle compressibility) of ionic microgels influences the width of the fluid-solid coexistence region, rendering it smaller for stiffer microgels \cite{PelaezFernandez:2015hu}. Thus, the size shrinkage with increasing concentration and the softness of ionic microgels must be included in a realistic modeling of the CP and cake layers. Work by the present authors on filtration modeling of polydisperse and charge-stabilized rigid colloidal particles and soft ionic microgels is in progress. Furthermore, we are currently involved in an experimental-theoretical study of cross-flow filtration of microgel dispersions \cite{SFB985B6}. The present theoretical results in particular for the concentration factor and energy efficiency indicators are used to analyze the performance and to design the filtration experiments. The outcome of this ongoing study will be communicated in a future publication.


\begin{acknowledgments}
We thank J. Riest and J. Dhont for helpful discussions about microgel dynamics, and B. Bl\"umich, P. Buzatu, M. K\"uppers, J. Viess, and M. Wessling for fruitful discussions about microgel filtration and dispersion flow visualization. Financial support by the Deutsche Forschungsgemeinschaft (SFB-985, Project B6) is gratefully acknowledged.
\end{acknowledgments}

\bibliographystyle{apsrev4-1}
\bibliography{filtration_refs.bib}

\begin{thebibliography}{86}%
\makeatletter
\providecommand \@ifxundefined [1]{%
 \@ifx{#1\undefined}
}%
\providecommand \@ifnum [1]{%
 \ifnum #1\expandafter \@firstoftwo
 \else \expandafter \@secondoftwo
 \fi
}%
\providecommand \@ifx [1]{%
 \ifx #1\expandafter \@firstoftwo
 \else \expandafter \@secondoftwo
 \fi
}%
\providecommand \natexlab [1]{#1}%
\providecommand \enquote  [1]{``#1''}%
\providecommand \bibnamefont  [1]{#1}%
\providecommand \bibfnamefont [1]{#1}%
\providecommand \citenamefont [1]{#1}%
\providecommand \href@noop [0]{\@secondoftwo}%
\providecommand \href [0]{\begingroup \@sanitize@url \@href}%
\providecommand \@href[1]{\@@startlink{#1}\@@href}%
\providecommand \@@href[1]{\endgroup#1\@@endlink}%
\providecommand \@sanitize@url [0]{\catcode `\\12\catcode `\$12\catcode
  `\&12\catcode `\#12\catcode `\^12\catcode `\_12\catcode `\%12\relax}%
\providecommand \@@startlink[1]{}%
\providecommand \@@endlink[0]{}%
\providecommand \url  [0]{\begingroup\@sanitize@url \@url }%
\providecommand \@url [1]{\endgroup\@href {#1}{\urlprefix }}%
\providecommand \urlprefix  [0]{URL }%
\providecommand \Eprint [0]{\href }%
\providecommand \doibase [0]{http://dx.doi.org/}%
\providecommand \selectlanguage [0]{\@gobble}%
\providecommand \bibinfo  [0]{\@secondoftwo}%
\providecommand \bibfield  [0]{\@secondoftwo}%
\providecommand \translation [1]{[#1]}%
\providecommand \BibitemOpen [0]{}%
\providecommand \bibitemStop [0]{}%
\providecommand \bibitemNoStop [0]{.\EOS\space}%
\providecommand \EOS [0]{\spacefactor3000\relax}%
\providecommand \BibitemShut  [1]{\csname bibitem#1\endcsname}%
\let\auto@bib@innerbib\@empty
\bibitem [{\citenamefont {Holmqvist}\ \emph {et~al.}(2012)\citenamefont
  {Holmqvist}, \citenamefont {Mohanty}, \citenamefont {N{\"a}gele},
  \citenamefont {Schurtenberger},\ and\ \citenamefont
  {Heinen}}]{Holmqvist:2012hv}%
  \BibitemOpen
  \bibfield  {author} {\bibinfo {author} {\bibfnamefont {P.}~\bibnamefont
  {Holmqvist}}, \bibinfo {author} {\bibfnamefont {P.~S.}\ \bibnamefont
  {Mohanty}}, \bibinfo {author} {\bibfnamefont {G.}~\bibnamefont {N{\"a}gele}},
  \bibinfo {author} {\bibfnamefont {P.}~\bibnamefont {Schurtenberger}}, \ and\
  \bibinfo {author} {\bibfnamefont {M.}~\bibnamefont {Heinen}},\ }\href@noop {}
  {\bibfield  {journal} {\bibinfo  {journal} {Phys. Rev. Lett.}\ }\textbf
  {\bibinfo {volume} {109}},\ \bibinfo {pages} {048302} (\bibinfo {year}
  {2012})}\BibitemShut {NoStop}%
\bibitem [{\citenamefont {Pan}\ \emph {et~al.}(2015)\citenamefont {Pan},
  \citenamefont {Torres}, \citenamefont {Hoare},\ and\ \citenamefont
  {Ghosh}}]{Pan:2015bb}%
  \BibitemOpen
  \bibfield  {author} {\bibinfo {author} {\bibfnamefont {S.}~\bibnamefont
  {Pan}}, \bibinfo {author} {\bibfnamefont {J.~M. G.~T.}\ \bibnamefont
  {Torres}}, \bibinfo {author} {\bibfnamefont {T.}~\bibnamefont {Hoare}}, \
  and\ \bibinfo {author} {\bibfnamefont {R.}~\bibnamefont {Ghosh}},\
  }\href@noop {} {\bibfield  {journal} {\bibinfo  {journal} {J. Membr. Sci.}\
  }\textbf {\bibinfo {volume} {479}},\ \bibinfo {pages} {141} (\bibinfo {year}
  {2015})}\BibitemShut {NoStop}%
\bibitem [{\citenamefont {Fernandez-Nieves}\ \emph {et~al.}(2011)\citenamefont
  {Fernandez-Nieves}, \citenamefont {Wyss}, \citenamefont {Mattsson},\ and\
  \citenamefont {Weitz}}]{FernandezNieves:2011uv}%
  \BibitemOpen
  \bibinfo {editor} {\bibfnamefont {A.}~\bibnamefont {Fernandez-Nieves}},
  \bibinfo {editor} {\bibfnamefont {H.}~\bibnamefont {Wyss}}, \bibinfo {editor}
  {\bibfnamefont {J.}~\bibnamefont {Mattsson}}, \ and\ \bibinfo {editor}
  {\bibfnamefont {D.~A.}\ \bibnamefont {Weitz}},\ eds.,\ \href@noop {} {\emph
  {\bibinfo {title} {{Microgel Suspensions: Fundamentals and Applications}}}}\
  (\bibinfo  {publisher} {Wiley-VCH},\ \bibinfo {address} {Weinheim},\ \bibinfo
  {year} {2011})\BibitemShut {NoStop}%
\bibitem [{\citenamefont {Siebenb{\"u}rger}\ \emph {et~al.}(2012)\citenamefont
  {Siebenb{\"u}rger}, \citenamefont {Fuchs},\ and\ \citenamefont
  {Ballauff}}]{Siebenburger:2012ex}%
  \BibitemOpen
  \bibfield  {author} {\bibinfo {author} {\bibfnamefont {M.}~\bibnamefont
  {Siebenb{\"u}rger}}, \bibinfo {author} {\bibfnamefont {M.}~\bibnamefont
  {Fuchs}}, \ and\ \bibinfo {author} {\bibfnamefont {M.}~\bibnamefont
  {Ballauff}},\ }\href@noop {} {\bibfield  {journal} {\bibinfo  {journal} {Soft
  Matter}\ }\textbf {\bibinfo {volume} {8}},\ \bibinfo {pages} {4014} (\bibinfo
  {year} {2012})}\BibitemShut {NoStop}%
\bibitem [{\citenamefont {Saunders}\ and\ \citenamefont
  {Vincent}(1999)}]{Saunders:1999ec}%
  \BibitemOpen
  \bibfield  {author} {\bibinfo {author} {\bibfnamefont {B.~R.}\ \bibnamefont
  {Saunders}}\ and\ \bibinfo {author} {\bibfnamefont {B.}~\bibnamefont
  {Vincent}},\ }\href@noop {} {\bibfield  {journal} {\bibinfo  {journal} {Adv.
  Colloid Interface Sci.}\ }\textbf {\bibinfo {volume} {80}},\ \bibinfo {pages}
  {1} (\bibinfo {year} {1999})}\BibitemShut {NoStop}%
\bibitem [{\citenamefont {Blatt}\ \emph {et~al.}(1970)\citenamefont {Blatt},
  \citenamefont {Dravid}, \citenamefont {Michaels},\ and\ \citenamefont
  {Nelsen}}]{Blatt:1970hq}%
  \BibitemOpen
  \bibfield  {author} {\bibinfo {author} {\bibfnamefont {W.~F.}\ \bibnamefont
  {Blatt}}, \bibinfo {author} {\bibfnamefont {A.}~\bibnamefont {Dravid}},
  \bibinfo {author} {\bibfnamefont {A.~S.}\ \bibnamefont {Michaels}}, \ and\
  \bibinfo {author} {\bibfnamefont {L.}~\bibnamefont {Nelsen}},\ }in\
  \href@noop {} {\emph {\bibinfo {booktitle} {Membrane Science and
  Technology}}}\ (\bibinfo  {publisher} {Springer US},\ \bibinfo {address}
  {Boston, MA},\ \bibinfo {year} {1970})\ pp.\ \bibinfo {pages}
  {47--97}\BibitemShut {NoStop}%
\bibitem [{\citenamefont {Bowen}\ and\ \citenamefont
  {Jenner}(1995{\natexlab{a}})}]{Bowen:1995do}%
  \BibitemOpen
  \bibfield  {author} {\bibinfo {author} {\bibfnamefont {W.~R.}\ \bibnamefont
  {Bowen}}\ and\ \bibinfo {author} {\bibfnamefont {F.}~\bibnamefont {Jenner}},\
  }\href@noop {} {\bibfield  {journal} {\bibinfo  {journal} {Adv. Colloid
  Interface Sci.}\ }\textbf {\bibinfo {volume} {56}},\ \bibinfo {pages} {141}
  (\bibinfo {year} {1995}{\natexlab{a}})}\BibitemShut {NoStop}%
\bibitem [{\citenamefont {Shannon}\ \emph {et~al.}(2008)\citenamefont
  {Shannon}, \citenamefont {Bohn}, \citenamefont {Elimelech}, \citenamefont
  {Georgiadis}, \citenamefont {Mari{\~n}as},\ and\ \citenamefont
  {Mayes}}]{Shannon:2008bk}%
  \BibitemOpen
  \bibfield  {author} {\bibinfo {author} {\bibfnamefont {M.~A.}\ \bibnamefont
  {Shannon}}, \bibinfo {author} {\bibfnamefont {P.~W.}\ \bibnamefont {Bohn}},
  \bibinfo {author} {\bibfnamefont {M.}~\bibnamefont {Elimelech}}, \bibinfo
  {author} {\bibfnamefont {J.~G.}\ \bibnamefont {Georgiadis}}, \bibinfo
  {author} {\bibfnamefont {B.~J.}\ \bibnamefont {Mari{\~n}as}}, \ and\ \bibinfo
  {author} {\bibfnamefont {A.~M.}\ \bibnamefont {Mayes}},\ }\href@noop {}
  {\bibfield  {journal} {\bibinfo  {journal} {Nature}\ }\textbf {\bibinfo
  {volume} {452}},\ \bibinfo {pages} {301} (\bibinfo {year}
  {2008})}\BibitemShut {NoStop}%
\bibitem [{\citenamefont {Nakamura}\ and\ \citenamefont
  {Matsumoto}(2013)}]{Nakamura:2013ck}%
  \BibitemOpen
  \bibfield  {author} {\bibinfo {author} {\bibfnamefont {K.}~\bibnamefont
  {Nakamura}}\ and\ \bibinfo {author} {\bibfnamefont {K.}~\bibnamefont
  {Matsumoto}},\ }\href@noop {} {\bibfield  {journal} {\bibinfo  {journal}
  {Membranes}\ }\textbf {\bibinfo {volume} {3}},\ \bibinfo {pages} {87}
  (\bibinfo {year} {2013})}\BibitemShut {NoStop}%
\bibitem [{\citenamefont {Noble}\ and\ \citenamefont
  {Stern}(1995)}]{Noble:1995tc}%
  \BibitemOpen
  \bibinfo {editor} {\bibfnamefont {R.~D.}\ \bibnamefont {Noble}}\ and\
  \bibinfo {editor} {\bibfnamefont {S.~A.}\ \bibnamefont {Stern}},\ eds.,\
  \href@noop {} {\emph {\bibinfo {title} {{Membrane Separations
  Technology:~Principles and Applications (Membrane Science and Technology,
  vol. 2) }}}}\ (\bibinfo  {publisher} {Elsevier},\ \bibinfo {address}
  {Amsterdam},\ \bibinfo {year} {1995})\BibitemShut {NoStop}%
\bibitem [{\citenamefont {Kozinski}\ and\ \citenamefont
  {Lightfoot}(1972)}]{Kozinski:1972fy}%
  \BibitemOpen
  \bibfield  {author} {\bibinfo {author} {\bibfnamefont {A.~A.}\ \bibnamefont
  {Kozinski}}\ and\ \bibinfo {author} {\bibfnamefont {E.~N.}\ \bibnamefont
  {Lightfoot}},\ }\href@noop {} {\bibfield  {journal} {\bibinfo  {journal}
  {AIChE Journal}\ }\textbf {\bibinfo {volume} {18}},\ \bibinfo {pages} {1030}
  (\bibinfo {year} {1972})}\BibitemShut {NoStop}%
\bibitem [{\citenamefont {Belfort}\ \emph {et~al.}(1994)\citenamefont
  {Belfort}, \citenamefont {Davis},\ and\ \citenamefont
  {Zydney}}]{Belfort:1993if}%
  \BibitemOpen
  \bibfield  {author} {\bibinfo {author} {\bibfnamefont {G.}~\bibnamefont
  {Belfort}}, \bibinfo {author} {\bibfnamefont {R.~H.}\ \bibnamefont {Davis}},
  \ and\ \bibinfo {author} {\bibfnamefont {A.~L.}\ \bibnamefont {Zydney}},\
  }\href@noop {} {\bibfield  {journal} {\bibinfo  {journal} {J. Membr. Sci.}\
  }\textbf {\bibinfo {volume} {96}},\ \bibinfo {pages} {1} (\bibinfo {year}
  {1994})}\BibitemShut {NoStop}%
\bibitem [{\citenamefont {Grimellec}\ \emph {et~al.}(1975)\citenamefont
  {Grimellec}, \citenamefont {Poujeol}, \citenamefont {Rouffignac},
  \citenamefont {Philippe},\ and\ \citenamefont {Malorey}}]{Grimellec:1975ib}%
  \BibitemOpen
  \bibfield  {author} {\bibinfo {author} {\bibfnamefont {C.}~\bibnamefont
  {Grimellec}}, \bibinfo {author} {\bibfnamefont {P.}~\bibnamefont {Poujeol}},
  \bibinfo {author} {\bibfnamefont {C.}~\bibnamefont {Rouffignac}}, \bibinfo
  {author} {\bibfnamefont {P.}~\bibnamefont {Philippe}}, \ and\ \bibinfo
  {author} {\bibfnamefont {P.}~\bibnamefont {Malorey}},\ }\href@noop {}
  {\bibfield  {journal} {\bibinfo  {journal} {Pflugers Arch}\ }\textbf
  {\bibinfo {volume} {354}},\ \bibinfo {pages} {117} (\bibinfo {year}
  {1975})}\BibitemShut {NoStop}%
\bibitem [{\citenamefont {Zydney}\ and\ \citenamefont
  {Colton}(1986)}]{Zydney:1986hv}%
  \BibitemOpen
  \bibfield  {author} {\bibinfo {author} {\bibfnamefont {A.~L.}\ \bibnamefont
  {Zydney}}\ and\ \bibinfo {author} {\bibfnamefont {C.~K.}\ \bibnamefont
  {Colton}},\ }\href@noop {} {\bibfield  {journal} {\bibinfo  {journal} {Chem.
  Eng. Commun.}\ }\textbf {\bibinfo {volume} {47}},\ \bibinfo {pages} {1}
  (\bibinfo {year} {1986})}\BibitemShut {NoStop}%
\bibitem [{\citenamefont {Romero}\ and\ \citenamefont
  {Davis}(1988)}]{Romero:1988vl}%
  \BibitemOpen
  \bibfield  {author} {\bibinfo {author} {\bibfnamefont {C.~A.}\ \bibnamefont
  {Romero}}\ and\ \bibinfo {author} {\bibfnamefont {R.~H.}\ \bibnamefont
  {Davis}},\ }\href@noop {} {\bibfield  {journal} {\bibinfo  {journal} {J.
  Membr. Sci.}\ }\textbf {\bibinfo {volume} {39}},\ \bibinfo {pages} {157}
  (\bibinfo {year} {1988})}\BibitemShut {NoStop}%
\bibitem [{\citenamefont {Pelekasis}(1998)}]{Pelekasis:1998jj}%
  \BibitemOpen
  \bibfield  {author} {\bibinfo {author} {\bibfnamefont {N.}~\bibnamefont
  {Pelekasis}},\ }\href@noop {} {\bibfield  {journal} {\bibinfo  {journal}
  {Chem. Eng. Sci.}\ }\textbf {\bibinfo {volume} {53}},\ \bibinfo {pages}
  {3469} (\bibinfo {year} {1998})}\BibitemShut {NoStop}%
\bibitem [{\citenamefont {Kromkamp}\ \emph {et~al.}(2005)\citenamefont
  {Kromkamp}, \citenamefont {Bastiaanse}, \citenamefont {Swarts}, \citenamefont
  {Brans}, \citenamefont {van~der Sman},\ and\ \citenamefont
  {Boom}}]{Kromkamp:2005kc}%
  \BibitemOpen
  \bibfield  {author} {\bibinfo {author} {\bibfnamefont {J.}~\bibnamefont
  {Kromkamp}}, \bibinfo {author} {\bibfnamefont {A.}~\bibnamefont
  {Bastiaanse}}, \bibinfo {author} {\bibfnamefont {J.}~\bibnamefont {Swarts}},
  \bibinfo {author} {\bibfnamefont {G.}~\bibnamefont {Brans}}, \bibinfo
  {author} {\bibfnamefont {R.~G.~M.}\ \bibnamefont {van~der Sman}}, \ and\
  \bibinfo {author} {\bibfnamefont {R.~M.}\ \bibnamefont {Boom}},\ }\href@noop
  {} {\bibfield  {journal} {\bibinfo  {journal} {J. Membr. Sci.}\ }\textbf
  {\bibinfo {volume} {253}},\ \bibinfo {pages} {67} (\bibinfo {year}
  {2005})}\BibitemShut {NoStop}%
\bibitem [{\citenamefont {Vollebregt}\ \emph {et~al.}(2010)\citenamefont
  {Vollebregt}, \citenamefont {van~der Sman},\ and\ \citenamefont
  {Boom}}]{Vollebregt:2010jk}%
  \BibitemOpen
  \bibfield  {author} {\bibinfo {author} {\bibfnamefont {H.~M.}\ \bibnamefont
  {Vollebregt}}, \bibinfo {author} {\bibfnamefont {R.~G.~M.}\ \bibnamefont
  {van~der Sman}}, \ and\ \bibinfo {author} {\bibfnamefont {R.~M.}\
  \bibnamefont {Boom}},\ }\href@noop {} {\bibfield  {journal} {\bibinfo
  {journal} {Soft Matter}\ }\textbf {\bibinfo {volume} {6}},\ \bibinfo {pages}
  {6052} (\bibinfo {year} {2010})}\BibitemShut {NoStop}%
\bibitem [{\citenamefont {Bacchin}\ \emph {et~al.}(1995)\citenamefont
  {Bacchin}, \citenamefont {Aimar},\ and\ \citenamefont
  {Sanchez}}]{Bacchin:1995hi}%
  \BibitemOpen
  \bibfield  {author} {\bibinfo {author} {\bibfnamefont {P.}~\bibnamefont
  {Bacchin}}, \bibinfo {author} {\bibfnamefont {P.}~\bibnamefont {Aimar}}, \
  and\ \bibinfo {author} {\bibfnamefont {V.}~\bibnamefont {Sanchez}},\
  }\href@noop {} {\bibfield  {journal} {\bibinfo  {journal} {AIChE Journal}\
  }\textbf {\bibinfo {volume} {41}},\ \bibinfo {pages} {368} (\bibinfo {year}
  {1995})}\BibitemShut {NoStop}%
\bibitem [{\citenamefont {Song}(1998)}]{Song:1998jg}%
  \BibitemOpen
  \bibfield  {author} {\bibinfo {author} {\bibfnamefont {L.~F.}\ \bibnamefont
  {Song}},\ }\href@noop {} {\bibfield  {journal} {\bibinfo  {journal} {J.
  Membr. Sci.}\ }\textbf {\bibinfo {volume} {139}},\ \bibinfo {pages} {183}
  (\bibinfo {year} {1998})}\BibitemShut {NoStop}%
\bibitem [{\citenamefont {Bacchin}\ \emph {et~al.}(2006)\citenamefont
  {Bacchin}, \citenamefont {Aimar},\ and\ \citenamefont
  {Field}}]{Bacchin:2006es}%
  \BibitemOpen
  \bibfield  {author} {\bibinfo {author} {\bibfnamefont {P.}~\bibnamefont
  {Bacchin}}, \bibinfo {author} {\bibfnamefont {P.}~\bibnamefont {Aimar}}, \
  and\ \bibinfo {author} {\bibfnamefont {R.~W.}\ \bibnamefont {Field}},\
  }\href@noop {} {\bibfield  {journal} {\bibinfo  {journal} {J. Membr. Sci.}\
  }\textbf {\bibinfo {volume} {281}},\ \bibinfo {pages} {42} (\bibinfo {year}
  {2006})}\BibitemShut {NoStop}%
\bibitem [{\citenamefont {Buetehorn}\ \emph {et~al.}(2011)\citenamefont
  {Buetehorn}, \citenamefont {Utiu}, \citenamefont {K{\"u}ppers}, \citenamefont
  {Bl{\"u}mich}, \citenamefont {Wintgens}, \citenamefont {Wessling},\ and\
  \citenamefont {Melin}}]{Buetehorn:2011dv}%
  \BibitemOpen
  \bibfield  {author} {\bibinfo {author} {\bibfnamefont {S.}~\bibnamefont
  {Buetehorn}}, \bibinfo {author} {\bibfnamefont {L.}~\bibnamefont {Utiu}},
  \bibinfo {author} {\bibfnamefont {M.}~\bibnamefont {K{\"u}ppers}}, \bibinfo
  {author} {\bibfnamefont {B.}~\bibnamefont {Bl{\"u}mich}}, \bibinfo {author}
  {\bibfnamefont {T.}~\bibnamefont {Wintgens}}, \bibinfo {author}
  {\bibfnamefont {M.}~\bibnamefont {Wessling}}, \ and\ \bibinfo {author}
  {\bibfnamefont {T.}~\bibnamefont {Melin}},\ }\href@noop {} {\bibfield
  {journal} {\bibinfo  {journal} {J. Membr. Sci.}\ }\textbf {\bibinfo {volume}
  {371}},\ \bibinfo {pages} {52} (\bibinfo {year} {2011})}\BibitemShut
  {NoStop}%
\bibitem [{\citenamefont {Shen}\ and\ \citenamefont
  {Probstein}(1977)}]{Shen:1977hm}%
  \BibitemOpen
  \bibfield  {author} {\bibinfo {author} {\bibfnamefont {J.~J.~S.}\
  \bibnamefont {Shen}}\ and\ \bibinfo {author} {\bibfnamefont {R.~F.}\
  \bibnamefont {Probstein}},\ }\href@noop {} {\bibfield  {journal} {\bibinfo
  {journal} {Ind. Eng. Chem. Fundam.}\ }\textbf {\bibinfo {volume} {16}},\
  \bibinfo {pages} {459} (\bibinfo {year} {1977})}\BibitemShut {NoStop}%
\bibitem [{\citenamefont {Trettin}\ and\ \citenamefont
  {Doshi}(1980)}]{Trettin:1980dm}%
  \BibitemOpen
  \bibfield  {author} {\bibinfo {author} {\bibfnamefont {D.~R.}\ \bibnamefont
  {Trettin}}\ and\ \bibinfo {author} {\bibfnamefont {M.~R.}\ \bibnamefont
  {Doshi}},\ }\href@noop {} {\bibfield  {journal} {\bibinfo  {journal} {Chem.
  Eng. Commun.}\ }\textbf {\bibinfo {volume} {4}},\ \bibinfo {pages} {507}
  (\bibinfo {year} {1980})}\BibitemShut {NoStop}%
\bibitem [{\citenamefont {Gill}\ \emph {et~al.}(1988)\citenamefont {Gill},
  \citenamefont {Wiley}, \citenamefont {Fell},\ and\ \citenamefont
  {Fane}}]{Gill:1988hp}%
  \BibitemOpen
  \bibfield  {author} {\bibinfo {author} {\bibfnamefont {W.~N.}\ \bibnamefont
  {Gill}}, \bibinfo {author} {\bibfnamefont {D.~E.}\ \bibnamefont {Wiley}},
  \bibinfo {author} {\bibfnamefont {C.~J.~D.}\ \bibnamefont {Fell}}, \ and\
  \bibinfo {author} {\bibfnamefont {A.~G.}\ \bibnamefont {Fane}},\ }\href@noop
  {} {\bibfield  {journal} {\bibinfo  {journal} {AIChE Journal}\ }\textbf
  {\bibinfo {volume} {34}},\ \bibinfo {pages} {1563} (\bibinfo {year}
  {1988})}\BibitemShut {NoStop}%
\bibitem [{\citenamefont {Wijmans}\ \emph {et~al.}(1984)\citenamefont
  {Wijmans}, \citenamefont {Nakao},\ and\ \citenamefont
  {Smolders}}]{Wijmans:1984fd}%
  \BibitemOpen
  \bibfield  {author} {\bibinfo {author} {\bibfnamefont {J.~G.}\ \bibnamefont
  {Wijmans}}, \bibinfo {author} {\bibfnamefont {S.}~\bibnamefont {Nakao}}, \
  and\ \bibinfo {author} {\bibfnamefont {C.~A.}\ \bibnamefont {Smolders}},\
  }\href@noop {} {\bibfield  {journal} {\bibinfo  {journal} {J. Membr. Sci.}\
  }\textbf {\bibinfo {volume} {20}},\ \bibinfo {pages} {115} (\bibinfo {year}
  {1984})}\BibitemShut {NoStop}%
\bibitem [{\citenamefont {Jonsson}(1984)}]{Jonsson:1984ee}%
  \BibitemOpen
  \bibfield  {author} {\bibinfo {author} {\bibfnamefont {G.}~\bibnamefont
  {Jonsson}},\ }\href@noop {} {\bibfield  {journal} {\bibinfo  {journal}
  {Desalination}\ }\textbf {\bibinfo {volume} {51}},\ \bibinfo {pages} {61}
  (\bibinfo {year} {1984})}\BibitemShut {NoStop}%
\bibitem [{\citenamefont {Ilias}\ and\ \citenamefont
  {Govind}(1993)}]{Ilias:1993gc}%
  \BibitemOpen
  \bibfield  {author} {\bibinfo {author} {\bibfnamefont {S.}~\bibnamefont
  {Ilias}}\ and\ \bibinfo {author} {\bibfnamefont {R.}~\bibnamefont {Govind}},\
  }\href@noop {} {\bibfield  {journal} {\bibinfo  {journal} {Separ. Sci.
  Technol.}\ }\textbf {\bibinfo {volume} {28}},\ \bibinfo {pages} {361}
  (\bibinfo {year} {1993})}\BibitemShut {NoStop}%
\bibitem [{\citenamefont {Elimelech}\ and\ \citenamefont
  {Bhattacharjee}(1998)}]{Elimelech:1998jm}%
  \BibitemOpen
  \bibfield  {author} {\bibinfo {author} {\bibfnamefont {M.}~\bibnamefont
  {Elimelech}}\ and\ \bibinfo {author} {\bibfnamefont {S.}~\bibnamefont
  {Bhattacharjee}},\ }\href@noop {} {\bibfield  {journal} {\bibinfo  {journal}
  {J. Membr. Sci.}\ }\textbf {\bibinfo {volume} {145}},\ \bibinfo {pages} {223}
  (\bibinfo {year} {1998})}\BibitemShut {NoStop}%
\bibitem [{\citenamefont {Bhattacharjee}\ \emph {et~al.}(1999)\citenamefont
  {Bhattacharjee}, \citenamefont {Kim},\ and\ \citenamefont
  {Elimelech}}]{Bhattacharjee:1999cz}%
  \BibitemOpen
  \bibfield  {author} {\bibinfo {author} {\bibfnamefont {S.}~\bibnamefont
  {Bhattacharjee}}, \bibinfo {author} {\bibfnamefont {A.~S.}\ \bibnamefont
  {Kim}}, \ and\ \bibinfo {author} {\bibfnamefont {M.}~\bibnamefont
  {Elimelech}},\ }\href@noop {} {\bibfield  {journal} {\bibinfo  {journal} {J.
  Colloid Interface Sci.}\ }\textbf {\bibinfo {volume} {212}},\ \bibinfo
  {pages} {81} (\bibinfo {year} {1999})}\BibitemShut {NoStop}%
\bibitem [{\citenamefont {Bowen}\ and\ \citenamefont
  {Williams}(2001)}]{Bowen:2001ib}%
  \BibitemOpen
  \bibfield  {author} {\bibinfo {author} {\bibfnamefont {W.~R.}\ \bibnamefont
  {Bowen}}\ and\ \bibinfo {author} {\bibfnamefont {P.~M.}\ \bibnamefont
  {Williams}},\ }\href@noop {} {\bibfield  {journal} {\bibinfo  {journal}
  {Chem. Eng. Sci.}\ }\textbf {\bibinfo {volume} {56}},\ \bibinfo {pages}
  {3083} (\bibinfo {year} {2001})}\BibitemShut {NoStop}%
\bibitem [{\citenamefont {Bacchin}\ \emph {et~al.}(2002)\citenamefont
  {Bacchin}, \citenamefont {Si-Hassen}, \citenamefont {Starov}, \citenamefont
  {Clifton},\ and\ \citenamefont {Aimar}}]{Bacchin:2002fm}%
  \BibitemOpen
  \bibfield  {author} {\bibinfo {author} {\bibfnamefont {P.}~\bibnamefont
  {Bacchin}}, \bibinfo {author} {\bibfnamefont {D.}~\bibnamefont {Si-Hassen}},
  \bibinfo {author} {\bibfnamefont {V.}~\bibnamefont {Starov}}, \bibinfo
  {author} {\bibfnamefont {M.~J.}\ \bibnamefont {Clifton}}, \ and\ \bibinfo
  {author} {\bibfnamefont {P.}~\bibnamefont {Aimar}},\ }\href@noop {}
  {\bibfield  {journal} {\bibinfo  {journal} {Chem. Eng. Sci.}\ }\textbf
  {\bibinfo {volume} {57}},\ \bibinfo {pages} {77} (\bibinfo {year}
  {2002})}\BibitemShut {NoStop}%
\bibitem [{\citenamefont {Batchelor}(1972)}]{Batchelor:1972ii}%
  \BibitemOpen
  \bibfield  {author} {\bibinfo {author} {\bibfnamefont {G.~K.}\ \bibnamefont
  {Batchelor}},\ }\href@noop {} {\bibfield  {journal} {\bibinfo  {journal} {J.
  Fluid Mech.}\ }\textbf {\bibinfo {volume} {52}},\ \bibinfo {pages} {245}
  (\bibinfo {year} {1972})}\BibitemShut {NoStop}%
\bibitem [{\citenamefont {Happel}(1958)}]{Happel:1958ip}%
  \BibitemOpen
  \bibfield  {author} {\bibinfo {author} {\bibfnamefont {J.}~\bibnamefont
  {Happel}},\ }\href@noop {} {\bibfield  {journal} {\bibinfo  {journal} {AIChE
  Journal}\ }\textbf {\bibinfo {volume} {4}},\ \bibinfo {pages} {197} (\bibinfo
  {year} {1958})}\BibitemShut {NoStop}%
\bibitem [{\citenamefont {Mewis}\ and\ \citenamefont
  {Wagner}(2011)}]{Mewis:2011bo}%
  \BibitemOpen
  \bibfield  {author} {\bibinfo {author} {\bibfnamefont {J.}~\bibnamefont
  {Mewis}}\ and\ \bibinfo {author} {\bibfnamefont {N.~J.}\ \bibnamefont
  {Wagner}},\ }\href@noop {} {\emph {\bibinfo {title} {{Colloidal Suspension
  Rheology}}}}\ (\bibinfo  {publisher} {Cambridge University Press},\ \bibinfo
  {address} {Cambridge},\ \bibinfo {year} {2011})\BibitemShut {NoStop}%
\bibitem [{\citenamefont {Bouchoux}\ \emph {et~al.}(2014)\citenamefont
  {Bouchoux}, \citenamefont {Qu}, \citenamefont {Bacchin},\ and\ \citenamefont
  {G{\'e}san-Guiziou}}]{Bouchoux:2014kb}%
  \BibitemOpen
  \bibfield  {author} {\bibinfo {author} {\bibfnamefont {A.}~\bibnamefont
  {Bouchoux}}, \bibinfo {author} {\bibfnamefont {P.}~\bibnamefont {Qu}},
  \bibinfo {author} {\bibfnamefont {P.}~\bibnamefont {Bacchin}}, \ and\
  \bibinfo {author} {\bibfnamefont {G.}~\bibnamefont {G{\'e}san-Guiziou}},\
  }\href@noop {} {\bibfield  {journal} {\bibinfo  {journal} {Langmuir}\
  }\textbf {\bibinfo {volume} {30}},\ \bibinfo {pages} {22} (\bibinfo {year}
  {2014})}\BibitemShut {NoStop}%
\bibitem [{\citenamefont {Abade}\ \emph
  {et~al.}(2010{\natexlab{a}})\citenamefont {Abade}, \citenamefont {Cichocki},
  \citenamefont {Ekiel-Je{\.{z}}ewska}, \citenamefont {N{\"a}gele},\ and\
  \citenamefont {Wajnryb}}]{Abade:2010gt}%
  \BibitemOpen
  \bibfield  {author} {\bibinfo {author} {\bibfnamefont {G.~C.}\ \bibnamefont
  {Abade}}, \bibinfo {author} {\bibfnamefont {B.}~\bibnamefont {Cichocki}},
  \bibinfo {author} {\bibfnamefont {M.~L.}\ \bibnamefont
  {Ekiel-Je{\.{z}}ewska}}, \bibinfo {author} {\bibfnamefont {G.}~\bibnamefont
  {N{\"a}gele}}, \ and\ \bibinfo {author} {\bibfnamefont {E.}~\bibnamefont
  {Wajnryb}},\ }\href@noop {} {\bibfield  {journal} {\bibinfo  {journal} {J.
  Chem. Phys.}\ }\textbf {\bibinfo {volume} {132}},\ \bibinfo {pages} {014503}
  (\bibinfo {year} {2010}{\natexlab{a}})}\BibitemShut {NoStop}%
\bibitem [{\citenamefont {Abade}\ \emph
  {et~al.}(2010{\natexlab{b}})\citenamefont {Abade}, \citenamefont {Cichocki},
  \citenamefont {Ekiel-Je{\.{z}}ewska}, \citenamefont {N{\"a}gele},\ and\
  \citenamefont {Wajnryb}}]{Abade:2010ev}%
  \BibitemOpen
  \bibfield  {author} {\bibinfo {author} {\bibfnamefont {G.~C.}\ \bibnamefont
  {Abade}}, \bibinfo {author} {\bibfnamefont {B.}~\bibnamefont {Cichocki}},
  \bibinfo {author} {\bibfnamefont {M.~L.}\ \bibnamefont
  {Ekiel-Je{\.{z}}ewska}}, \bibinfo {author} {\bibfnamefont {G.}~\bibnamefont
  {N{\"a}gele}}, \ and\ \bibinfo {author} {\bibfnamefont {E.}~\bibnamefont
  {Wajnryb}},\ }\href@noop {} {\bibfield  {journal} {\bibinfo  {journal} {J.
  Chem. Phys.}\ }\textbf {\bibinfo {volume} {133}},\ \bibinfo {pages} {084906}
  (\bibinfo {year} {2010}{\natexlab{b}})}\BibitemShut {NoStop}%
\bibitem [{\citenamefont {Abade}\ \emph {et~al.}(2012)\citenamefont {Abade},
  \citenamefont {Cichocki}, \citenamefont {Ekiel-Je{\.{z}}ewska}, \citenamefont
  {N{\"a}gele},\ and\ \citenamefont {Wajnryb}}]{Abade:2012em}%
  \BibitemOpen
  \bibfield  {author} {\bibinfo {author} {\bibfnamefont {G.~C.}\ \bibnamefont
  {Abade}}, \bibinfo {author} {\bibfnamefont {B.}~\bibnamefont {Cichocki}},
  \bibinfo {author} {\bibfnamefont {M.~L.}\ \bibnamefont
  {Ekiel-Je{\.{z}}ewska}}, \bibinfo {author} {\bibfnamefont {G.}~\bibnamefont
  {N{\"a}gele}}, \ and\ \bibinfo {author} {\bibfnamefont {E.}~\bibnamefont
  {Wajnryb}},\ }\href@noop {} {\bibfield  {journal} {\bibinfo  {journal} {J.
  Chem. Phys.}\ }\textbf {\bibinfo {volume} {136}},\ \bibinfo {pages} {104902}
  (\bibinfo {year} {2012})}\BibitemShut {NoStop}%
\bibitem [{\citenamefont {Riest}\ \emph {et~al.}(2015)\citenamefont {Riest},
  \citenamefont {Eckert}, \citenamefont {Richtering},\ and\ \citenamefont
  {N{\"a}gele}}]{Riest:2015jo}%
  \BibitemOpen
  \bibfield  {author} {\bibinfo {author} {\bibfnamefont {J.}~\bibnamefont
  {Riest}}, \bibinfo {author} {\bibfnamefont {T.}~\bibnamefont {Eckert}},
  \bibinfo {author} {\bibfnamefont {W.}~\bibnamefont {Richtering}}, \ and\
  \bibinfo {author} {\bibfnamefont {G.}~\bibnamefont {N{\"a}gele}},\
  }\href@noop {} {\bibfield  {journal} {\bibinfo  {journal} {Soft Matter}\
  }\textbf {\bibinfo {volume} {11}},\ \bibinfo {pages} {2821} (\bibinfo {year}
  {2015})}\BibitemShut {NoStop}%
\bibitem [{\citenamefont {Eckert}\ and\ \citenamefont
  {Richtering}(2008)}]{Eckert:2008hp}%
  \BibitemOpen
  \bibfield  {author} {\bibinfo {author} {\bibfnamefont {T.}~\bibnamefont
  {Eckert}}\ and\ \bibinfo {author} {\bibfnamefont {W.}~\bibnamefont
  {Richtering}},\ }\href@noop {} {\bibfield  {journal} {\bibinfo  {journal} {J.
  Chem. Phys.}\ }\textbf {\bibinfo {volume} {129}},\ \bibinfo {pages} {124902}
  (\bibinfo {year} {2008})}\BibitemShut {NoStop}%
\bibitem [{\citenamefont {Brady}(1993)}]{Brady:1993hu}%
  \BibitemOpen
  \bibfield  {author} {\bibinfo {author} {\bibfnamefont {J.~F.}\ \bibnamefont
  {Brady}},\ }\href@noop {} {\bibfield  {journal} {\bibinfo  {journal} {J.
  Chem. Phys.}\ }\textbf {\bibinfo {volume} {99}},\ \bibinfo {pages} {567}
  (\bibinfo {year} {1993})}\BibitemShut {NoStop}%
\bibitem [{\citenamefont {Probstein}(1989)}]{Probstein:2005vy}%
  \BibitemOpen
  \bibfield  {author} {\bibinfo {author} {\bibfnamefont {R.~F.}\ \bibnamefont
  {Probstein}},\ }\href@noop {} {\emph {\bibinfo {title} {{Physicochemical
  Hydrodynamics}}}}\ (\bibinfo  {publisher} {Butterworths},\ \bibinfo {address}
  {London},\ \bibinfo {year} {1989})\BibitemShut {NoStop}%
\bibitem [{\citenamefont {Yurkovetsky}\ and\ \citenamefont
  {Morris}(2008)}]{Yurkovetsky:2008ev}%
  \BibitemOpen
  \bibfield  {author} {\bibinfo {author} {\bibfnamefont {Y.}~\bibnamefont
  {Yurkovetsky}}\ and\ \bibinfo {author} {\bibfnamefont {J.~F.}\ \bibnamefont
  {Morris}},\ }\href@noop {} {\bibfield  {journal} {\bibinfo  {journal} {J.
  Rheol.}\ }\textbf {\bibinfo {volume} {52}},\ \bibinfo {pages} {141} (\bibinfo
  {year} {2008})}\BibitemShut {NoStop}%
\bibitem [{\citenamefont {Romero}\ and\ \citenamefont
  {Davis}(1990)}]{Romero:1990ei}%
  \BibitemOpen
  \bibfield  {author} {\bibinfo {author} {\bibfnamefont {C.~A.}\ \bibnamefont
  {Romero}}\ and\ \bibinfo {author} {\bibfnamefont {R.~H.}\ \bibnamefont
  {Davis}},\ }\href@noop {} {\bibfield  {journal} {\bibinfo  {journal} {Chem.
  Eng. Sci.}\ }\textbf {\bibinfo {volume} {45}},\ \bibinfo {pages} {13}
  (\bibinfo {year} {1990})}\BibitemShut {NoStop}%
\bibitem [{\citenamefont {De}\ and\ \citenamefont
  {Bhattacharya}(1996)}]{De:1996hq}%
  \BibitemOpen
  \bibfield  {author} {\bibinfo {author} {\bibfnamefont {S.}~\bibnamefont
  {De}}\ and\ \bibinfo {author} {\bibfnamefont {P.~K.}\ \bibnamefont
  {Bhattacharya}},\ }\href@noop {} {\bibfield  {journal} {\bibinfo  {journal}
  {J. Membr. Sci.}\ }\textbf {\bibinfo {volume} {109}},\ \bibinfo {pages} {109}
  (\bibinfo {year} {1996})}\BibitemShut {NoStop}%
\bibitem [{\citenamefont {Kierzenka}\ and\ \citenamefont
  {Shampine}(2001)}]{Kierzenka:2001bsb}%
  \BibitemOpen
  \bibfield  {author} {\bibinfo {author} {\bibfnamefont {J.}~\bibnamefont
  {Kierzenka}}\ and\ \bibinfo {author} {\bibfnamefont {L.~F.}\ \bibnamefont
  {Shampine}},\ }\href@noop {} {\bibfield  {journal} {\bibinfo  {journal} {ACM
  Trans. Math. Softw.}\ }\textbf {\bibinfo {volume} {27}},\ \bibinfo {pages}
  {299} (\bibinfo {year} {2001})}\BibitemShut {NoStop}%
\bibitem [{\citenamefont {Rintoul}\ and\ \citenamefont
  {Torquato}(1996)}]{Rintoul:1996fs}%
  \BibitemOpen
  \bibfield  {author} {\bibinfo {author} {\bibfnamefont {M.}~\bibnamefont
  {Rintoul}}\ and\ \bibinfo {author} {\bibfnamefont {S.}~\bibnamefont
  {Torquato}},\ }\href@noop {} {\bibfield  {journal} {\bibinfo  {journal}
  {Phys. Rev. Lett.}\ }\textbf {\bibinfo {volume} {77}},\ \bibinfo {pages}
  {4198} (\bibinfo {year} {1996})}\BibitemShut {NoStop}%
\bibitem [{\citenamefont {Brinkman}(1947)}]{Brinkman:1947wz}%
  \BibitemOpen
  \bibfield  {author} {\bibinfo {author} {\bibfnamefont {H.~C.}\ \bibnamefont
  {Brinkman}},\ }\href@noop {} {\bibfield  {journal} {\bibinfo  {journal}
  {Appl. sci. Res.}\ }\textbf {\bibinfo {volume} {A1}},\ \bibinfo {pages} {27}
  (\bibinfo {year} {1947})}\BibitemShut {NoStop}%
\bibitem [{\citenamefont {Debye}\ and\ \citenamefont
  {Bueche}(1948)}]{Debye:1948if}%
  \BibitemOpen
  \bibfield  {author} {\bibinfo {author} {\bibfnamefont {P.}~\bibnamefont
  {Debye}}\ and\ \bibinfo {author} {\bibfnamefont {A.~M.}\ \bibnamefont
  {Bueche}},\ }\href@noop {} {\bibfield  {journal} {\bibinfo  {journal} {J.
  Chem. Phys.}\ }\textbf {\bibinfo {volume} {16}},\ \bibinfo {pages} {573}
  (\bibinfo {year} {1948})}\BibitemShut {NoStop}%
\bibitem [{\citenamefont {Duits}\ \emph {et~al.}(2001)\citenamefont {Duits},
  \citenamefont {Nommensen}, \citenamefont {van~den Ende},\ and\ \citenamefont
  {Mellema}}]{Duits:2001ij}%
  \BibitemOpen
  \bibfield  {author} {\bibinfo {author} {\bibfnamefont {M.}~\bibnamefont
  {Duits}}, \bibinfo {author} {\bibfnamefont {P.~A.}\ \bibnamefont
  {Nommensen}}, \bibinfo {author} {\bibfnamefont {D.}~\bibnamefont {van~den
  Ende}}, \ and\ \bibinfo {author} {\bibfnamefont {J.}~\bibnamefont
  {Mellema}},\ }\href@noop {} {\bibfield  {journal} {\bibinfo  {journal}
  {Colloids Surf., A}\ }\textbf {\bibinfo {volume} {183-185}},\ \bibinfo
  {pages} {335} (\bibinfo {year} {2001})}\BibitemShut {NoStop}%
\bibitem [{\citenamefont {N{\"a}gele}(1996)}]{Nagele:1996hr}%
  \BibitemOpen
  \bibfield  {author} {\bibinfo {author} {\bibfnamefont {G.}~\bibnamefont
  {N{\"a}gele}},\ }\href@noop {} {\bibfield  {journal} {\bibinfo  {journal}
  {Phys. Rep.}\ }\textbf {\bibinfo {volume} {272}},\ \bibinfo {pages} {216}
  (\bibinfo {year} {1996})}\BibitemShut {NoStop}%
\bibitem [{\citenamefont {N{\"a}gele}(2013)}]{Nagele:2013co}%
  \BibitemOpen
  \bibfield  {author} {\bibinfo {author} {\bibfnamefont {G.}~\bibnamefont
  {N{\"a}gele}},\ }in\ \href@noop {} {\emph {\bibinfo {booktitle} {Physics of
  Complex Colloids}}},\ \bibinfo {editor} {edited by\ \bibinfo {editor}
  {\bibfnamefont {C.}~\bibnamefont {Bechinger}}, \bibinfo {editor}
  {\bibfnamefont {F.}~\bibnamefont {Sciortino}}, \ and\ \bibinfo {editor}
  {\bibfnamefont {P.}~\bibnamefont {Ziherl}}}\ (\bibinfo  {publisher} {IOS
  Press},\ \bibinfo {address} {Amsterdam},\ \bibinfo {year} {2013})\ pp.\
  \bibinfo {pages} {507--601}\BibitemShut {NoStop}%
\bibitem [{\citenamefont {Wajnryb}\ \emph {et~al.}(2004)\citenamefont
  {Wajnryb}, \citenamefont {Szymczak},\ and\ \citenamefont
  {Cichocki}}]{Wajnryb:2004di}%
  \BibitemOpen
  \bibfield  {author} {\bibinfo {author} {\bibfnamefont {E.}~\bibnamefont
  {Wajnryb}}, \bibinfo {author} {\bibfnamefont {P.}~\bibnamefont {Szymczak}}, \
  and\ \bibinfo {author} {\bibfnamefont {B.}~\bibnamefont {Cichocki}},\
  }\href@noop {} {\bibfield  {journal} {\bibinfo  {journal} {Physica A}\
  }\textbf {\bibinfo {volume} {335}},\ \bibinfo {pages} {339} (\bibinfo {year}
  {2004})}\BibitemShut {NoStop}%
\bibitem [{\citenamefont {Dhont}(1996)}]{Dhont:1996ub}%
  \BibitemOpen
  \bibfield  {author} {\bibinfo {author} {\bibfnamefont {J.~K.~G.}\
  \bibnamefont {Dhont}},\ }\href@noop {} {\emph {\bibinfo {title} {{An
  Introduction to Dynamics of Colloids}}}}\ (\bibinfo  {publisher} {Elsevier},\
  \bibinfo {address} {Amsterdam},\ \bibinfo {year} {1996})\BibitemShut
  {NoStop}%
\bibitem [{\citenamefont {Benes}\ \emph {et~al.}(2007)\citenamefont {Benes},
  \citenamefont {Tong},\ and\ \citenamefont {Ackerson}}]{Benes:2007kv}%
  \BibitemOpen
  \bibfield  {author} {\bibinfo {author} {\bibfnamefont {K.}~\bibnamefont
  {Benes}}, \bibinfo {author} {\bibfnamefont {P.}~\bibnamefont {Tong}}, \ and\
  \bibinfo {author} {\bibfnamefont {B.}~\bibnamefont {Ackerson}},\ }\href@noop
  {} {\bibfield  {journal} {\bibinfo  {journal} {Phys. Rev. E}\ }\textbf
  {\bibinfo {volume} {76}},\ \bibinfo {pages} {056302} (\bibinfo {year}
  {2007})}\BibitemShut {NoStop}%
\bibitem [{\citenamefont {Ladd}(1990)}]{Ladd:1990gr}%
  \BibitemOpen
  \bibfield  {author} {\bibinfo {author} {\bibfnamefont {A.~J.~C.}\
  \bibnamefont {Ladd}},\ }\href@noop {} {\bibfield  {journal} {\bibinfo
  {journal} {J. Chem. Phys.}\ }\textbf {\bibinfo {volume} {93}},\ \bibinfo
  {pages} {3484} (\bibinfo {year} {1990})}\BibitemShut {NoStop}%
\bibitem [{\citenamefont {Cichocki}\ \emph {et~al.}(2002)\citenamefont
  {Cichocki}, \citenamefont {Ekiel-Je{\.{z}}ewska}, \citenamefont {Szymczak},\
  and\ \citenamefont {Wajnryb}}]{Cichocki:2002ft}%
  \BibitemOpen
  \bibfield  {author} {\bibinfo {author} {\bibfnamefont {B.}~\bibnamefont
  {Cichocki}}, \bibinfo {author} {\bibfnamefont {M.~L.}\ \bibnamefont
  {Ekiel-Je{\.{z}}ewska}}, \bibinfo {author} {\bibfnamefont {P.}~\bibnamefont
  {Szymczak}}, \ and\ \bibinfo {author} {\bibfnamefont {E.}~\bibnamefont
  {Wajnryb}},\ }\href@noop {} {\bibfield  {journal} {\bibinfo  {journal} {J.
  Chem. Phys.}\ }\textbf {\bibinfo {volume} {117}},\ \bibinfo {pages} {1231}
  (\bibinfo {year} {2002})}\BibitemShut {NoStop}%
\bibitem [{\citenamefont {Banchio}\ and\ \citenamefont
  {N{\"a}gele}(2008)}]{Banchio:2008gt}%
  \BibitemOpen
  \bibfield  {author} {\bibinfo {author} {\bibfnamefont {A.~J.}\ \bibnamefont
  {Banchio}}\ and\ \bibinfo {author} {\bibfnamefont {G.}~\bibnamefont
  {N{\"a}gele}},\ }\href@noop {} {\bibfield  {journal} {\bibinfo  {journal} {J.
  Chem. Phys.}\ }\textbf {\bibinfo {volume} {128}},\ \bibinfo {pages} {104903}
  (\bibinfo {year} {2008})}\BibitemShut {NoStop}%
\bibitem [{\citenamefont {Russel}\ \emph {et~al.}(1989)\citenamefont {Russel},
  \citenamefont {Saville},\ and\ \citenamefont {Schowalter}}]{Russel:1989vm}%
  \BibitemOpen
  \bibfield  {author} {\bibinfo {author} {\bibfnamefont {W.~B.}\ \bibnamefont
  {Russel}}, \bibinfo {author} {\bibfnamefont {D.~A.}\ \bibnamefont {Saville}},
  \ and\ \bibinfo {author} {\bibfnamefont {W.~R.}\ \bibnamefont {Schowalter}},\
  }\href@noop {} {\emph {\bibinfo {title} {{Colloidal Dispersions}}}}\
  (\bibinfo  {publisher} {Cambridge University Press},\ \bibinfo {address}
  {Cambridge},\ \bibinfo {year} {1989})\BibitemShut {NoStop}%
\bibitem [{\citenamefont {Heinen}\ \emph {et~al.}(2012)\citenamefont {Heinen},
  \citenamefont {Zanini}, \citenamefont {Roosen-Runge}, \citenamefont
  {Fedunov{\'a}}, \citenamefont {Zhang}, \citenamefont {Hennig}, \citenamefont
  {Seydel}, \citenamefont {Schweins}, \citenamefont {Sztucki}, \citenamefont
  {Antal{\'\i}k}, \citenamefont {Schreiber},\ and\ \citenamefont
  {N{\"a}gele}}]{Heinen:2012iy}%
  \BibitemOpen
  \bibfield  {author} {\bibinfo {author} {\bibfnamefont {M.}~\bibnamefont
  {Heinen}}, \bibinfo {author} {\bibfnamefont {F.}~\bibnamefont {Zanini}},
  \bibinfo {author} {\bibfnamefont {F.}~\bibnamefont {Roosen-Runge}}, \bibinfo
  {author} {\bibfnamefont {D.}~\bibnamefont {Fedunov{\'a}}}, \bibinfo {author}
  {\bibfnamefont {F.}~\bibnamefont {Zhang}}, \bibinfo {author} {\bibfnamefont
  {M.}~\bibnamefont {Hennig}}, \bibinfo {author} {\bibfnamefont
  {T.}~\bibnamefont {Seydel}}, \bibinfo {author} {\bibfnamefont
  {R.}~\bibnamefont {Schweins}}, \bibinfo {author} {\bibfnamefont
  {M.}~\bibnamefont {Sztucki}}, \bibinfo {author} {\bibfnamefont
  {M.}~\bibnamefont {Antal{\'\i}k}}, \bibinfo {author} {\bibfnamefont
  {F.}~\bibnamefont {Schreiber}}, \ and\ \bibinfo {author} {\bibfnamefont
  {G.}~\bibnamefont {N{\"a}gele}},\ }\href@noop {} {\bibfield  {journal}
  {\bibinfo  {journal} {Soft Matter}\ }\textbf {\bibinfo {volume} {8}},\
  \bibinfo {pages} {1404} (\bibinfo {year} {2012})}\BibitemShut {NoStop}%
\bibitem [{\citenamefont {Segr{\`e}}\ \emph {et~al.}(1995)\citenamefont
  {Segr{\`e}}, \citenamefont {Behrend},\ and\ \citenamefont
  {Pusey}}]{Segre:1995uu}%
  \BibitemOpen
  \bibfield  {author} {\bibinfo {author} {\bibfnamefont {P.}~\bibnamefont
  {Segr{\`e}}}, \bibinfo {author} {\bibfnamefont {O.}~\bibnamefont {Behrend}},
  \ and\ \bibinfo {author} {\bibfnamefont {P.}~\bibnamefont {Pusey}},\
  }\href@noop {} {\bibfield  {journal} {\bibinfo  {journal} {Phys. Rev. E}\
  }\textbf {\bibinfo {volume} {52}},\ \bibinfo {pages} {5070} (\bibinfo {year}
  {1995})}\BibitemShut {NoStop}%
\bibitem [{\citenamefont {Weiss}\ \emph {et~al.}(1998)\citenamefont {Weiss},
  \citenamefont {Dingenouts}, \citenamefont {Ballauff}, \citenamefont {Senff},\
  and\ \citenamefont {Richtering}}]{Weiss:1998hd}%
  \BibitemOpen
  \bibfield  {author} {\bibinfo {author} {\bibfnamefont {A.}~\bibnamefont
  {Weiss}}, \bibinfo {author} {\bibfnamefont {N.}~\bibnamefont {Dingenouts}},
  \bibinfo {author} {\bibfnamefont {M.}~\bibnamefont {Ballauff}}, \bibinfo
  {author} {\bibfnamefont {H.}~\bibnamefont {Senff}}, \ and\ \bibinfo {author}
  {\bibfnamefont {W.}~\bibnamefont {Richtering}},\ }\href@noop {} {\bibfield
  {journal} {\bibinfo  {journal} {Langmuir}\ }\textbf {\bibinfo {volume}
  {14}},\ \bibinfo {pages} {5083} (\bibinfo {year} {1998})}\BibitemShut
  {NoStop}%
\bibitem [{\citenamefont {Foss}\ and\ \citenamefont
  {Brady}(2000)}]{Foss:2000kj}%
  \BibitemOpen
  \bibfield  {author} {\bibinfo {author} {\bibfnamefont {D.~R.}\ \bibnamefont
  {Foss}}\ and\ \bibinfo {author} {\bibfnamefont {J.~F.}\ \bibnamefont
  {Brady}},\ }\href@noop {} {\bibfield  {journal} {\bibinfo  {journal} {J.
  Fluid Mech.}\ }\textbf {\bibinfo {volume} {407}},\ \bibinfo {pages} {167}
  (\bibinfo {year} {2000})}\BibitemShut {NoStop}%
\bibitem [{\citenamefont {Ohshima}(2009)}]{Ohshima:2009ij}%
  \BibitemOpen
  \bibfield  {author} {\bibinfo {author} {\bibfnamefont {H.}~\bibnamefont
  {Ohshima}},\ }\href@noop {} {\bibfield  {journal} {\bibinfo  {journal}
  {Colloids Surf., A}\ }\textbf {\bibinfo {volume} {347}},\ \bibinfo {pages}
  {33} (\bibinfo {year} {2009})}\BibitemShut {NoStop}%
\bibitem [{\citenamefont {Ruiz-Reina}\ \emph {et~al.}(2003)\citenamefont
  {Ruiz-Reina}, \citenamefont {Carrique}, \citenamefont {Rubio-Hern{\'a}ndez},
  \citenamefont {G{\'o}mez-Merino},\ and\ \citenamefont
  {Garc{\'\i}a-S{\'a}nchez}}]{RuizReina:2003dc}%
  \BibitemOpen
  \bibfield  {author} {\bibinfo {author} {\bibfnamefont {E.}~\bibnamefont
  {Ruiz-Reina}}, \bibinfo {author} {\bibfnamefont {F.}~\bibnamefont
  {Carrique}}, \bibinfo {author} {\bibfnamefont {F.~J.}\ \bibnamefont
  {Rubio-Hern{\'a}ndez}}, \bibinfo {author} {\bibfnamefont {A.~I.}\
  \bibnamefont {G{\'o}mez-Merino}}, \ and\ \bibinfo {author} {\bibfnamefont
  {P.}~\bibnamefont {Garc{\'\i}a-S{\'a}nchez}},\ }\href@noop {} {\bibfield
  {journal} {\bibinfo  {journal} {J. Phys. Chem. B}\ }\textbf {\bibinfo
  {volume} {107}},\ \bibinfo {pages} {9528} (\bibinfo {year}
  {2003})}\BibitemShut {NoStop}%
\bibitem [{\citenamefont {de~Kruif}\ \emph {et~al.}(1985)\citenamefont
  {de~Kruif}, \citenamefont {van Iersel}, \citenamefont {Vrij},\ and\
  \citenamefont {Russel}}]{deKruif:1985eh}%
  \BibitemOpen
  \bibfield  {author} {\bibinfo {author} {\bibfnamefont {C.~G.}\ \bibnamefont
  {de~Kruif}}, \bibinfo {author} {\bibfnamefont {E.~M.~F.}\ \bibnamefont {van
  Iersel}}, \bibinfo {author} {\bibfnamefont {A.}~\bibnamefont {Vrij}}, \ and\
  \bibinfo {author} {\bibfnamefont {W.~B.}\ \bibnamefont {Russel}},\
  }\href@noop {} {\bibfield  {journal} {\bibinfo  {journal} {J. Chem. Phys.}\
  }\textbf {\bibinfo {volume} {83}},\ \bibinfo {pages} {4717} (\bibinfo {year}
  {1985})}\BibitemShut {NoStop}%
\bibitem [{\citenamefont {Sarkar}\ \emph {et~al.}(2008)\citenamefont {Sarkar},
  \citenamefont {DasGupta},\ and\ \citenamefont {De}}]{Sarkar:2008de}%
  \BibitemOpen
  \bibfield  {author} {\bibinfo {author} {\bibfnamefont {B.}~\bibnamefont
  {Sarkar}}, \bibinfo {author} {\bibfnamefont {S.}~\bibnamefont {DasGupta}}, \
  and\ \bibinfo {author} {\bibfnamefont {S.}~\bibnamefont {De}},\ }\href@noop
  {} {\bibfield  {journal} {\bibinfo  {journal} {J. Colloid Interface Sci.}\
  }\textbf {\bibinfo {volume} {319}},\ \bibinfo {pages} {236} (\bibinfo {year}
  {2008})}\BibitemShut {NoStop}%
\bibitem [{\citenamefont {Denisov}(1994)}]{Denisov:1994jh}%
  \BibitemOpen
  \bibfield  {author} {\bibinfo {author} {\bibfnamefont {G.~A.}\ \bibnamefont
  {Denisov}},\ }\href@noop {} {\bibfield  {journal} {\bibinfo  {journal} {J.
  Membr. Sci.}\ }\textbf {\bibinfo {volume} {91}},\ \bibinfo {pages} {173}
  (\bibinfo {year} {1994})}\BibitemShut {NoStop}%
\bibitem [{\citenamefont {Manth}\ \emph {et~al.}(2002)\citenamefont {Manth},
  \citenamefont {Gabor},\ and\ \citenamefont {Oklejas}}]{Manth:2002fr}%
  \BibitemOpen
  \bibfield  {author} {\bibinfo {author} {\bibfnamefont {T.}~\bibnamefont
  {Manth}}, \bibinfo {author} {\bibfnamefont {M.}~\bibnamefont {Gabor}}, \ and\
  \bibinfo {author} {\bibfnamefont {E.}~\bibnamefont {Oklejas}},\ }\href@noop
  {} {\bibfield  {journal} {\bibinfo  {journal} {Desalination}\ }\textbf
  {\bibinfo {volume} {157}},\ \bibinfo {pages} {9} (\bibinfo {year}
  {2002})}\BibitemShut {NoStop}%
\bibitem [{\citenamefont {Zhu}\ \emph {et~al.}(2009{\natexlab{a}})\citenamefont
  {Zhu}, \citenamefont {Christofides},\ and\ \citenamefont
  {Cohen}}]{Zhu:2009et}%
  \BibitemOpen
  \bibfield  {author} {\bibinfo {author} {\bibfnamefont {A.}~\bibnamefont
  {Zhu}}, \bibinfo {author} {\bibfnamefont {P.~D.}\ \bibnamefont
  {Christofides}}, \ and\ \bibinfo {author} {\bibfnamefont {Y.}~\bibnamefont
  {Cohen}},\ }\href@noop {} {\bibfield  {journal} {\bibinfo  {journal} {Ind.
  Eng. Chem. Res.}\ }\textbf {\bibinfo {volume} {48}},\ \bibinfo {pages} {6010}
  (\bibinfo {year} {2009}{\natexlab{a}})}\BibitemShut {NoStop}%
\bibitem [{\citenamefont {Zhu}\ \emph {et~al.}(2009{\natexlab{b}})\citenamefont
  {Zhu}, \citenamefont {Christofides},\ and\ \citenamefont
  {Cohen}}]{Zhu:2009ir}%
  \BibitemOpen
  \bibfield  {author} {\bibinfo {author} {\bibfnamefont {A.}~\bibnamefont
  {Zhu}}, \bibinfo {author} {\bibfnamefont {P.~D.}\ \bibnamefont
  {Christofides}}, \ and\ \bibinfo {author} {\bibfnamefont {Y.}~\bibnamefont
  {Cohen}},\ }\href@noop {} {\bibfield  {journal} {\bibinfo  {journal} {J.
  Membr. Sci.}\ }\textbf {\bibinfo {volume} {339}},\ \bibinfo {pages} {126}
  (\bibinfo {year} {2009}{\natexlab{b}})}\BibitemShut {NoStop}%
\bibitem [{\citenamefont {Li}(2010)}]{Li:2010ho}%
  \BibitemOpen
  \bibfield  {author} {\bibinfo {author} {\bibfnamefont {M.}~\bibnamefont
  {Li}},\ }\href@noop {} {\bibfield  {journal} {\bibinfo  {journal} {Ind. Eng.
  Chem. Res.}\ }\textbf {\bibinfo {volume} {49}},\ \bibinfo {pages} {1822}
  (\bibinfo {year} {2010})}\BibitemShut {NoStop}%
\bibitem [{\citenamefont {Li}(2011)}]{Li:2011fr}%
  \BibitemOpen
  \bibfield  {author} {\bibinfo {author} {\bibfnamefont {M.}~\bibnamefont
  {Li}},\ }\href@noop {} {\bibfield  {journal} {\bibinfo  {journal}
  {Desalination}\ }\textbf {\bibinfo {volume} {276}},\ \bibinfo {pages} {128}
  (\bibinfo {year} {2011})}\BibitemShut {NoStop}%
\bibitem [{\citenamefont {Sharif}\ \emph {et~al.}(2009)\citenamefont {Sharif},
  \citenamefont {Merdaw}, \citenamefont {Al-Bahadili}, \citenamefont {Al-Taee},
  \citenamefont {Al-Aibi}, \citenamefont {Rahal},\ and\ \citenamefont
  {Derwish}}]{Sharif:2009vk}%
  \BibitemOpen
  \bibfield  {author} {\bibinfo {author} {\bibfnamefont {A.~O.}\ \bibnamefont
  {Sharif}}, \bibinfo {author} {\bibfnamefont {A.~A.}\ \bibnamefont {Merdaw}},
  \bibinfo {author} {\bibfnamefont {H.}~\bibnamefont {Al-Bahadili}}, \bibinfo
  {author} {\bibfnamefont {A.}~\bibnamefont {Al-Taee}}, \bibinfo {author}
  {\bibfnamefont {S.}~\bibnamefont {Al-Aibi}}, \bibinfo {author} {\bibfnamefont
  {Z.}~\bibnamefont {Rahal}}, \ and\ \bibinfo {author} {\bibfnamefont
  {G.~A.~W.}\ \bibnamefont {Derwish}},\ }\href@noop {} {\bibfield  {journal}
  {\bibinfo  {journal} {Desalination and Water Treatment}\ }\textbf {\bibinfo
  {volume} {3}},\ \bibinfo {pages} {111} (\bibinfo {year} {2009})}\BibitemShut
  {NoStop}%
\bibitem [{\citenamefont {Sarkar}\ \emph {et~al.}(2009)\citenamefont {Sarkar},
  \citenamefont {DasGupta},\ and\ \citenamefont {De}}]{Sarkar:2009de}%
  \BibitemOpen
  \bibfield  {author} {\bibinfo {author} {\bibfnamefont {B.}~\bibnamefont
  {Sarkar}}, \bibinfo {author} {\bibfnamefont {S.}~\bibnamefont {DasGupta}}, \
  and\ \bibinfo {author} {\bibfnamefont {S.}~\bibnamefont {De}},\ }\href@noop
  {} {\bibfield  {journal} {\bibinfo  {journal} {J. Membr. Sci.}\ }\textbf
  {\bibinfo {volume} {331}},\ \bibinfo {pages} {75} (\bibinfo {year}
  {2009})}\BibitemShut {NoStop}%
\bibitem [{\citenamefont {van~der Sman}(2008)}]{vanderSman:2008gj}%
  \BibitemOpen
  \bibfield  {author} {\bibinfo {author} {\bibfnamefont {R.~G.~M.}\
  \bibnamefont {van~der Sman}},\ }\href@noop {} {\bibfield  {journal} {\bibinfo
   {journal} {J. Food Eng.}\ }\textbf {\bibinfo {volume} {85}},\ \bibinfo
  {pages} {243} (\bibinfo {year} {2008})}\BibitemShut {NoStop}%
\bibitem [{\citenamefont {Jaffrin}(2012{\natexlab{a}})}]{Jaffrin:2012fo}%
  \BibitemOpen
  \bibfield  {author} {\bibinfo {author} {\bibfnamefont {M.~Y.}\ \bibnamefont
  {Jaffrin}},\ }\href@noop {} {\bibfield  {journal} {\bibinfo  {journal} {Annu.
  Rev. Fluid Mech.}\ }\textbf {\bibinfo {volume} {44}},\ \bibinfo {pages} {77}
  (\bibinfo {year} {2012}{\natexlab{a}})}\BibitemShut {NoStop}%
\bibitem [{\citenamefont {Jaffrin}(2012{\natexlab{b}})}]{Jaffrin:2012jx}%
  \BibitemOpen
  \bibfield  {author} {\bibinfo {author} {\bibfnamefont {M.~Y.}\ \bibnamefont
  {Jaffrin}},\ }\href@noop {} {\bibfield  {journal} {\bibinfo  {journal}
  {Current Opinion in Chemical Engineering}\ }\textbf {\bibinfo {volume} {1}},\
  \bibinfo {pages} {171} (\bibinfo {year} {2012}{\natexlab{b}})}\BibitemShut
  {NoStop}%
\bibitem [{\citenamefont {van~der Sman}\ and\ \citenamefont
  {Vollebregt}(2013)}]{vanderSman:2013fv}%
  \BibitemOpen
  \bibfield  {author} {\bibinfo {author} {\bibfnamefont {R.~G.~M.}\
  \bibnamefont {van~der Sman}}\ and\ \bibinfo {author} {\bibfnamefont {H.~M.}\
  \bibnamefont {Vollebregt}},\ }\href@noop {} {\bibfield  {journal} {\bibinfo
  {journal} {J. Membr. Sci.}\ }\textbf {\bibinfo {volume} {435}},\ \bibinfo
  {pages} {21} (\bibinfo {year} {2013})}\BibitemShut {NoStop}%
\bibitem [{\citenamefont {van~der Sman}\ and\ \citenamefont
  {Vollebregt}(2012)}]{vanderSman:2012df}%
  \BibitemOpen
  \bibfield  {author} {\bibinfo {author} {\bibfnamefont {R.~G.~M.}\
  \bibnamefont {van~der Sman}}\ and\ \bibinfo {author} {\bibfnamefont {H.~M.}\
  \bibnamefont {Vollebregt}},\ }\href@noop {} {\bibfield  {journal} {\bibinfo
  {journal} {Adv. Colloid Interface Sci.}\ }\textbf {\bibinfo {volume}
  {185-186}},\ \bibinfo {pages} {1} (\bibinfo {year} {2012})}\BibitemShut
  {NoStop}%
\bibitem [{\citenamefont {Wang}\ \emph {et~al.}(2015)\citenamefont {Wang},
  \citenamefont {Heinen},\ and\ \citenamefont {Brady}}]{Wang:2015gf}%
  \BibitemOpen
  \bibfield  {author} {\bibinfo {author} {\bibfnamefont {M.}~\bibnamefont
  {Wang}}, \bibinfo {author} {\bibfnamefont {M.}~\bibnamefont {Heinen}}, \ and\
  \bibinfo {author} {\bibfnamefont {J.~F.}\ \bibnamefont {Brady}},\ }\href@noop
  {} {\bibfield  {journal} {\bibinfo  {journal} {J. Chem. Phys.}\ }\textbf
  {\bibinfo {volume} {142}},\ \bibinfo {pages} {064905} (\bibinfo {year}
  {2015})}\BibitemShut {NoStop}%
\bibitem [{\citenamefont {Bowen}\ and\ \citenamefont
  {Jenner}(1995{\natexlab{b}})}]{Bowen:1995ft}%
  \BibitemOpen
  \bibfield  {author} {\bibinfo {author} {\bibfnamefont {W.~R.}\ \bibnamefont
  {Bowen}}\ and\ \bibinfo {author} {\bibfnamefont {F.}~\bibnamefont {Jenner}},\
  }\href@noop {} {\bibfield  {journal} {\bibinfo  {journal} {Chem. Eng. Sci.}\
  }\textbf {\bibinfo {volume} {50}},\ \bibinfo {pages} {1707} (\bibinfo {year}
  {1995}{\natexlab{b}})}\BibitemShut {NoStop}%
\bibitem [{\citenamefont {Kim}\ \emph {et~al.}(2006)\citenamefont {Kim},
  \citenamefont {Marion}, \citenamefont {Jeong},\ and\ \citenamefont
  {Hoek}}]{Kim:2006en}%
  \BibitemOpen
  \bibfield  {author} {\bibinfo {author} {\bibfnamefont {S.}~\bibnamefont
  {Kim}}, \bibinfo {author} {\bibfnamefont {M.}~\bibnamefont {Marion}},
  \bibinfo {author} {\bibfnamefont {B.-H.}\ \bibnamefont {Jeong}}, \ and\
  \bibinfo {author} {\bibfnamefont {E.~M.~V.}\ \bibnamefont {Hoek}},\
  }\href@noop {} {\bibfield  {journal} {\bibinfo  {journal} {J. Membr. Sci.}\
  }\textbf {\bibinfo {volume} {284}},\ \bibinfo {pages} {361} (\bibinfo {year}
  {2006})}\BibitemShut {NoStop}%
\bibitem [{\citenamefont {Pelaez-Fernandez}\ \emph {et~al.}(2015)\citenamefont
  {Pelaez-Fernandez}, \citenamefont {Souslov}, \citenamefont {Lyon},
  \citenamefont {Goldbart},\ and\ \citenamefont
  {Fernandez-Nieves}}]{PelaezFernandez:2015hu}%
  \BibitemOpen
  \bibfield  {author} {\bibinfo {author} {\bibfnamefont {M.}~\bibnamefont
  {Pelaez-Fernandez}}, \bibinfo {author} {\bibfnamefont {A.}~\bibnamefont
  {Souslov}}, \bibinfo {author} {\bibfnamefont {L.~A.}\ \bibnamefont {Lyon}},
  \bibinfo {author} {\bibfnamefont {P.~M.}\ \bibnamefont {Goldbart}}, \ and\
  \bibinfo {author} {\bibfnamefont {A.}~\bibnamefont {Fernandez-Nieves}},\
  }\href@noop {} {\bibfield  {journal} {\bibinfo  {journal} {Phys. Rev. Lett.}\
  }\textbf {\bibinfo {volume} {114}},\ \bibinfo {pages} {098303} (\bibinfo
  {year} {2015})}\BibitemShut {NoStop}%
\bibitem [{\citenamefont {Buzatu}\ \emph {et~al.}()\citenamefont {Buzatu},
  \citenamefont {Roa}, \citenamefont {Vie{\ss}}, \citenamefont {K{\"u}ppers},
  \citenamefont {Bl{\"u}mich}, \citenamefont {Dhont}, \citenamefont
  {N{\"a}gele},\ and\ \citenamefont {Wessling}}]{SFB985B6}%
  \BibitemOpen
  \bibfield  {author} {\bibinfo {author} {\bibfnamefont {P.}~\bibnamefont
  {Buzatu}}, \bibinfo {author} {\bibfnamefont {R.}~\bibnamefont {Roa}},
  \bibinfo {author} {\bibfnamefont {J.}~\bibnamefont {Vie{\ss}}}, \bibinfo
  {author} {\bibfnamefont {M.}~\bibnamefont {K{\"u}ppers}}, \bibinfo {author}
  {\bibfnamefont {B.}~\bibnamefont {Bl{\"u}mich}}, \bibinfo {author}
  {\bibfnamefont {J.~K.~G.}\ \bibnamefont {Dhont}}, \bibinfo {author}
  {\bibfnamefont {G.}~\bibnamefont {N{\"a}gele}}, \ and\ \bibinfo {author}
  {\bibfnamefont {M.}~\bibnamefont {Wessling}},\ }\href@noop {} {\bibinfo
  {journal} {work in progress}\ }\BibitemShut {NoStop}%
\end{thebibliography}%


\newpage\phantom{new}
\newpage

\setcounter{page}{1}
\renewcommand{\thepage}{S-\arabic{page}}

\appendix*
\setcounter{equation}{0}
\renewcommand{\theequation}{S.\arabic{equation}}
\setcounter{figure}{0}
\renewcommand{\thefigure}{S\arabic{figure}}

\section*{Supplementary Information: \\ process indicators}

\begin{figure}[h!]
\vspace{0.5cm}
\begin{center}
\includegraphics[width=1\linewidth]{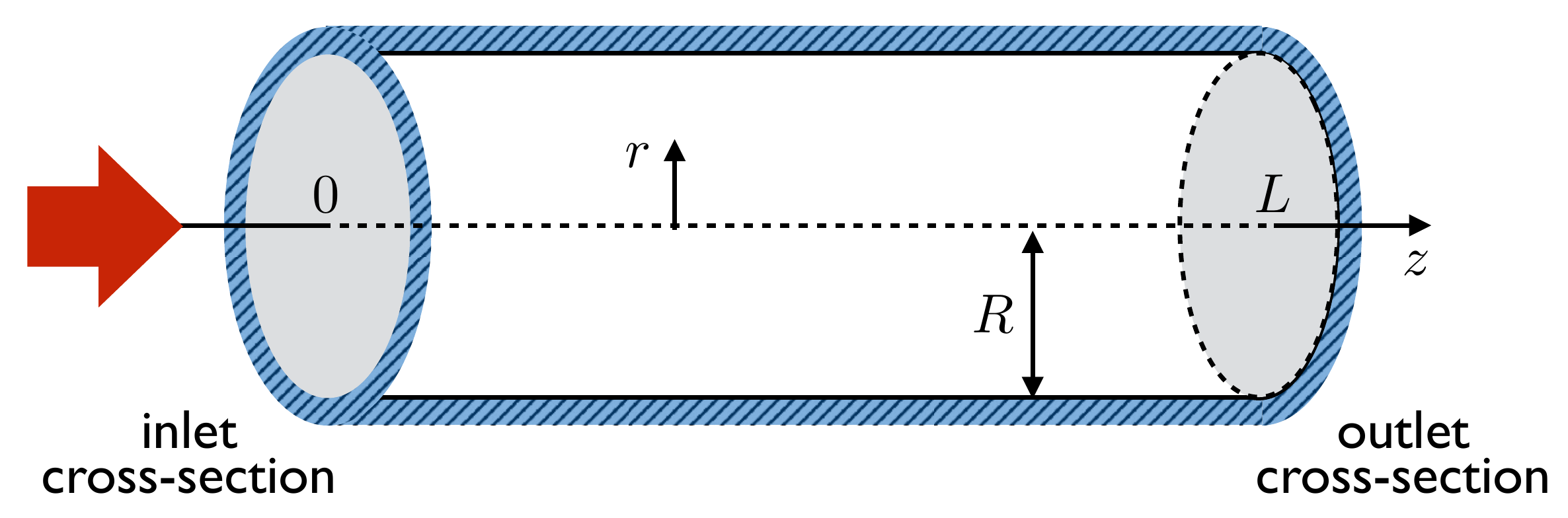}
\caption{Global cylindrical coordinate system $(r,z)$. The $z$-coordinate line extends along the axis of the hollow cylindrical fiber membrane of length $L$ and inner radius $R$.}
\label{figS1}
\end{center}
\end{figure}

We discuss here in more detail the process indicators introduced in Subsec. \ref{sec:indicators}. 
These are functions of the adjustable operation parameters and here in particular of the volume fraction, $\phi_0$, of colloidal particles in the homogeneous feed dispersion. For inside-out cross-flow UF in a hollow cylindrical fiber membrane (see Fig. \ref{figS1}) the indicators are obtained from the calculated spatial distributions of the local particle volume fraction, $\phi(r,z)$, and the dispersion velocity, 
\begin{equation}\label{eqa1}
\mathbf{v}(r,z)=u_r(r,z)\;\!{\bf e}_r + u_z(r,z)\;\!{\bf e}_z \;\!,
\end{equation}
expressed in cylindrical coordinates, $(r,z)$, with the radial velocity component $u_r$ along the unit vector ${\bf e}_r$ and the axial velocity component $u_z$ along the unit vector ${\bf e}_z$. In this global coordinate frame, the velocity components at the inner membrane surface, $r=R$, are $u_r(R,z)= u(x,y=0)$ and $u_z(R,z)=v(x,y=0)=-v_w(x)$ for $x=z$, where $u(x,y)$ and $v(x,y)$ are the longitudinal and axial velocity components measured, respectively, in the membrane surface anchored local coordinate frame used in our CP layer analysis (see Eq. (\ref{localvelocity}) and Fig. \ref{fig2}).

\subsection{Final product efficiency}

The final product is identified in the present work with the retentate dispersion of mean volume concentration $\phi_f$, flowing through the outlet cross-section at $z=L$ into a final reservoir. The particle advective flux through the outlet cross-section is $\int_S dS\;\!\phi(r,z)u_z(r,L)$ where $\int_SdS = 2\pi \int_0^R r dr$. Hence, $\phi_f$ is given by the ratio of the longitudinal particle advective to the longitudinal dispersion volume flux,
\begin{equation}\label{eqa2}
\phi_f=\frac{\int_S dS\;\!\phi(r,L)u_z(r,L)}{\int_SdS\;\! u_z(r,L)} \;\!,
\end{equation}
where we have neglected the very small longitudinal diffusion flux contribution.  

For a steady-state UF process where $\nabla\cdot {\bf J}=0$ is valid according to Eq. (\ref{continuity0}) 
and for a fully particle retentive membrane where Eq. (\ref{bcnoflux}) applies to, the longitudinal particle fluxes through different fiber cross-sections should be equal. In particular, 
\begin{equation}\label{eqa3}
\int_S dS\;\!\phi(r,L)u_z(r,L)=\phi_0\int_S dS\;\!u_z(r,0) 
\end{equation}
where we have ignored once again the small longitudinal diffusion flux contribution. The right-hand-side of Eq. (\ref{eqa3}) is the inlet particle flux of the feed dispersion of uniform volume fraction $\phi(r,0)=\phi_0$. By combining Eqs. (\ref{eqa2}) and (\ref{eqa3}), the expression
\begin{equation}\label{eqa4}
\alpha=\frac{\phi_f}{\phi_0}=\frac{\int_SdS\;\!u_z(r,0)}{\int_SdS\;\!u_z(r,L)}
\end{equation}
for the Degree of Concentration factor $\alpha$ is obtained, equal to the ratio of inlet to outlet dispersion volume flux. 

Another expression for $\alpha$, particularly suitable for the presented boundary layer analysis, is obtained from dispersion volume conservation ($\nabla \cdot {\bf v}=0$), 
\begin{eqnarray}\label{eqa5}
\int_S dS\;\!u_z(r,0)  &=& \int_S dS\;\!u_z(r,L) \nonumber \\
  && + \;2\pi R \int_0^L\!\!dz\;\!u_r(R,z) \;\!,
\end{eqnarray}
stating that the volume inflow through the inlet cross-section is equal to the sum of the volume outflows through the outlet cross-section and the cylindrical membrane of area $2\pi R L$. The combination of Eqs. (\ref{eqa4}) and (\ref{eqa5}) yields
\begin{equation}\label{eqa6}
\alpha=\frac{1}{1-\beta} \;\!,
\end{equation}
with the Solvent Recovery indicator, 
\begin{equation}\label{eqa7}
\beta=\frac{2\pi R\int_0^Ldz\;\!u_r(R,z)}{\int_S dS\;\!u_z(r,0)} \;\!,
\end{equation}
equal to the fraction of initial dispersion volume recovered in the permeate compartment. Using $u_r(R,z)=-v_w(x)$ together with Eq. (\ref{vwaverage}) for $\langle v_w \rangle $ and the cylindrical Pouiseuille flow profile for $u_z(r,0)$, the Eq. (\ref{beta}) for $\beta$ in the main text is obtained. 

The logical way of calculating the final product efficiency indicators is to determine first $\beta$ using Eq. (\ref{eqa7}), with $\alpha$ and $\phi_f$ determined subsequently using Eq. (\ref{eqa6}).

\subsection{Productivity per unit membrane area}

The indicator of Productivity per Unit Membrane Area, $\theta$, is defined here as the retentate flux divided by the membrane area,
\begin{equation}\label{eqa14}
\theta=\frac{\int_S dS\;\!u_z(r,L)}{2\pi R L} \;\!,
\end{equation}
giving this indicator the dimension of a velocity.  
Using Eqs. (\ref{eqa4})-(\ref{eqa6}), $\theta$ can be related to $\beta$ by
\begin{equation}\label{eqa15}
\theta=\frac{\int_S dS\;\!u_z(r,0)}{(1-\beta)2\pi R L} = 
\frac{\int_0^Ldz\;\!u_r(R,z)}{\beta(1-\beta)L} \;\!.
\end{equation}

The integral in the second equality is proportional to the fiber-length-averaged permeate velocity. Using Eqs. (\ref{vwaverage}) and (\ref{alpha}) in the previous expression, Eq. (\ref{theta}) in the main text is obtained.

\subsection{Energy cost}

We discuss first the indicator of Specific Energy Consumption, $\omega$, defined as the energy consumed to produce a unit volume of final product which, in our case, is the retentate dispersion of volume fraction $\phi_f$. Thus, for a steady-state process, $\omega$ is equal to the ratio of consumed power to the outlet volume per unit time. Assuming that basically the whole external power is spent in pressing the solvent through the membrane, this implies
\begin{equation}\label{eqa8}
\omega=\frac{2\pi R \int_0^Ldz\;\!\Delta P(z) u_r(R,z)}{\int_S dS\;\!u_z(r,L)} \;\!,
\end{equation}
where $\Delta P(z)$ is the local transmembrane pressure difference which in general depends on the axial distance $z$ from the inlet. In our calculations, the TMP has been taken as constant. Note that the physical dimension of $\omega$ is energy per volume.   

In general, one can expect that variations of $\Delta P(z)$ along the hollow fiber membrane are much smaller than the mean pressure value, denoted here as $\Delta P$, so that $\Delta P(z)$ approximated by $\Delta P$ can be shifted out of the integral in Eq. (\ref{eqa8}). Consequently, by using in addition Eqs. (\ref{eqa4}) and (\ref{eqa6}), Eq. (\ref{omega}) in the main text is obtained wherein $\omega$ is expressed by the product of the fiber-length-averaged TMP $\Delta P$ and $(\alpha-1)$.

As discussed in Subsec. \ref{sec:indicators}, the Specific Energy Efficiency indicator, $\epsilon$, is defined as the ratio of the thermodynamically necessary minimal reverse-osmosis compression work, $\omega_\mathrm{min}$, required to produce a unity of retentate dispersion to the Specific Energy Consumption $\omega$. According to Eq. (\ref{epsilon}), $\epsilon$ can be expressed in terms of $\phi_0$, $\Delta P$, and osmotic pressure $\Pi(z)$ along the membrane surface as  
\begin{equation}\label{eqa10}
\epsilon = \frac{\alpha\;\!\phi_0}{\left(\alpha-1\right)\Delta P}\int_{\phi_0}^{\phi_f}
d\phi\;\!\left(\frac{\Pi(\phi)}{\phi^2}\right) \;\!.
\end{equation}

Substituting $\Pi(\phi)$ as described by the CS Eq. (\ref{Zcarnahanstarling}) for colloidal hard spheres into Eq. (\ref{eqa10}), we obtain the analytic expression
\begin{equation}\label{epsilonexplicit}
\epsilon = \frac{3\;\!k_BT}{4\pi a^3}\frac{\alpha\;\!\phi_0}{(\alpha-1)\;\!\Delta P} \left[ 
\ln\left(\alpha\right) +\frac{3 - 2\phi_f}{\left(1-\phi_f\right)^2} - \frac{3 -2 \phi_0}{\left(1 - \phi_0 \right)^2}
\right]
\end{equation}
\newline
for $\epsilon$ in terms of the input parameters $\phi_0$, $\Delta P$, and hard-core radius $a$, and also in terms of $\alpha=\phi_f/\phi_0$ which must be determined prior to $\epsilon$. The factor to the left of the bracket in Eq. (\ref{epsilonexplicit}) can be recast into the suggestive form $\left(k_B T n_b/\Delta P\right)\left(\alpha/(\alpha-1)\right)$ invoking the ratio of the van't Hoff pressure $k_B T n_b$ and the TMP, where $n_b$ is the number density of the injected feed dispersion. Note further the explicit $1/a^3$ dependence of $\epsilon$ which renders this indicator very small for larger colloidal particles where the osmotic pressure buildup along the membrane is small.

\end{document}